\documentclass[a4paper,11pt]{article}

\usepackage{classeJBArxive}

\newcommand{\jqUnit}{\ \bigl[\,\mathsf{W.\,m^{\,-2}}\,\bigr]}

\newcommand{\rhoUnit}{\ \bigl[\,\mathsf{kg.\,m^{\,-3}}\,\bigr]}

\newcommand{\DthetaUnit}{\ \bigl[\,\mathsf{m^{\,2}.\,s^{\,-1}}\,\bigr]} 
\newcommand{\DTUnit}{\ \bigl[\,\mathsf{m^{\,2}.\,s^{\,-1}.\,K^{\,-1}}\,\bigr]}

\newcommand{\kQUnit}{\ \bigl[\,\mathsf{W.\,m^{\,-1}.\,K^{\,-1}}\,\bigr]} 
\newcommand{\kTMUnit}{\ \bigl[\,\mathsf{kg.\,m^{\,-1}.\,s^{\,-1}}\,\bigr]} 

\newcommand{\cQUnit}{\ \bigl[\,\mathsf{J.\,m^{\,-3}.\,K^{\,-1}}\,\bigr]} 
\newcommand{\cOUnit}{\ \bigl[\,\mathsf{J.\,kg^{\,-1}.\,K^{\,-1}}\,\bigr]} 
\newcommand{\hQUnit}{\ \bigl[\,\mathsf{W.\,m^{\,-2}.\,K^{\,-1}}\,\bigr]} 
\newcommand{\EUnit}{\ \bigl[\,\mathsf{J.\,m^{\,-2}}\,\bigr]}

\newcommand{\RUnit}{\ \bigl[\,\mathsf{J.\,kg^{\,-1}.\,K^{\,-1}}\,\bigr]} 
\newcommand{\hUnit}{\ \bigl[\,\mathsf{J.\,kg^{\,-1}}\,\bigr]}

\newcommand{\BiT}{\mathrm{Bi}_{\,T}}

\newcommand{\cTs}{c_{\,T}}

\newcommand{\Fo}{\mathrm{Fo}}
\newcommand{\FoT}{\mathrm{Fo}_{\,T}}
\newcommand{\FoM}{\mathrm{Fo}_{\,M}}

\newcommand{\kTs}{k_{\,T}}
\newcommand{\kTMs}{k_{\,TM}}
\newcommand{\ts}{t}\newcommand{\uinf}{u_{\,\infty}}
\newcommand{\uinfL}{u_{\,\infty}^{\,L}}
\newcommand{\uinfR}{u_{\,\infty}^{\,R}}

\newcommand{\xs}{x}

\usepackage{arydshln}
\title{\vspace{-1.5cm}
	Average reduced model to simulate solutions for heat and mass transfer through porous material. 
	\vspace{4pt}}

\author{Julien Berger\textsuperscript{a}$^{\ast}$, Madina Abdykarim\textsuperscript{b}} 
\date{\empty}

\begin{document}
	\maketitle
	\begin{center}
		\small
		\textsuperscript{a} Laboratoire des Sciences del’Ingénieur pour l’Environnement (LaSIE), UMR 7356 CNRS, La Rochelle Université, 17000 La Rochelle, France  \\
		\textsuperscript{b} Univ. Savoie Mont Blanc, CNRS, LOCIE, 73000 Chambéry, France \\
		$^{\ast}$Corresponding author, e-mail address : julien.berger@univ-lr.fr\\
		
	\end{center}
	\abstract{
		The design of numerical tools to model the behavior of building materials is a challenging task. The crucial point is to save computational cost and maintain high accuracy of predictions. There are two main limitations on the time scale choice, which put an obstacle to solve the above issues. First one is the numerical restriction. A number of research is dedicated to overcome this limitation and it is shown that it can be relaxed with innovative numerical schemes. The second one is the physical restriction. It is imposed by properties of a material, phenomena itself and corresponding boundary conditions. This work is focused on the study of a methodology that enables to overcome the physical restriction on the time grid. So-called Average Reduced Model (ARM) is suggested. It is based on smoothing the time-dependent boundary conditions. Besides, the approximate solution is decomposed into average and fluctuating components. The primer is obtained by integrating the equations over time, whereas the latter is an user-defined empirical model. The methodology is investigated for both heat diffusion and coupled heat and mass transfer. It is demonstrated that the signal core of the boundary conditions is preserved and the physical restriction can be relaxed. The model proved to be reliable, accurate and efficient also in comparison with the experimental data of two years.   		The implementation of the scarce time-step of $1 \, \sf{h}$ is justified. 
		It is shown, that by maintaining the tolerable error it is possible to cut computational effort up to almost four times in comparison with the complete model with the same time grid.
		} 

	\section{Introduction}\label{sec:Introduction}
Thermal performance of building materials is studied for more than half a century and still remains an actual topic. 
Apart from experimental approaches, which are costly and never at the full scale, another method to estimate such parameters is numerical simulations. 
Existing numerical tools are collected into various Building Performance Simulation (BPS) programs and the development of innovative programs is an ongoing process with constant improvements and alterations. 
Numerous studies are concentrated on evolution, moderation and replacement of classical numerical approaches \citep{mendes2017numerical, gasparin2019solving}.
The fidelity and the efficiency of a numerical method is crucial for a long-term modeling \citep{clark2010ancient, kavetski2010ancient}. 
In the recent review of state--of--the--art \citep{sasic2014multi, clarke2015vision, hong2018} it was indicated that despite wide range of programs, there are still some drawbacks in terms of accuracy, ease-of-use and the high computational cost. 

Building simulation programs are mostly based on several mathematical models. 
Among others, the heat and mass transfer in building porous materials is one of the most relevant to estimate the thermal performance \citep{hagentoft2004assessment, woloszyn2008tools}. 
In this model based on diffusion processes, the \textsc{Fourier} number is of major importance.
It characterizes the influence of a (heat or mass) diffusion transfer through the material. 
It is a key parameter in the numerical model. 
 
First of all, it is directly related to the stability of a numerical scheme when an explicit approach is implemented. 
Thereby, the numerical model based on a conditionally stable time scheme is restricted to a very fine time grid. The latter is defined by time instant where the solution requires to be computed to reach a defined accuracy and consistency.
An illustration of its importance is provided in \citep{Dos_santos_2004} for the model of interests. 
Nevertheless, it is possible to overcome this \emph{numerical} restriction. 
A wide variety of methods have been proposed in the field, such as the basic \textsc{Euler} implicit \citep{steeman2009coupled}, the \textsc{{Du\,Fort}--Frankel} \citep{gasparin2017improved, gasparin2017stable} and the Super--Time--Stepping (STS) \citep{alexiades1996, meyer2014, abdykarim2019} for recent examples. 
Note that implicit scheme do not have stability restrictions. 
However, due to the high nonlinearities of the problem and the requirement of sub-iterations to treat them, the computational costs is similar to explicit approaches.

Secondly, the \textsc{Fourier} number is also related to the characteristic time of the problem. 
It corresponds to the ratio between the characteristic time of the physical phenomena (of heat or mass transfer in the material) and the characteristic time of the investigations. 
The primer depends on the length of the material and its properties. 
The latter is set up by the user according to the horizon of simulation and the time scales of the boundary conditions. 
In building physics, the characteristic time of the investigations generally equals to one hour since the boundary climate variations are given for such time step. 
Therefore, to produce accurate predictions, the numerical time-step size of the simulation cannot be higher than one hour too. 
As demonstrated in \citep{gasparin2017stable}, if this \emph{physical} restriction is not respected, the predictions of the physical phenomena are not reliable.

Thus, as it is possible to overcome the \emph{numerical} restriction by employing innovative methods, it remains a so-called \emph{physical} restriction on the computational time grid. 
To overcome this restriction, the application of an Averaged Reduced Model (ARM) is investigated in the given article. 
The method is novel to the field and adopted from modelling in fluid dynamics. 
It is assumed that the ARM shall permit to implement the time grid bigger than the one imposed by the physical phenomena. 
In addition to that, it can save considerable amount of numerical effort to estimate parameters such as heat flux, conduction loads and thermal resistance.

A design of the model consists of three important stages. 
First, the boundary conditions are filtered and smoothed using a certain time-averaging period. 
In this way, it is assumed that the signal core will be preserved and the restriction on the characteristic time will be relaxed. 
Secondly, the approximate solution of the physical problem is decomposed in two parts with the average and fluctuating components. 
The averaging part is obtained by integrating the equations over the time period, whereas the fluctuating one is defined with an empirical model. The last stage consists in defining the empirical model. Candidates model are proposed which coefficient are determined through an identification procedure.

The article is organized as follows. 
The ARM first tested and implemented to heat diffusion equation in Section~\ref{sec:heat_eq}. 
The detailed description of the methodology is given in Section~\ref{sec:reduced_model_Heat}. 
After briefly presenting the numerical method and error metrics in Section~\ref{sec:num_methods}, the ARM is applied to the heat transfer case study in Section~\ref{sec:case_study_heat}. 
Thereafter, the methodology is tested on the coupled heat and mass transfer, described in Section~\ref{sec:HM_transfer}. 
Following the extensions of both average reduced and numerical models in Sections~\ref{sec:reduced_model_HM} and \ref{sec:num_methods_HM}, the case study is presented in Section~\ref{sec:case_study_HM}. 
Both offline and online procedures are discussed in details in Sections~\ref{sec:HM_empir_model} and \ref{sec:HM_results}. 
The ARM is further compared to experimental data, applied to predict the physical properties and advantages of the methodology is presented. 
The summary is made in Conclusion~\ref{sec:conclusion}.
\section{Heat transfer}\label{sec:heat_eq}
This section deals with effectiveness studies of the proposed methodology in the case of a heat transfer. 
First, a simple diffusion equation is described in a physical and then in a dimensionless forms. 
Obtained dimensionless mathematical model shall be called \emph{complete model}. 
Afterwards, the \emph{average reduced model} is described. 
\subsection{Complete Model}\label{sec:math_model}
Initial-boundary value problem, namely simple diffusion equation, in one dimension is considered. 
The equation is defined by the spacial $\Omega_{\, x} \egal [ \, 0, \, \ell \, ]$ and time $\Omega_{\, t} \egal [ \, 0, \, \mathds{T}\, ]$ domains, where $\ell\ \bigl[\,\mathsf{m}\,\bigr]$ is the thickness of a material and $\mathds{T} \ \bigl[\,\mathsf{h}\,\bigr]$ is the final time. 
\begin{equation}\label{eq:diffusion_equation}
c_{\, T} \cdot \pd{\, T}{\, t} \egal \pd{}{\, x} \, \Biggl( \, k_{\, T} \cdot \pd{\, T}{\, x}  \, \Biggr) \,,
\end{equation}
where $k_{\, T}$ $\kQUnit$ is the thermal conductivity, which is taken as the temperature dependent first order polynomial with constant coefficients $k_{\,0}$ and $k_{\,1}$:
\begin{equation*}
k_{\,T} \ : \ T \ \mapsto \ k_{\,0} \plus k_{\,1} \cdot T\, ,
\end{equation*}  
and $c_{\, T}$  $\cQUnit$ is the volumetric thermal capacity, which is also first order temperature dependent:
\begin{equation*}
c_{\,T} \ : \ T \ \mapsto \ c_{\,0} \plus c_{\,1} \cdot T\, , 
\end{equation*}
with constant coefficients $c_{\,0}$ and $c_{\,1}$. 
The initial condition is $T \, \bigl(\, x , \, t \egal 0\, \bigr) \egal T_{\, 0} \, \left(\, x\,\right)$. 
The \textsc{Robin}--type boundary conditions are given for $x \egal 0$ as:
\begin{equation*}\label{eq:diffusion_equation_LBC}
k_{\, T} \cdot \pd{\, T}{\, x} \egal h_{\, T}^{\, L} \cdot \Bigl( \, T \ - \ T_{\, \infty}^{\, L} \,  \Bigr) \moins \alpha \cdot g_{\, \infty}^{\, L} \, ,
\end{equation*}
and for $x \egal \ell $  as 
\begin{equation*}\label{eq:diffusion_equation_RBC}
k_{\, T} \cdot \pd{\, T}{\, x} \egal - \, h_{\, T}^{\, R} \cdot \Bigl( \, T \ - \ T_{\, \infty}^{\, R} \,  \Bigr)\, ,
\end{equation*}
where 
 $h_{\,T}^{\, L, \, R} \ \hQUnit$ is the surface heat transfer coefficient, 
 $\alpha \cdot g_{\, \infty}^{\, L} \ \jqUnit$ is the absorbed short-wave radiation,  
 $T_{\, \infty}^{\, L}$ $\bigl[\,^{\, \circ} \sf{C}\,\bigr] $ is the outside air temperature and  
$T_{\, \infty}^{\, R}$ $\bigl[\,^{\, \circ} \sf{C}\,\bigr] $ is the inside air temperature, given with periodic function in the range between $[\, 18 - 23^{\, \circ} \, \sf{C} \, ]$:
\begin{equation*}
T_{\,\infty}^{\, L} \ : \ \ts \ \mapsto \ T_{\,\infty}^{\, L} \, \left(\,\ts\,\right) \, , \qquad 
T_{\,\infty}^{\, R} \ : \ \ts \ \mapsto \ T_{\,\infty}^{\, R} \, \left(\,\ts\,\right) \, , \qquad
g_{\, \infty}^{\, L} \ : \ \ts \ \mapsto \ g_{\, \infty}^{\, L} \, \left(\,\ts\,\right) \, .
\end{equation*} 

In addition to the above, following two interesting outputs of the building physics framework are considered. 
The first one is the sensible heat flux $J_{\, q} \ \jqUnit$, which is defined as:
\begin{equation}\label{eq:heat_flux}
J_{\, q} \ \eqdef \, \moins k_{\, T} \cdot \pd{\, T}{\, x} \, \Bigr|_{\, x \egal L} \,.
\end{equation}
The second one is the conduction loads $E \ \EUnit$, which represents the sum of the heat fluxes at a building envelope internal surface and, in terms of the energy density, it can be evaluated as:
\begin{equation}
E \ \eqdef \ \int_{t_{\,1}}^{t_{\,2}} \, J_{\, q} \, \mathrm{d} \,t\, ,
\end{equation}
where $(\, t_{\,2} \moins t_{\,1}\, )$ is equal to one month. 

The governing equation~\eqref{eq:diffusion_equation} together with boundary conditions is solved numerically in a dimensionless form. 
The solution in the dimensionless formulation has advantages such as the application to a class of problems sharing the same scaling parameters (e.g. \textsc{Fourier} and \textsc{Biot} numbers) \citep{berger2017, trabelsi2018response}, simplification of a problem using asymptotic methods \citep{nayfeh2008perturbation} and restriction of round-off errors. 
For these purposes, the following dimensionless quantities are defined:
\begin{equation*}
u \ \eqdef \ \frac{T}{T^{\,\circ}}\,,  \qquad
\uinf \ \eqdef \ \frac{T_{\,\infty}}{T^{\,\circ}}\,, \qquad
g_{\, \infty}^{\,\star} \ \eqdef \ \frac{\ell \cdot g_{\, \infty}}{T^{\,\circ} \cdot \kTs^{\,\circ}}\,, 
\end{equation*}
where the superscripts $^{\,\circ}$ and $^{\,\star}$ represent a characteristic value of a dimensionless variable.
Dimensionless quantities of the space and time domains are:
\begin{equation*}
\begin{split}
\xs^{\,\star} &\ \eqdef \ \frac{\xs}{\ell}\,, 	   \ 
\end{split}
\quad \quad
\begin{split}
\ts^{\,\star} &\ \eqdef \  \frac{\ts}{\ts^{\,\circ}}\,. 	\ 
\end{split}
\end{equation*}
Material properties can be written in the dimensionless form as follows:
\begin{equation*}
\cTs^{\,\star} \ \eqdef \ \frac{\cTs}{\cTs^{\,\circ}}\,,   \qquad
\kTs^{\,\star} \ \eqdef \ \frac{\kTs}{\kTs^{\,\circ}}\,.  
\end{equation*}
The \textsc{Fourier} number is defined as:
\begin{equation*}
\FoT \ \eqdef \ \frac{t^{\,\circ} \cdot \, \kTs^{\,\circ}}{\ell^2 \cdot \, \cTs^{\,\circ}}\,. 	
\end{equation*}
The \textsc{Biot} numbers can be expressed as:
\begin{equation*}
\BiT^{\, L, \, R}\ \eqdef \  \frac{\ell \, \cdot \, h_{\,T}^{\, L, \, R} }{\kTs^{\,\circ}} \,.
\end{equation*}
As a result, the model~\eqref{eq:diffusion_equation} can be written in the dimensionless form for $x^{\, \star} \, \in \, \bigl[\,0 \,,\, 1\,\bigr]$ and $t^{\, \star} \, \in \, \bigl[\, 0 \,,\, \mathds{T} \,\bigr]$:
\begin{equation}\label{eq:dimless_diffusion_equation}
c_{\, T}^{\, \star} \cdot \pd{\, u}{\, t^{\, \star}} \egal \FoT \cdot \pd{}{\, x} \, \Biggl( \, \kTs^{\, \star} \cdot \pd{\, u}{\, x^{\, \star}}  \, \Biggr) \,,
\end{equation}
together with the initial condition $u \, \bigl(\, x^{\, \star}, \, 0\,\, \bigr) \egal u_{\, 0} \, (\, x^{\, \star}\,)$ and the following boundary conditions:
\begin{subequations}
	\label{eq:BC_dimless_model}
	\begin{align}
	\kTs^{\, \star} \cdot \pd{\, u}{\, x^{\, \star}} &\egal \BiT^{\, L} \cdot \left( \, u \ - \ \uinf^{\, L} \,  \right) \moins \alpha \cdot g_{\, \infty}^{\, \star} \, ,
 \\[4pt]
	\kTs^{\, \star} \cdot \pd{\, u}{\, x^{\, \star}} &\egal - \, \BiT^{\,R}\cdot \Bigl(\, u \moins \uinf^{\, R} \,\Bigr)\,. 	
	\end{align}
\end{subequations}
In the next section the \textit{average reduced model} is discussed. 
It starts with the general methodology followed by the derivation of a model itself and the description of an empirical model. 

\subsection{Average Reduced Model}\label{sec:reduced_model_Heat}
\begin{figure}[!ht]
	\centering
	\includegraphics[width=0.7\textwidth]{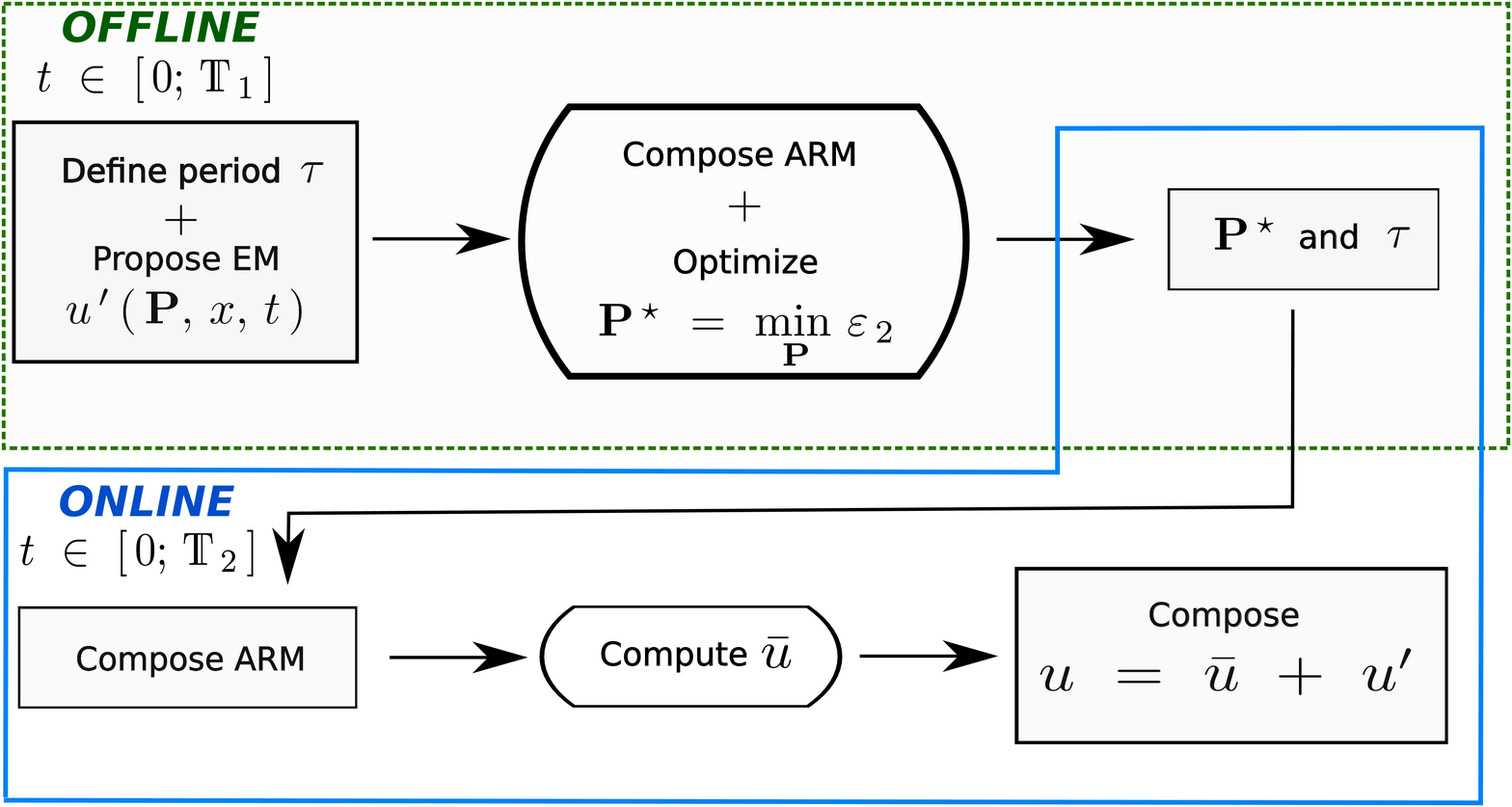}
	\caption{\small Schematic representation of offline and online procedures to build the ARM noting that $\mathds{T}_{\, 1} \ \ll \ \mathds{T}_{\, 2}$. }
	\label{fig:drawing_ARM}
\end{figure}

Average Reduced Model (ARM) is an \emph{a posteriori} method based on a preliminary learning procedure. 
The latter is carried on for given time-averaging periods $\tau$, and is denoted as \emph{offline}. 
This stage is performed for a representative time period $\mathds{T}_{\, 1}$ in order to cut a computational effort and to obtain parameters $\mathbf{P}^{\, \star}$ of an Empirical Model (EM). 
Later, those parameters $\mathbf{P}^{\, \star}$ are used for a full scale numerical investigations in an \emph{online} procedure for a wider time horizon $\mathds{T}_{\, 2}$. 
It should be noted that $\mathds{T}_{\, 1} \ \ll \ \mathds{T}_{\, 2}$ is necessary to attain efficiency of the method.  
The scheme of the methodology is presented on Figure~\ref{fig:drawing_ARM}.    

\subsubsection{Obtaining an ARM}\label{sec:obtaining_RM}
The idea is to relax the physical restriction on a time scale imposed by an hourly climatic data.
For this purpose, the solution $u \, \left(\,x, \, t\,\right) $ of the \emph{complete model} (CM), given in Eq.~\eqref{eq:dimless_diffusion_equation} is taken to be approximately equal to $\widetilde{u}\, \left(\,x, \, t\,\right) $ in a following average reduced form:
\begin{equation}\label{eq:decomposed_solution}
u \ \simeq \ \widetilde{u} \egal \overline{u} \plus u'\, ,
\end{equation}
where $\overline{u} \, \left(\,x, \, t\,\right) $ is the time-averaged mean value and $u' \, \left(\,x, \, t\,\right) $ is the fluctuating (high frequency) value \citep{argyropoulos2015recent}. 
The chosen averaging method takes the mean values at a fixed place in space and averaged over a time span, that is large enough for the mean values to be independent of it:
\begin{equation*}\label{eq:mean}
\overline{u} \ \eqdef \ \dfrac{1}{\tau} \, \int\limits_{0}^{\tau} u \, \left(\,x, \, t\,\right)  \, \mathrm{d} \, t\, ,
\end{equation*}
where $\overline{(\, \cdot \,)}$ is the time-averaging operator and $\tau$ is the chosen time interval or \emph{period}. 
The shape of fluctuating values $u'$ is discussed later. 
At this stage, it should be noted that the time-averaged values of $u'$ are defined to be zero:
\begin{equation}\label{eq:mean_u'}
\overline{u'} \ \eqdef \ 0 \, .
\end{equation}
The CM, given in Eq.~\eqref{eq:dimless_diffusion_equation} for $u$ is averaged as a reduced model for $\widetilde{u}$ in the following way with Eq.~\eqref{eq:decomposed_solution}:
\begin{equation}\label{eq:reduced_model_big}
\overline{c^{\, \star}\,\left(\,\overline{u} \plus u' \,\right) \cdot \pd{\, \left(\,\overline{u} \plus u' \,\right)}{\, t^{\, \star}}} \egal \overline{\Fo \cdot \pd{}{\, x} \, \Biggl( \, k^{\, \star}\, \left(\,\overline{u} \plus u' \,\right) \cdot \pd{\, \left(\,\overline{u} \plus u' \,\right)}{\, x^{\, \star}}  \, \Biggr)} \, .
\end{equation}
The dimensionless material properties $c^{\, \star} \, \left(\,u\,\right) $ and $k^{\, \star} \, \left(\,u\,\right)$ are the temperature dependent first order polynomials:
\begin{equation*}
c^{\, \star} \ : \ u \ \mapsto \ c^{\, \star}_{\,0} \plus c^{\, \star}_{\,1} \cdot u \, , \qquad
k^{\, \star} \ : \ u \ \mapsto \ k^{\, \star}_{\,0} \plus k^{\, \star}_{\,1} \cdot u \, ,
\end{equation*} 
where $c^{\, \star}_{\,0}, \, c^{\, \star}_{\,1}, \, k^{\, \star}_{\,0} \ \text{and} \ k^{\, \star}_{\,1} \, \in \mathds{R}$. 
By taking into account rules of time-averaging \citep{reynolds1895iv, monin1971statistical}, the above reduced model, given in Eq.~\eqref{eq:decomposed_solution} is transformed as shown below. 
Firstly, one can transform the left-hand-side of Eq.~\eqref{eq:decomposed_solution}:
\begin{align*}\label{eq:LHS}
\overline{c^{\, \star}\,\left(\,\overline{u} \plus u' \,\right) \cdot \pd{\, \left(\,\overline{u} \plus u' \,\right)}{\, t^{\, \star}}} 
\egal& 
\overline{\left(\, 1 \plus c^{\, \star}_{\,1} \cdot \overline{u} \plus c^{\, \star}_{\,1} \cdot u' \,\right) \cdot \left(\, \pd{\,\overline{u}}{\,t^{\, \star}} \plus \pd{\,u'}{\,t^{\, \star}}\,\right)} \, , \\\nonumber
\egal& 
c^{\, \star}\,\left(\,\overline{u}\,\right) \cdot \pd{\, \overline{u} }{\, t^{\, \star}} \plus c^{\, \star}_{\,1} \cdot \overline{ u'\cdot \pd{\,u'}{\,t^{\, \star}} } \, .
\end{align*}
Likewise, the right-hand-side can be transformed into the following form:
\begin{equation*}\label{eq:RHS}
\overline{\Fo \cdot \pd{}{\, x} \, \Biggl( \, k^{\, \star}\, \left(\,\overline{u} \plus u' \,\right) \cdot \pd{\, \left(\,\overline{u} \plus u' \,\right)}{\, x^{\, \star}}  \, \Biggr)} 
\egal 
\Fo \cdot \pd{}{\, x} \, \left(\,k^{\, \star}\, \left(\,\overline{u} \,\right) \cdot \pd{\,\overline{u}}{\, x^{\, \star}} \,\right) 
\plus 
\Fo \cdot \pd{}{\, x} \, \left(\, k^{\, \star}_{\,1} \cdot \overline{ u' \cdot \pd{\,u'}{\, x^{\, \star}}} \,\right) \,.
\end{equation*}
Hence, in general, the reduced model, given in Eq.~\eqref{eq:reduced_model_big} can be written as:
\begin{equation*}\label{eq:reduced_model_short}
c^{\, \star}\,\left(\,\overline{u}\,\right) \cdot \pd{\, \overline{u} }{\, t^{\, \star}} 
\egal 
\Fo \cdot \pd{}{\, x} \, \left(\,k^{\, \star}\, \left(\,\overline{u} \,\right) \cdot \pd{\,\overline{u}}{\, x^{\, \star}} \,\right) 
\plus 
\mathds{S} \,\left(\,x, \, t\,\right) \,,
\end{equation*} 
where 
\begin{align}\label{eq:S_term_full}
\mathds{S} \,\left(\,x, \, t\,\right) 
\ \eqdef & 
\moins  c^{\, \star}_{\,1} \cdot \overline{ u'\cdot \pd{\,u'}{\,t^{\, \star}} } 
\plus 
\Fo \cdot \pd{}{\, x} \, \left(\, k^{\, \star}_{\,1} \cdot \overline{ u' \cdot \pd{\,u'}{\, x^{\, \star}}} \,\right)  \\ \nonumber
\egal &
\dfrac{1}{\tau} \, \int\limits_{0}^{\tau} 
\Biggl(\, \moins c^{\, \star}_{\,1} \cdot u'\cdot \pd{\,u'}{\,t^{\, \star}} 
\plus 
\Fo \cdot \pd{}{\, x} \, \left(\, k^{\, \star}_{\,1} \cdot u' \cdot \pd{\,u'}{\, x^{\, \star}} \,\right) 
\,\Biggr) \ \mathrm{d} \, t^{\, \star} \, .
\end{align}
The boundary conditions can be written in the reduced form too: 
\begin{subequations}
	\label{eq:BC_reduced_model}
	\begin{align*}
	k^{\, \star} \,\left(\,\overline{u}\,\right) \cdot \pd{\,\overline{u}}{\, x^{\, \star}} &\egal 
	\BiT^{\, L} \cdot \left( \, \overline{u} \moins \overline{\uinf^{\, L}} \,  \right) \moins \alpha \cdot \overline{g_{\, \infty}^{\, \star} } 
	\moins \mathds{S}_{\, k} \,\left(\,x, \, t\,\right)  \, ,
	\\[4pt]
	k^{\, \star} \,\left(\,\overline{u}\,\right) \cdot \pd{\,\overline{u}}{\, x^{\, \star}} &\egal - \, \BiT^{\,R}\cdot \Bigl(\, \overline{u} \ - \ \overline{\uinf^{\, R}}  \,\Bigr) \moins \mathds{S}_{\, k} \,\left(\,x, \, t\,\right)\,,	
	\end{align*}
\end{subequations}
where 
\begin{equation}\label{eq:Sk_term_full}
\mathds{S}_{\, k} \,\left(\,x, \, t\,\right) \ \eqdef \ k^{\, \star}_{\,1} \cdot \overline{u' \cdot \pd{\,u'}{\, x^{\, \star}}} \, .
\end{equation}
Now, the \textit{average reduced model} is written and the next step is to propose a formulation of an \emph{empirical model} for the fluctuating values $u'$, which are unknown for the given case.

\subsubsection{Empirical Model for fluctuating values}\label{sec:Methodology_empir_model}
This section presents fluctuating values in a form of a mathematical model in a general form as:
\begin{equation}\label{eq:gen_empirical_model}
u^{\, \prime}  \ :\ \left(\,\mathbf{P}, \, x, \, t\,\right) \ \mapsto \ f \, \left(\,\mathbf{P}, \, x, \, t\,\right) \, ,
\end{equation}
where $\mathbf{P}$ is a set of parameters. 
The function $f \, \left(\, \cdot \, \right)$ should be designed in such a way, that the averaging of it should be in accordance with Eq.~\eqref{eq:mean_u'}:
\begin{equation*}
\overline{f \, \left(\, \cdot \, \right)} \egal 0 \, .
\end{equation*}
\paragraph{Optimization of coefficients.}
The goal is to solve the minimization problem $\mathbf{P}^{\, \star} \egal \min\limits_{\, \mathbf{P}} \, \varepsilon_{\, 2}$, where $\varepsilon_{\, 2} \ \eqdef \ \bigl|\bigl|\, \,u_{\, \text{ref}} \, (\,\xs,\,\ts\,) \moins \widetilde{u}\, (\,\xs,\,\ts\,) \, \bigr|\bigr|_{\, 2}$ is the error between simulation results $u \,\left(\,x, \, t\,\right)$ of complete model and $\widetilde{u}\,\left(\,x, \, t\,\right)$ of average reduced model. 
The optimization is carried out for coefficients $\mathbf{P}$ of an empirical model. As presented in \cite{Mohebbi_2017}, several approaches can be used to solve this inverse problem and retrieve the unknown coefficients of the empirical model. As starting investigations, the \textsc{Levenberg}--\textsc{Marquardt} method is used as nonlinear least-squares algorithms. The optimized values ensure the lowest error possible for the given empirical model and time-averaging period.
Based on the assumption that the high frequency fluctuating values $u^\prime$ have a periodic nature, the optimization can be made for a certain part of the simulation only. 
This is also done in order to avoid an extra computational time effort. 
One can then assume that the optimized parameters enable to obtain the results which are true for the whole time horizon. 

\subsection{Numerical Methods}\label{sec:num_methods}
	The space and time domain are discretized in the following way. 
	A uniform discretization of the space interval $\Omega_{\, x}  \rightsquigarrow \Omega_{\, h} $ is written as:
	\begin{equation*}
	\Omega_{\, h} \egal \bigcup_{\, j \egal 1}^{\, N_{\, x}} [\, x_{\, j}, \, x_{\, j \plus 1}\, ], \quad x_{\, j \plus 1} \moins x_{\, j} \ \equiv \ \Delta \, x, \, \, \forall \, j \, \in \, \{ \, 1, \ldots, N_{\, x}\, \} \,. 
	\end{equation*}
	Time layers are spaced uniformly as well $t^{\, n} \egal n \, \Delta \, t, \, \Delta \, t = \textnormal{const} \ > \ 0, \, \, \forall \, n \, \in \, \{ \, 0, \ldots, N_{\, t} \, \}\,. $
	The values of the solution function $u\,(\, x,\, t\, )$ are defined at discrete nodes and denoted by $u_{\,j}^{\,n} \ := \ u\, (\, x_{\,j},\, t^{\,n} \,)$. 
	\subsubsection{The Super--Time--Stepping Method}\label{sec:STS_Methods}
	In this article, only one Super--Time--Stepping (STS) approach is considered based on results of previous works \citep{abdykarim2019, abdykarim2019b}. 
	The method is based on a family of orthogonal shifted \textsc{Legendre} polynomials of the first order \citep{meyer2014}.
	One can express discretization of Eq.~\eqref{eq:dimless_diffusion_equation} in the following way for $n \egal 0\,, 1\,, \ldots\, N_{\, STS}$:
	\begin{equation}\label{eq:STS_general_heat_equation_discretisation}
	u^{\, n+1} \egal \Biggl( \, \mathsf{P}_{\, N} \bigl(\, \Delta \, t_{\,\mathrm{S}} \,, \mathds{A} \, \bigr) \, \Biggr) \cdot u^{\, n} \,,
	\end{equation}
	where $\mathsf{P}_{\, N_{\,\mathrm{S}}}$ is the stability polynomial, 
	$N_{\, \text{STS}} \eqdef \dfrac{\tau}{\Delta\, t_{\, \mathrm{S}}}$ is the number of approximate solutions and 
	matrix $\mathds{A}$ is constructed according to the chosen explicit space discretization. 
	The solutions are found with the scheme involving a so-called super-time-step $\Delta \, t_{\,\mathrm{S}}$, which should satisfy \textsc{Legendre} stability polynomial.
	Time-step $\Delta \, t_{\,\mathrm{S}}$ is chosen according to the number of super-time-steps $N_{\, \mathrm{S}}$ and explicit time-step $\Delta \, t_{\, \text{exp}} \eqdef \frac{2}{\lambda_{\, \max}}\,$, where $\lambda_{\,\max} \egal  \dfrac{4 \cdot \FoT}{\Delta \,x^{\, 2}}$ \citep{alexiades1996}. 
	
	Thereby for STS \textsc{Runge--Kutta--Legendre} (RKL) method of the first order the super-time-step should satisfy following condition: 
		\begin{equation*}
		\Delta \, t_{\, \mathrm{S}} \ \leqslant \ \frac{N_{\, \mathrm{S}}^{\, 2} \plus N_{\, \mathrm{S}}}{2} \cdot \Delta \, t_{\, \text{exp}}\,. 
		\label{eq:STS_RKL_delta_tS}
		\end{equation*}
	It can be seen that it is now possible to implement the time-step at least $\mathcal{O} \, (\,N_{\, \mathrm{S}}^{\, 2}\,)$ bigger than a time-step required by the explicit \textsc{Euler} scheme due to \textsc{Courant--Friedrichs--Lewy} (CFL) \citep{courant1967partial} stability condition. 
	This "widening" technique allows to perform much faster calculations and more details can be found in the previous works \citep{abdykarim2019, abdykarim2019b}.
\subsubsection{Error metrics}\label{sec:methodology}
One wants to examine the maximum of the root-mean-square error to quantify the discrepancies between two solutions. 
The $\mathcal{L}_{\,2}$ error between a solution  $\widetilde{u} \, \left(\,x, \, t\,\right)$ with average reduced model and a reference solution $u_{\, \text{ref}} \, \left(\,x, \, t\,\right)$ with complete model (or experimental data):
\begin{equation*}\label{eq:L2_error}
	\varepsilon_{\,2} \, (\,\xs\,)\ 
	\eqdef \ \sqrt{\dfrac{1}{N_{\, \ts}} \cdot \sum_{j\,=\,1}^{N_{\, \ts}} \Big(\,  u_{\,j,\, \text{ref}}\,(\,\xs \,,\ts_{\,j}\,) \moins \widetilde{u}_{\,j}\,(\,\xs \,,\ts_{\,j}\,) \,\Big)^{\, 2}  } \,,
\end{equation*}
where $N_{\, \ts}$ is the total number of temporal steps. 
The global uniform error $\mathcal{L}_{\,\infty}$ is defined as:
\begin{equation*}\label{eq:Linfty_error}
\varepsilon_{\,\infty} \ \eqdef \max_{\, \xs \, \in \,  \bigl[\, 0, \, \ell \, \bigr]} \varepsilon_{\,2} \, (\,\xs\,)\ .
\end{equation*}
The relative error $\eta_{\,2}$ $[\, \varnothing \, ]$ is used to compare the discrepancies of a result from a reference value, and defined as follows:
\begin{equation*}\label{eq:eta2_error}
	\eta_{\,2} \, (\,\xs\,) \egal \sqrt{\dfrac{1}{N_{\, \ts}} \cdot \sum_{j\,=\,1}^{N_{\, \ts}} \dfrac{ \Big(\,  u_{\,j,\, \text{ref}}\,(\,\xs \,,\ts_{\,j}\,) \moins \widetilde{u}_{\,j}\,(\,\xs \,,\ts_{\,j}\,) \,\Big)^{\, 2}}{\Delta \,  u_{\,j,\, \text{ref}}\,(\,\xs \,,\ts_{\,j}\,)}  } \,,
\end{equation*}
where $\Delta \,  u_{\, \text{ref}}$ is taken as the difference between the maximum and minimum values to avoid division by zero for some cases:
\begin{equation*}
\Delta \,  u_{\, \text{ref}}\,(\,\xs \,,\ts\,)\  \eqdef \ \max \, \left\{ \, u_{\, \text{ref}}\,(\,\xs \,,\ts\,) \, \right\} 
\moins \min\, \left\{ \, u_{\, \text{ref}}\,(\,\xs \,,\ts\,)\, \right\} \, .
\end{equation*}
The global relative error $\eta_{\,\infty} \ [\, \%\, ]$ is defined as:
\begin{equation*}\label{eq:eta_infty_error}
\eta_{\,\infty} \ \eqdef \ \max_{\, \xs \, \in \,  \bigl[\, 0, \,\ell \, \bigr]} \eta_{\,2} \, (\,\xs\,)\ .
\end{equation*}
In this article also the attention is paid to the difference in the computational (CPU) run time of numerical simulations. 
The ratio, $\varrho_{\, \text{CPU}} \ \bigl[\,\%\,\bigr]$, shows how computational cost can be shortened compared to the complete model (CM):
\begin{equation*}\label{eq:cpur}
\varrho_{\, \text{CPU}} \ \eqdef \ \dfrac{t_{\, \text{CPU}}^{\, \text{ARM}}}{t_{\, \text{CPU}}^{\, \textsc{CM}}} \cdot 100 \, \% \,,
\end{equation*}
where $t_{\, \text{CPU}}^{\, \text{ARM}} \ \bigl[\,\sf s\,\bigr]$ and $t_{\, \text{CPU}}^{\, \textsc{CM}} \ \bigl[\,\sf s\,\bigr]$ are computational times required by the ARM and by the CM respectively.
In order to evaluate how many seconds are required to perform the simulation for one astronomical day, the computational time ratio per day is computed as follows:
\begin{equation}\label{eq:varrho_CPUday}
\varrho_{\, \text{\tiny CPU}}^{\, \text{\tiny day}} \ \eqdef \ \dfrac{t_{\, \text{\tiny CPU}}^{\, \text{\tiny scheme}}}{\tau_{\, \sf d}} \ \bigl[\,\sf s/d\,\bigr] \, .
\end{equation} 
The CPU is measured on a computer with \texttt{Intel Core i7} and 16 GB of RAM, and the numerical models are run on the $\texttt{Matlab}^{\texttt{TM}}$ environment. 

\subsection{Case study}\label{sec:case_study_heat}
The case study below is for the heat transfer described in the Section~\ref{sec:heat_eq}.
It should be taken into account that this study is preliminary case before the heat and mass transfer study, which is discussed later in the  Section~\ref{sec:HM_transfer}.
\subsubsection{Presentation of the case study}
All necessary dimensionless parameters of the mathematical model are presented below.
\textsc{Fourier} number is equal to $\FoT  \egal 5.1 \, \cdot 10^{\, -2}.$
Dimensionless material properties are taken as follows:
\begin{equation*}
\cTs^{\, \star} \, \left(\, u \,\right) \egal 0.2 \plus 0.1 \cdot u \, , \qquad
\kTs^{\, \star} \, \left(\, u \,\right) \egal 0.051 \plus 0.01 \cdot u.
\end{equation*}
\textsc{Biot} numbers are expressed as parameters for the boundary conditions and are taken to be equal to:
\begin{equation*}
x \egal 0: \ \ \BiT^{\,T, \,L} \egal 2.65\,, \qquad
x \egal 1: \ \ \BiT^{\,T, \,R} \egal 1.42\,.
\end{equation*}
Additional flux parameter is taken as $\alpha \egal 0$. 
The left boundary data $\uinfL$ is taken from the weather records. 
Variation of the right boundary signal is set to obey the following periodic function:
\begin{equation*}
x  \egal 1: \ \  \uinfR \, \left(\, t^{\, \star} \,\right) \egal 1 \plus  1.4 \, \cdot 10^{\, -2} \cdot \, \sin \left( 2\pi \cdot \frac{\ts^{\, \star}}{1.92 \cdot 10^{\, 4}}\right) \,. 
\end{equation*}
The initial temperature is taken as a linear function over the material and its dimensionless formulation is given by:
\begin{equation*}
u_{\, 0} \, \left(\, x^{\, \star} \,\right) \egal  0.95  \plus 4.5 \, \cdot 10^{\, -2} \cdot x^{\, \star}\, .
\end{equation*}
The reference width of the material is $\ell^{\,\circ} \egal 20 \ \mathsf{cm}$. 
Other reference parameters are $T^{\,\circ} \egal 293.15 \ \mathsf{K}$, $t^{\,\circ} \egal 3600 \ \mathsf{s}$, $\cTs^{\,\circ} \egal 2 \times 10^{\, 6} \ \mathsf{J.\,m^{\,-3}.\,K^{\,-1}} $, $\kTs^{\,\circ} \egal 1.13 \ \mathsf{W.\,m^{\,-1}.\,K^{\,-1}}$. 
The space discretization parameter is $\Delta \, x^{\, \star} \egal 10^{\,-2}$ for all cases. 
\begin{figure}[!ht]
	\centering
	\includegraphics[width=0.5\textwidth]{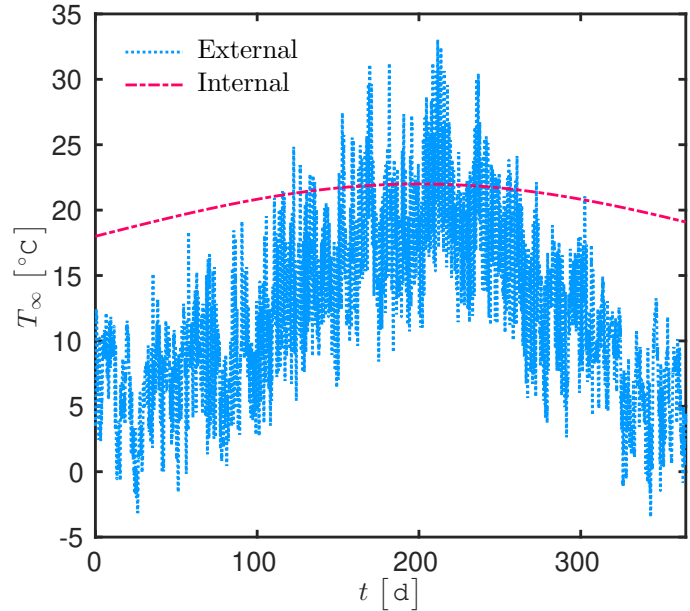}
	\caption{\small Boundary conditions for external and internal surfaces. }
	\label{fig:BC_T}
\end{figure}
\subsubsection{Candidates for an Empirical Model, offline procedure}\label{sec:Heat_empir_model}
Following the methodology presented in the Section~\ref{sec:Methodology_empir_model}, two models are proposed below. 
Both of them are potentially good fits for the fluctuating values.
The first model, called as Model I, is defined as:
\begin{equation}\label{eq:empirical_model_I}
u'  \ :\ \left(\,x, \, t\,\right) \ \mapsto \ \, u_{\, o} \cdot \Biggl(\, 1 \moins \dfrac{x}{\ell_{\, o}}\,\Biggr) \cdot \sin \,\Biggl(\,\frac{2 \pi}{\tau}  \cdot t\,\Biggr) \, ,
\end{equation}
where $u_{\, o}, \, \ell_{\, o} \, \in \,\mathds{R}$ are some coefficients to be estimated for each time-averaging period $\tau$.
Another possible way to model the fluctuating values $u^\prime$ is written as follows and dedicated as Model II:
\begin{equation}\label{eq:empirical_model_II}
u'  \ :\ \left(\,x, \, t\,\right) \ \mapsto \ e^{\moins \dfrac{\bigl( \, x \moins x_{\, o} \, \bigr)}{\ell_{\, o}} } \cdot \sin \,\Biggl(\,\frac{2 \pi}{\tau}  \cdot t\,\Biggr) \, ,
\end{equation}
where parameter $x_{\, o}\, \in \,\mathds{R}$ is also optimized according to the period $\tau$.
Given the empirical model, one can find values for the source terms $\mathds{S} \,\left(\,x, \, t\,\right)$, given in Eq.~\eqref{eq:S_term_full} and $\mathds{S}_{\, k} \,\left(\,x, \, t\,\right)$, given in Eq.~\eqref{eq:Sk_term_full} by substituting $u' \, \left(\,x, \, t\,\right)$ with the functions given in Eqs.~\eqref{eq:empirical_model_I} or \eqref{eq:empirical_model_II}. 

\begin{figure}[!ht]
	\centering
		\includegraphics[width=0.7\textwidth]{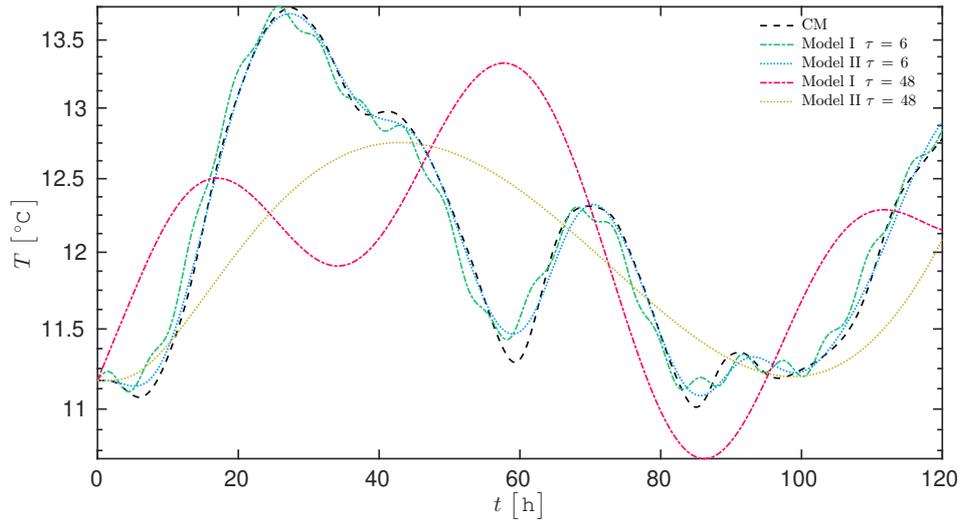}
	\caption{\small Time evolutions of the temperature with complete and average reduced models, results of which are obtained with empirical models I and II for $u^{\, \prime}$ for the simulation time $\mathds{T} \egal 120 \,\mathsf{h}$. }
	\label{fig:EM_T_120}
\end{figure}

Before evaluating the reliability of the average reduced model (ARM), it is important to choose the empirical model (EM) first. 
By following the logic proposed in the methodology of the ARM, one starts with the \emph{offline} procedure to determine the parameters to an optimal EM. 
The choice of EM can be made by comparing simulation results. 
As a reference solution one can take the results, obtained with the complete model (CM) and compare them to the two results of the average reduced model with different empirical models for fluctuating values $u^{\, \prime} \, \left(\,x, \, t\,\right)$. 
The optimization processes have been carried out for the simulation time $\mathds{T} \egal 120 \,\mathsf{h}$. 
For the sake of clarity, the parameters of both models are optimized for the time-averaging periods $\tau \egal [\,6, \, 12, \, 24, \, 48\,] \,\mathsf{h}$ only. 
The evolutions of the temperature for the representative simulation time with both CM and ARM with different empirical models are shown in Figure~\ref{fig:EM_T_120}. 
It can be seen that the results of ARM are closer for a smaller time-averaging period. 
Indeed, it is consistent since $\displaystyle \lim_{\tau \, \to \, 0} \, \tilde{u} \egal u$. 
It can also be noted that for $\tau \, \gg \, 1$ the model is out of the reality. 
The error and the optimized parameters for Model I and Model II are collected in Table~\ref{tab:parameters_T_period}. 
The obtained coefficients are then implied to empirical models and subsequently to the average reduced model. 
%
\begin{table}[!ht]                                                               
	\centering                                                               
	\begin{tabular}{|l|c|c|c|c|c|}  
		\cline{3-6}                                                                                                          
		\multicolumn{2}{c|}{} & $\tau \egal 6 \,\mathsf{h} $ & $\tau \egal 12 \,\mathsf{h} $ & $\tau \egal 24 \,\mathsf{h} $ & $\tau \egal 48 \,\mathsf{h}$ \\                                                  
		\hline\hline                                                                                                                                   
		& $x_{\, o}$ & $1.9 \times 10^{\, -4}$ & $ 1.7 \times 10^{\, -3}$ & $-5.4 \times 10^{\, -3}$ &$7.4 \times 10^{\, -5}$\\
		\cline{2-6}  
		\it{Model I} & $\ell_{\, o}$ & $7.4 \times 10^{\, -7}$ & $5.6 \times 10^{\, -3} $ &  $-1.2 \times 10^{\, -2}$& $-5.5 \times 10^{\, -3}$\\
		\cline{2-6}  
		&	$\varepsilon_{\, 2}$ & $\mathbf{6.87 \times 10^{\, -1}}$ & $\mathbf{1.47}$ &  $\mathbf{2.03}$& $\mathbf{2.97}$\\
		\hline\hline			
		& $x_{\, o}$ & $-9.5 \times 10^{\, -1}$ & $ -3.7 \times 10^{\, -1}$ & $-3.7 \times 10^{\, -1}$ &$-8.8 \times 10^{\, -1}$\\
		\cline{2-6}  
		\it{Model II} & $\ell_{\, o}$ & $1.7 \times 10^{\, 2}$ & $1.5 \times 10 $ &  $1.5 \times 10^{\, 2}$& $4.8 \times 10$\\
		\cline{2-6}  
		&	$\varepsilon_{\, 2}$ & $\mathbf{6.90 \times 10^{\, -1}}$ & $\mathbf{1.46}$ &  $\mathbf{2.67}$& $\mathbf{3.12}$\\
		\hline\hline	                                                        
	\end{tabular}   
	\bigskip  \caption{\small  Parameters of Model I, given in Eq.~\eqref{eq:empirical_model_I} and Model II, given in Eq.~\eqref{eq:empirical_model_II} in relation to time-averaging periods $\tau$ and $\varepsilon_{\, 2}$ error between simulation results $u \,\left(\,x, \, t\,\right)$ of complete model and $\widetilde{u}\,\left(\,x, \, t\,\right)$ of average reduced model.} 					
	\label{tab:parameters_T_period}                                                       
\end{table} 

\subsubsection{Results and discussion for online procedure}\label{sec:Results_heat}
\begin{figure}[!ht]
	\centering
	\subfigure[]{\includegraphics[width=.45\textwidth]{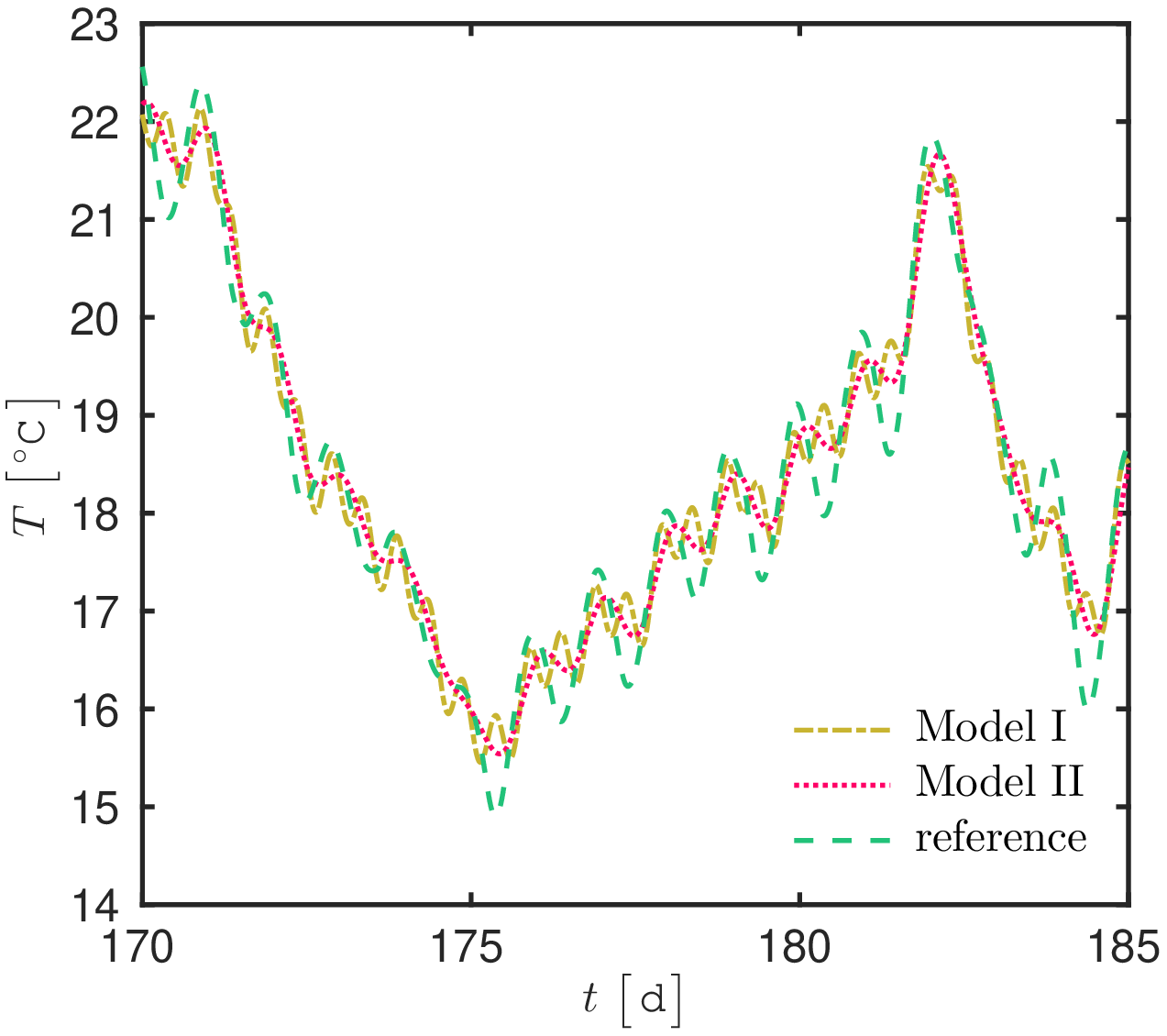}}\hspace{0.2cm}
	\subfigure[]{\includegraphics[width=.45\textwidth]{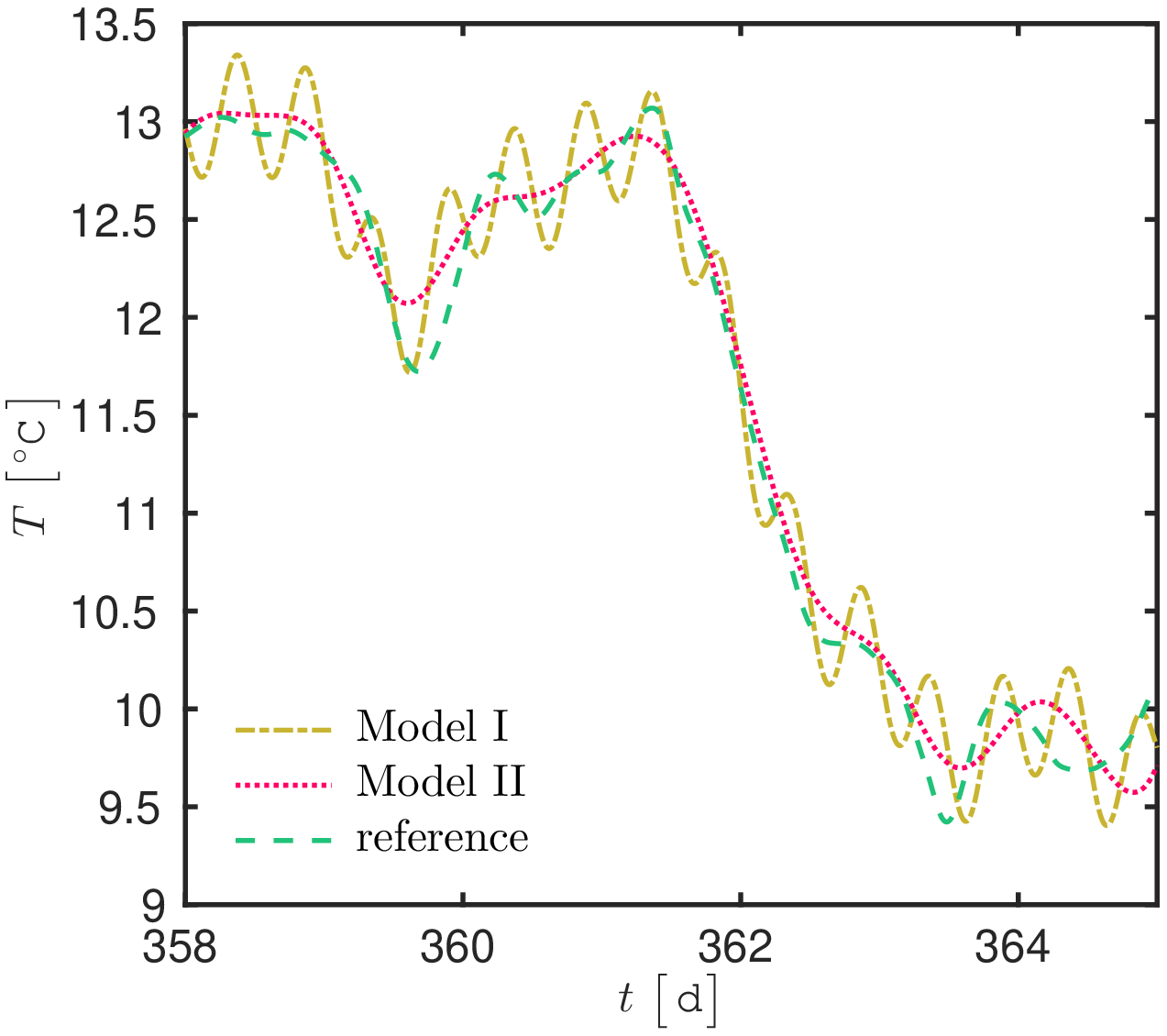}}\\
	\subfigure[]{\includegraphics[width=.45\textwidth]{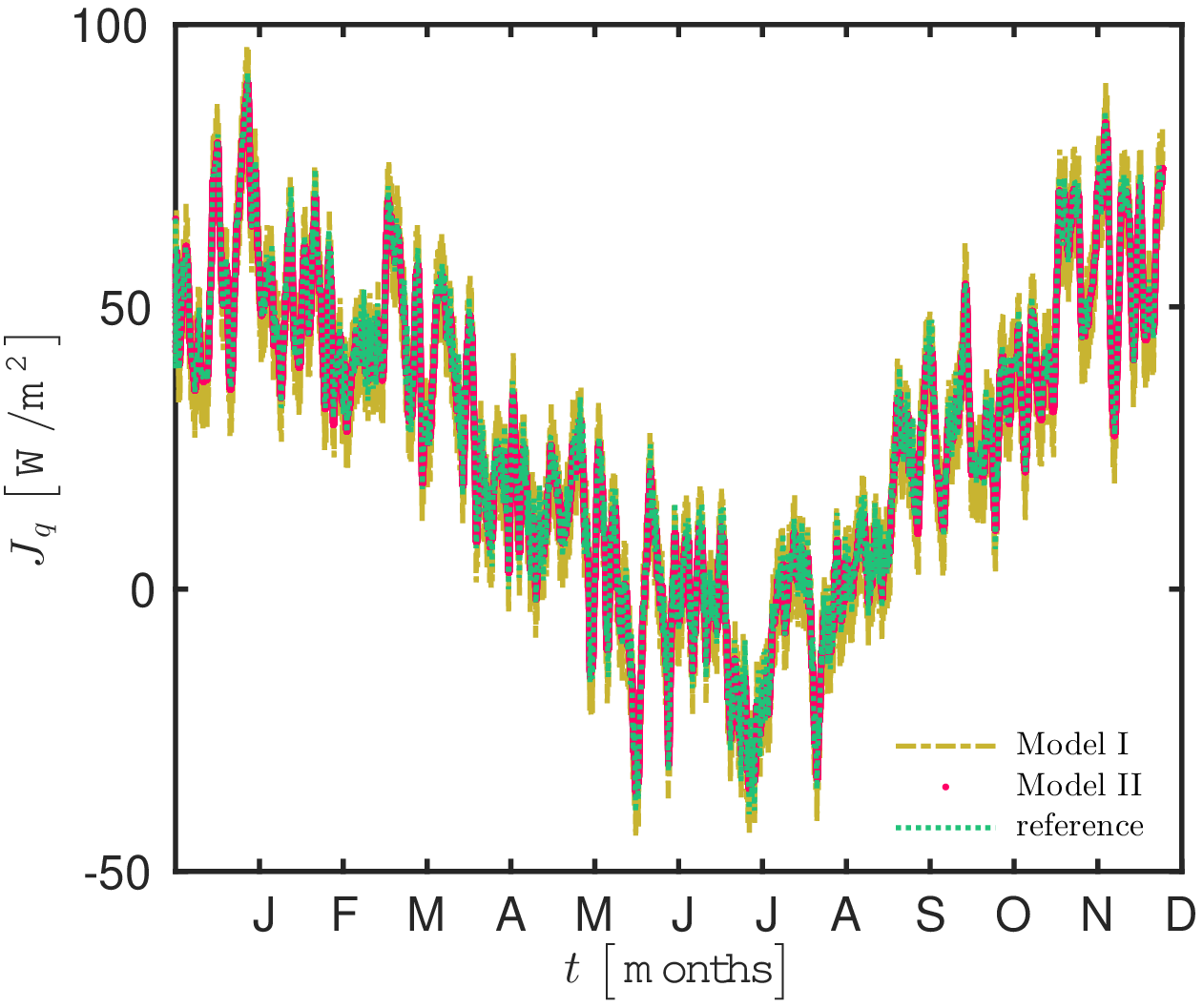}}\hspace{0.2cm}
	\subfigure[]{\includegraphics[width=.45\textwidth]{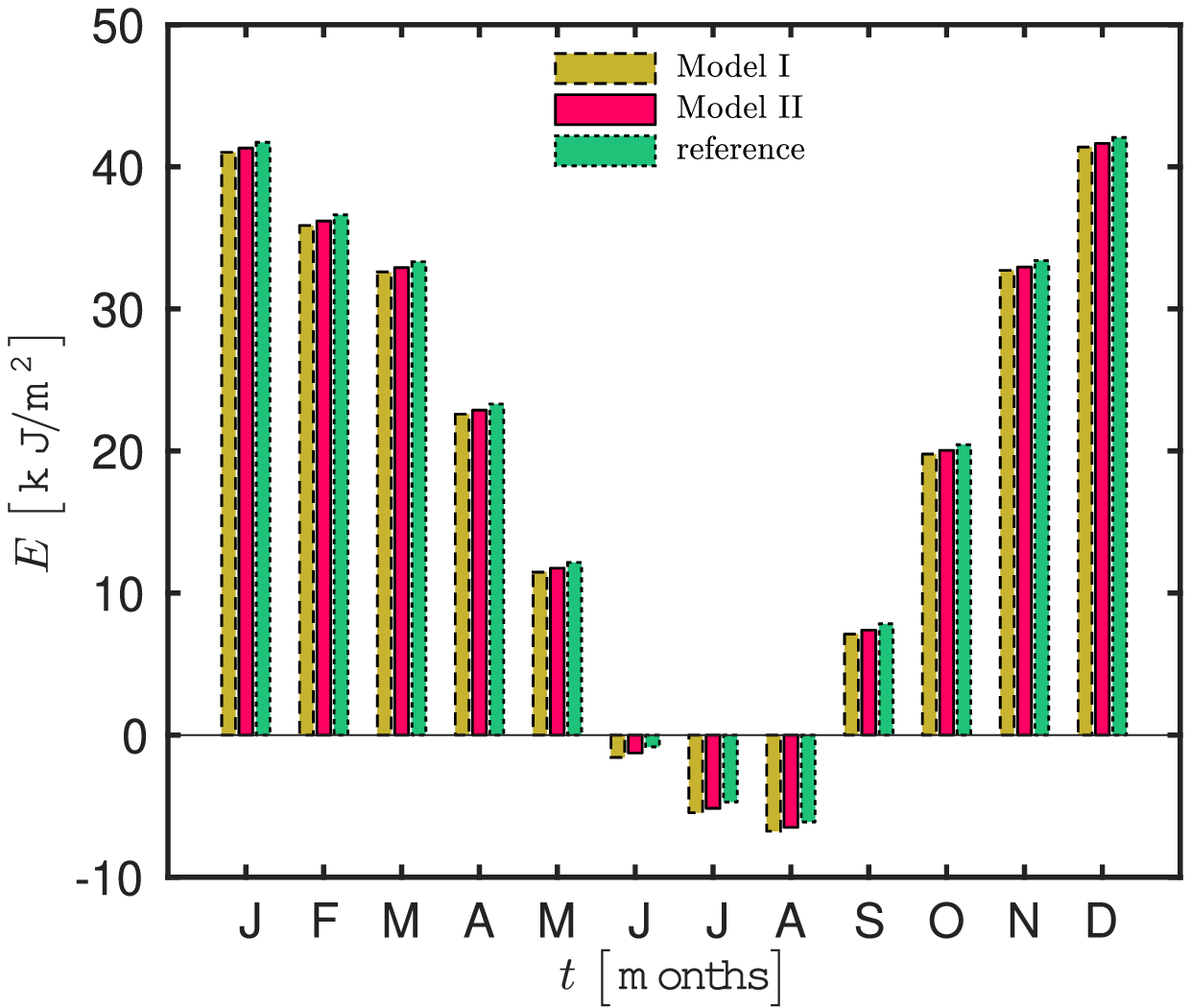}}
	\caption{\small ARM results obtained with empirical models I and II for $u^{\, \prime}$. 
		Comparison of the temperature evolutions is plotted for two weeks in the summer period (\emph{a}) and at the end of the simulation time (\emph{b}).
		The heat flux (\emph{c}) and the conduction loads (\emph{d}) computed with the ARM are presented with the complete model results as a reference solution. 
		Time-averaging period is $\tau \egal 12 \,\mathsf{h}$.}
	\label{fig:Heat_Empir_Model}
\end{figure}

\begin{figure}[!ht]
	\centering
	\subfigure[]{\includegraphics[width=.45\textwidth]{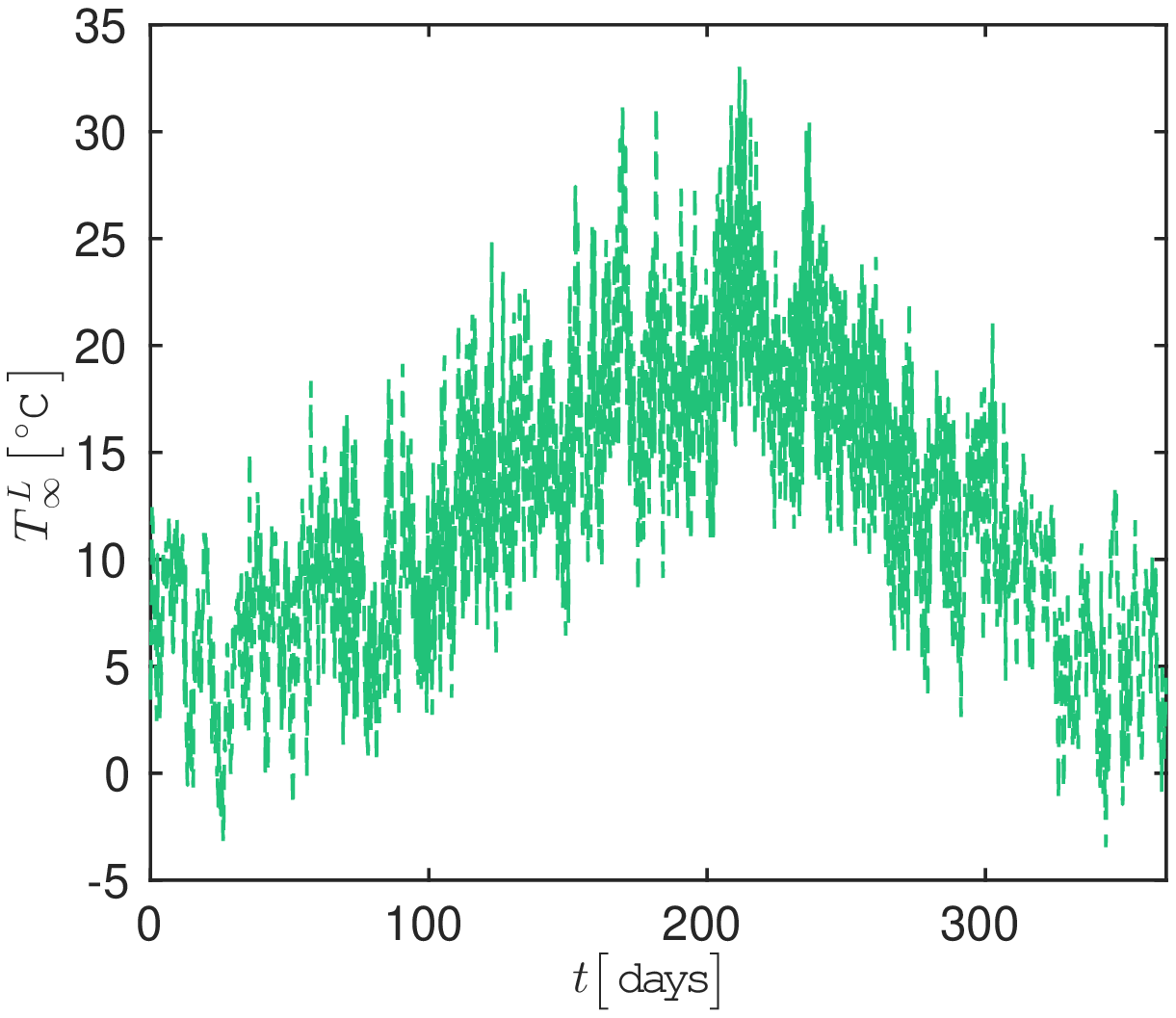}} \hspace{0.2cm}
	\subfigure[]{\includegraphics[width=.45\textwidth]{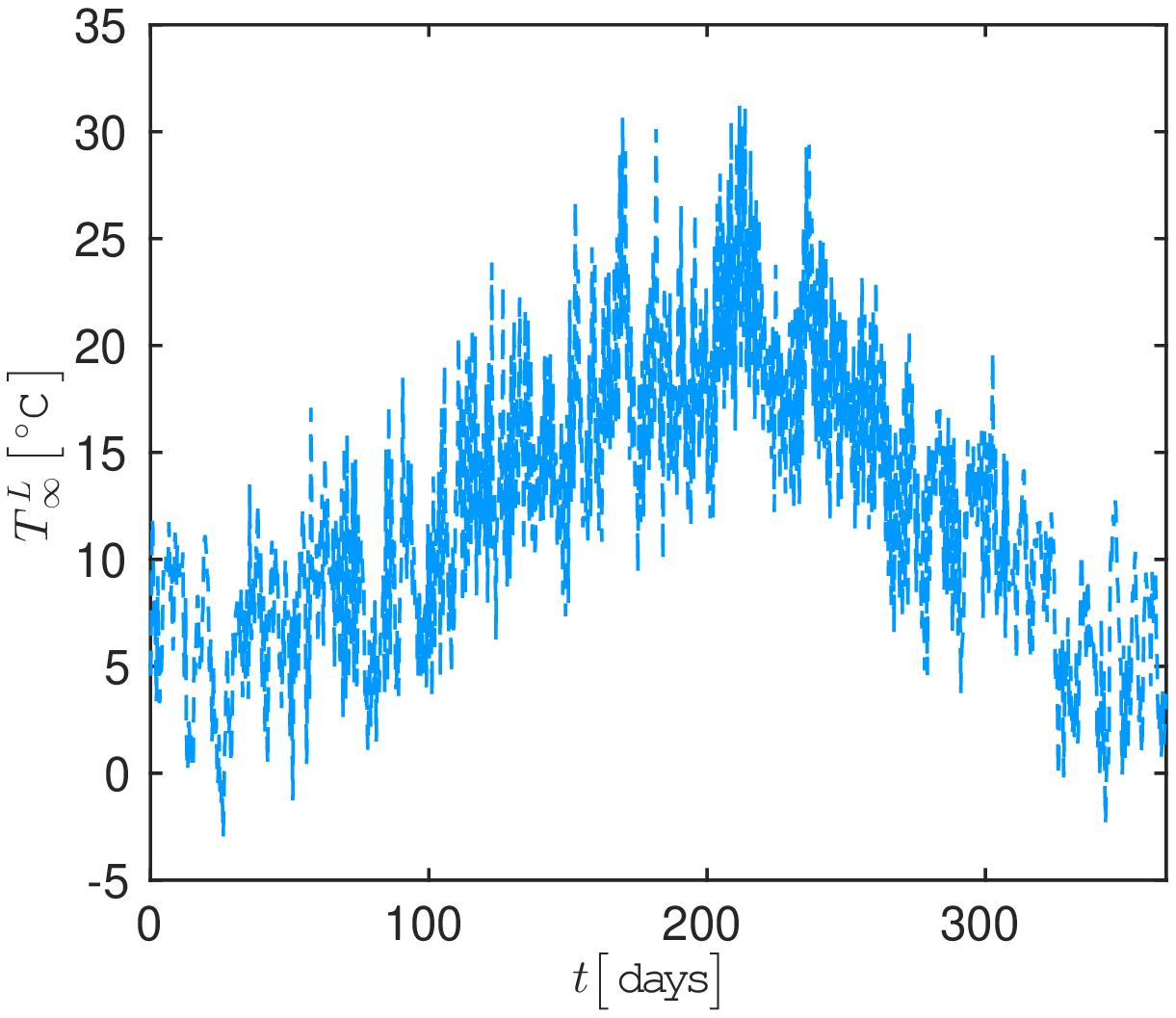}} \\
	\subfigure[]{\includegraphics[width=.45\textwidth]{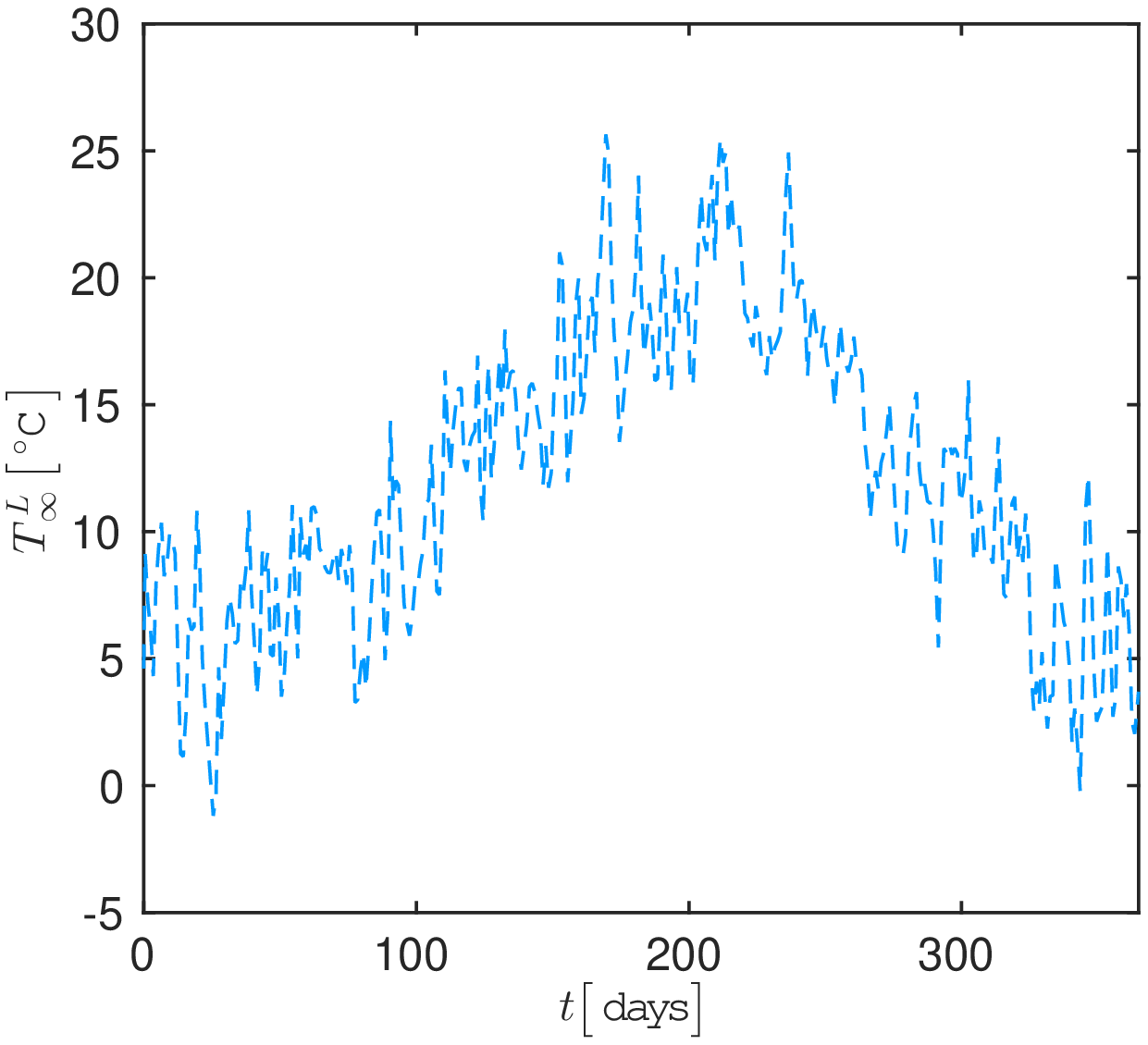}} \hspace{0.2cm}
	\subfigure[]{\includegraphics[width=.45\textwidth]{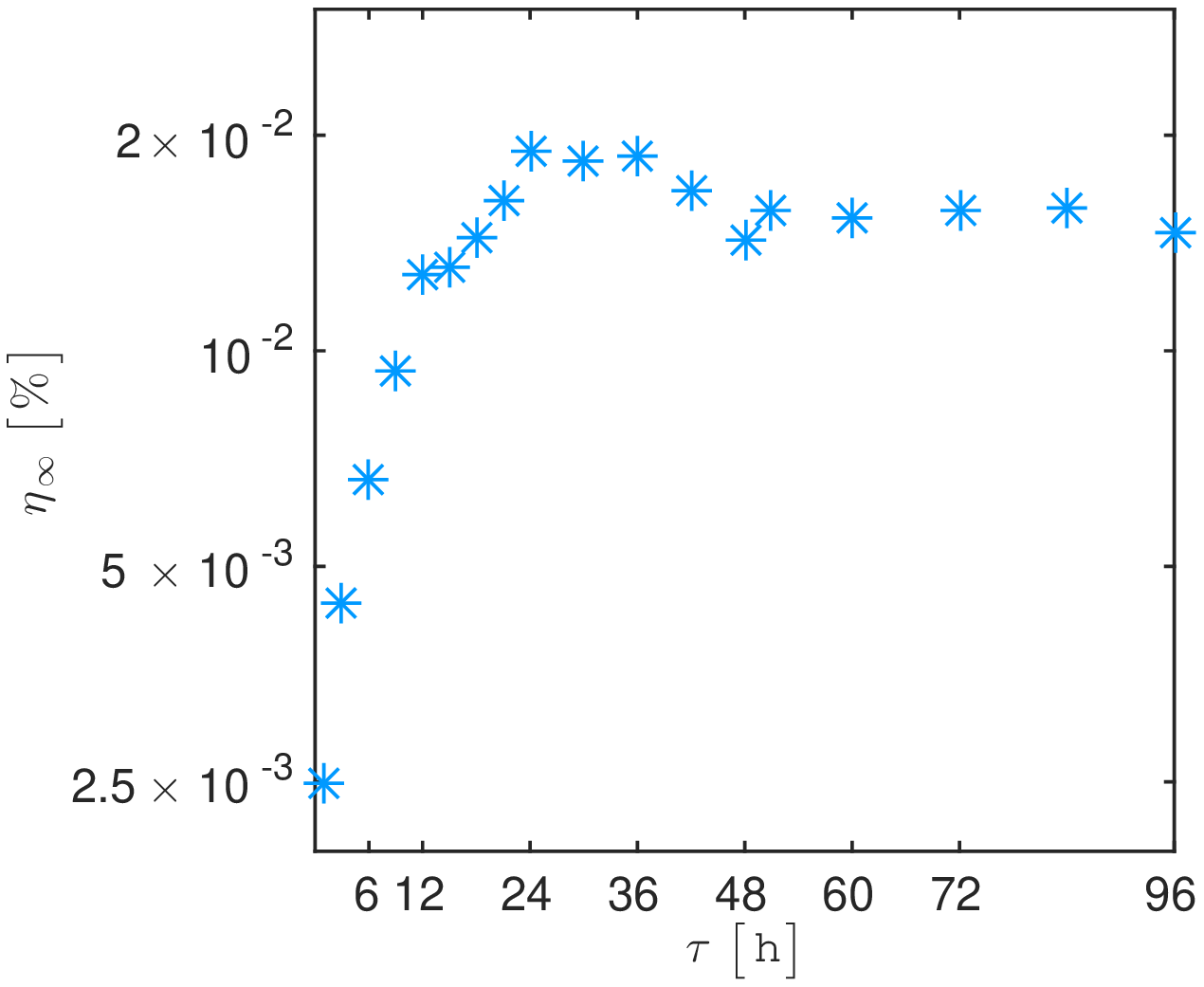}} 
	\caption{\small Full (\emph{a}) and time-averaged boundary conditions (\emph{b, c}), where $\tau$ is the time-averaging period and the error $\eta_{\, \infty}$ between them (\emph{d}). }
	\label{fig:BC_eta_tau}
\end{figure}

\begin{figure}[!ht]
	\centering
	\subfigure[]{\includegraphics[width=.45\textwidth]{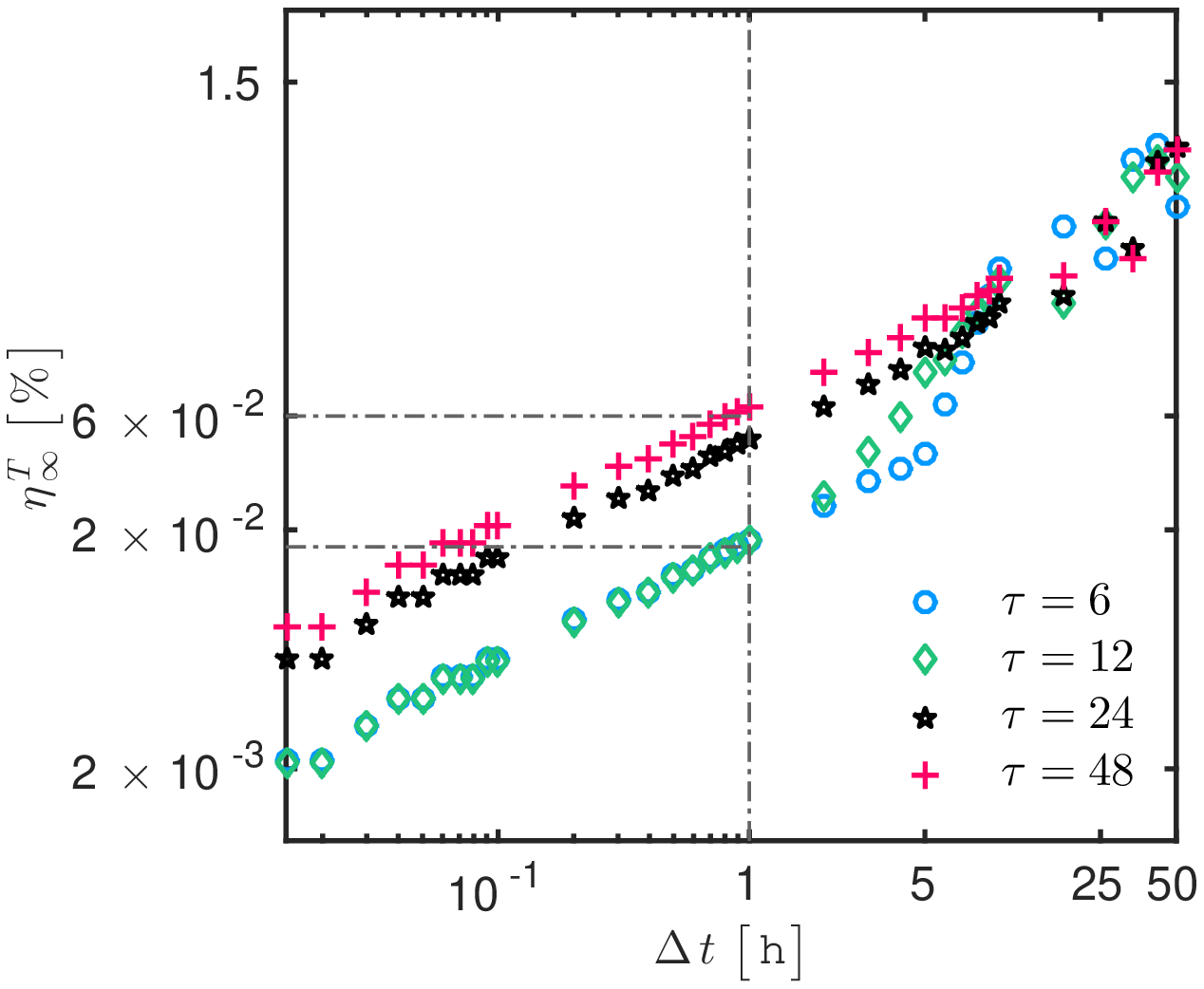}} \hspace{0.2cm}
	\subfigure[]{\includegraphics[width=.45\textwidth]{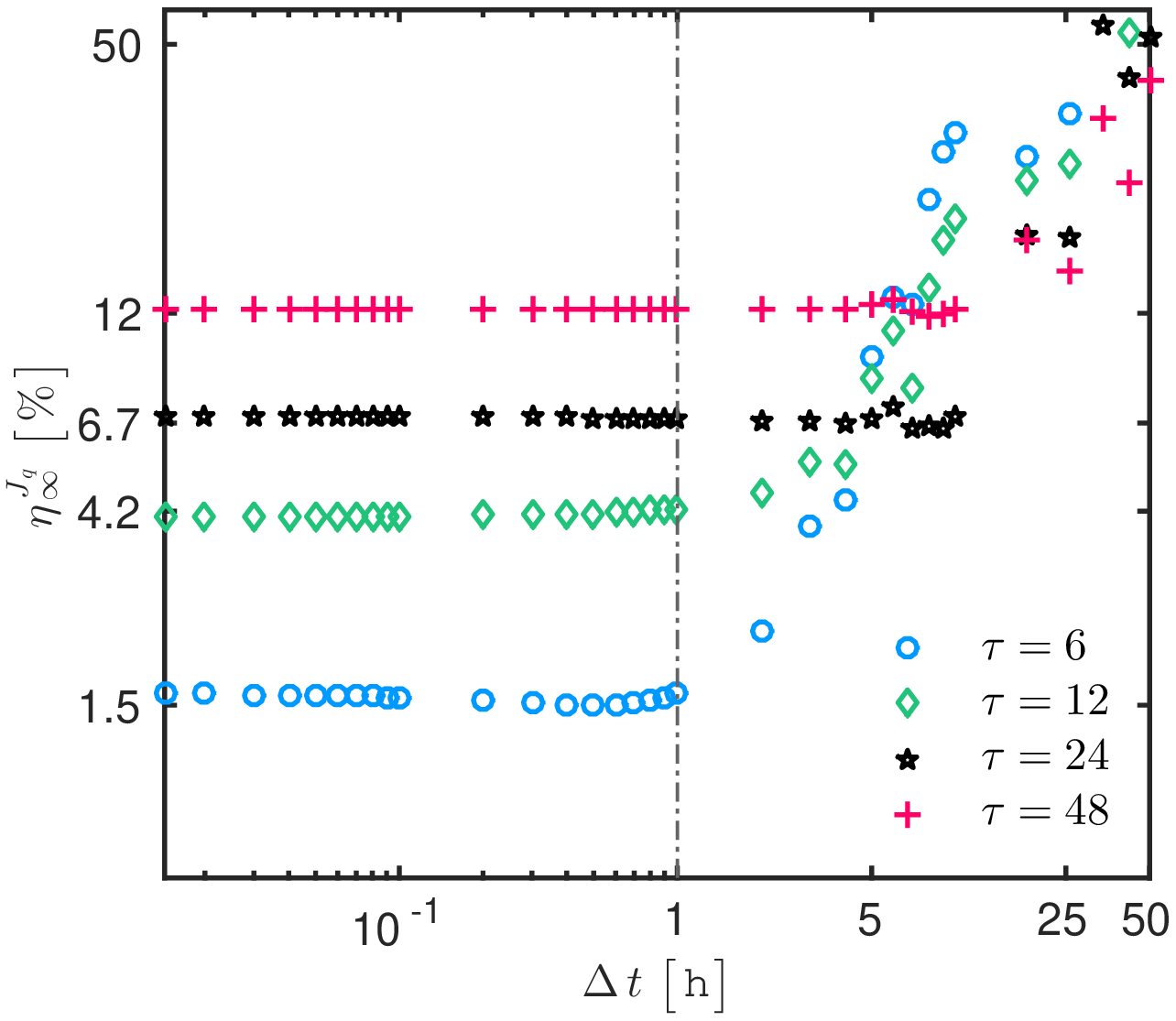}} \\
	\subfigure[]{\includegraphics[width=.45\textwidth]{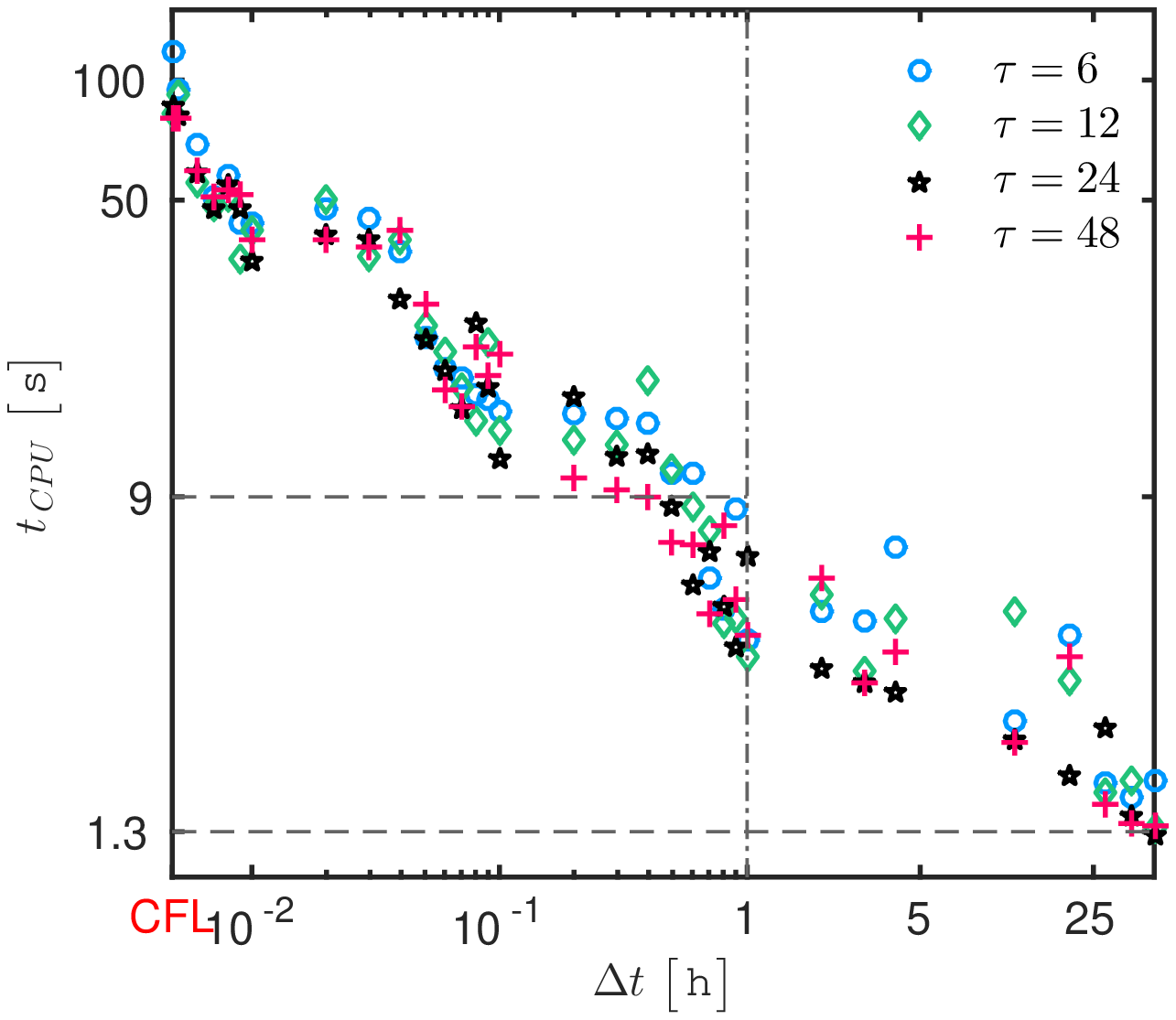}}
	\caption{\small ARM:  $\eta_{\, \infty}$ errors of temperature (\emph{a}) and heat flux (\emph{b}). 
		CPU time to solve average reduced heat equation (\emph{c}). }
	\label{fig:Reduced_RKL1_err}
\end{figure}

After collecting all necessary parameters in \emph{offline} procedure, the study now can be expanded for bigger simulation time in \emph{online} procedure.  
It is decided, that first of all it is necessary to choose one empirical model. 
For this purpose, a comparison between two models is carried out for the time-averaging period $\tau \egal 12 \ \mathsf{h}$. 
The choice is simply based on the lower error and medium coefficients. 
The results of simulations for the $\mathds{T} \egal 1 \ \mathsf{year}$ can be seen in Figure~\ref{fig:Heat_Empir_Model} .  
By comparing the temperature evolutions, the sensible heat fluxes $J_{\, q} \ \jqUnit$ and the conduction loads $E \ \EUnit$ it can be noticed that Model II is closer to the reference solution and more accurately presents the long term phenomena in comparison to the Model I.
Thereby, Model II is chosen for further studies of the benefits of proposed average reduced model. 
The climatic boundary conditions, implemented in the case study, are provided as an hourly data.
To see the influence of the time-averaging to the boundary conditions the Figure~\ref{fig:BC_eta_tau} is plotted. 
It can be seen that the signal becomes smoother as the time-averaging period increases, while the nature of the signal is preserved. 
Moreover, from the Figure~\ref{fig:BC_eta_tau} (\emph{d}) it can be seen that the error between the full and time-averaged boundary conditions is in the tolerable range. 
With $\tau \egal 24 \ \mathsf{h}$ the error is approximately $\eta_{\, \infty} \approx 2 \times 10^{\, -2}$ on the signal of the boundary condition only.
Later, it can be seen in Figure~\ref{fig:Reduced_RKL1_err} (\emph{a}) that the ARM $\tau \egal 24 \ \mathsf{h}$ enables computations with $\eta_{\, \infty} \approx 4 \times 10^{\, -2}$. 
The mathematical model of ARM does not increase the error from the smoothed signal. 
Thereby, it can be concluded that the restriction to a very fine time grid can be relaxed. 

Now, since the EM is chosen, the parametric study can be conducted for the important parameters of model.  
The reliability and efficiency of the average reduced model are evaluated with STS RKL scheme. 
The total simulation time for the first part of the study is taken as $\mathds{T} \egal 100 \ \mathsf{days}$.  
The results are compared for four $\tau \egal [\,6, \, 12, \, 24, \, 48\,] \ \mathsf{h}$ averaging periods and different $\Delta \, t$ time-step sizes. 
The Figure~\ref{fig:Reduced_RKL1_err} illustrates the evolution of the global relative error and computational time.  
It can be seen that the optimal time-step is $\Delta \, t \egal 1 \ \mathsf{h} $, where the errors for both temperature and the heat flux are stable. 
When time-step is more than $5 \ \mathsf{h} $, the error grows sharply. 
In terms of $\tau$ the time-averaging period size, it can be noticed that the error of the ARM for temperature, displayed in Figure~\ref{fig:Reduced_RKL1_err} (\emph{a}), is close to the error of time-averaging the signal for $\Delta \, t \egal 1 \ \mathsf{h} $ (see Figure~\ref{fig:BC_eta_tau} (\emph{d})) . 
Hence, the error does not increase because of ARM formulation. 
This point proves the reliability of the ARM, whereas the origin of the error might arise both from time-averaging period and bigger time steps. 
It can be also noticed, that the lowest error is reachable with the $\tau \egal 6 \ \mathsf{h}$ as the boundary conditions (see Figure~\ref{fig:BC_eta_tau} (\emph{a})) are less smoothed and hence the difference between complete and average reduced models is minimal. 

\begin{figure}[!ht]
	\centering
	\subfigure[]{\includegraphics[width=.45\textwidth]{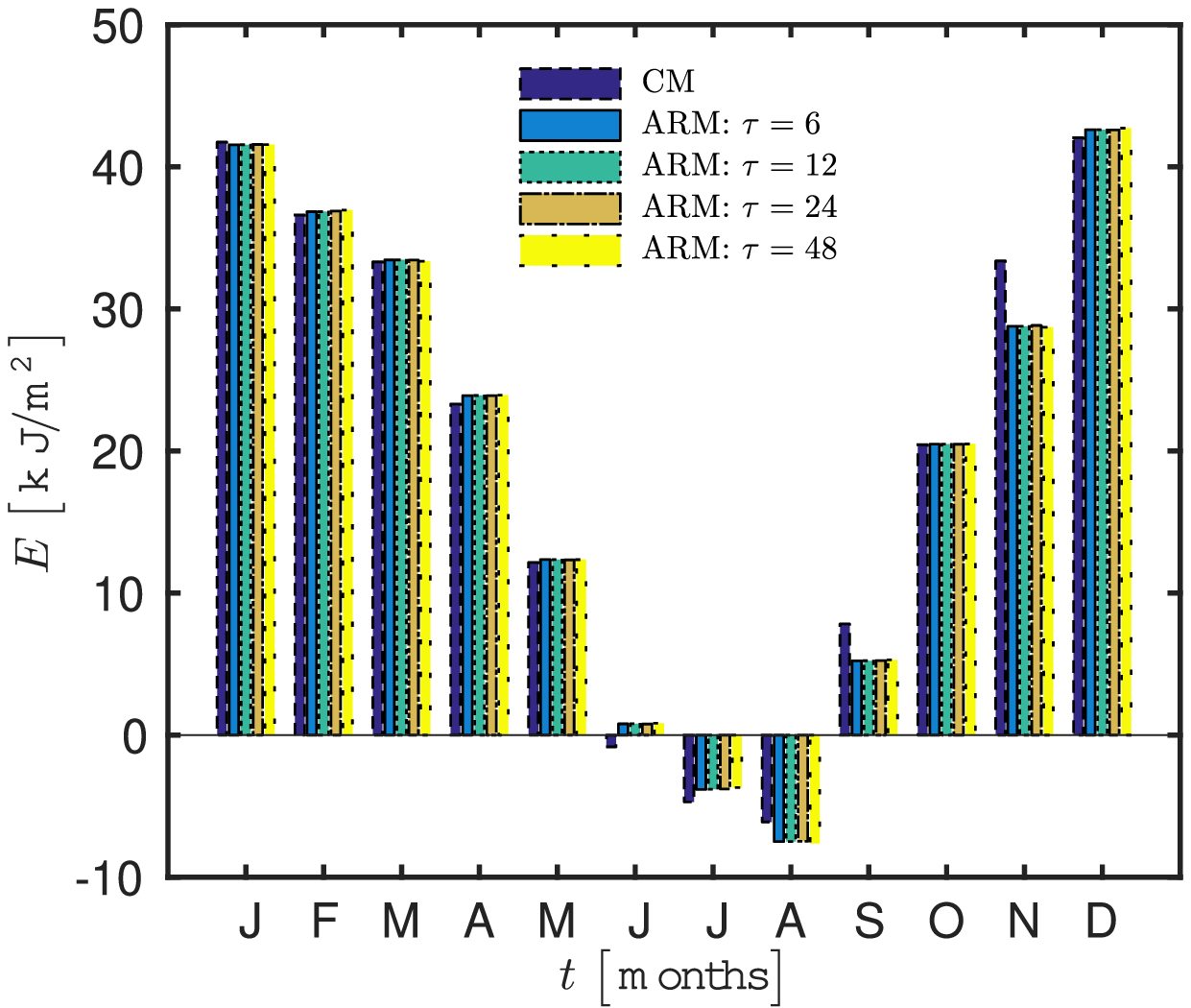}} \hspace{0.2cm}
	\subfigure[]{\includegraphics[width=.45\textwidth]{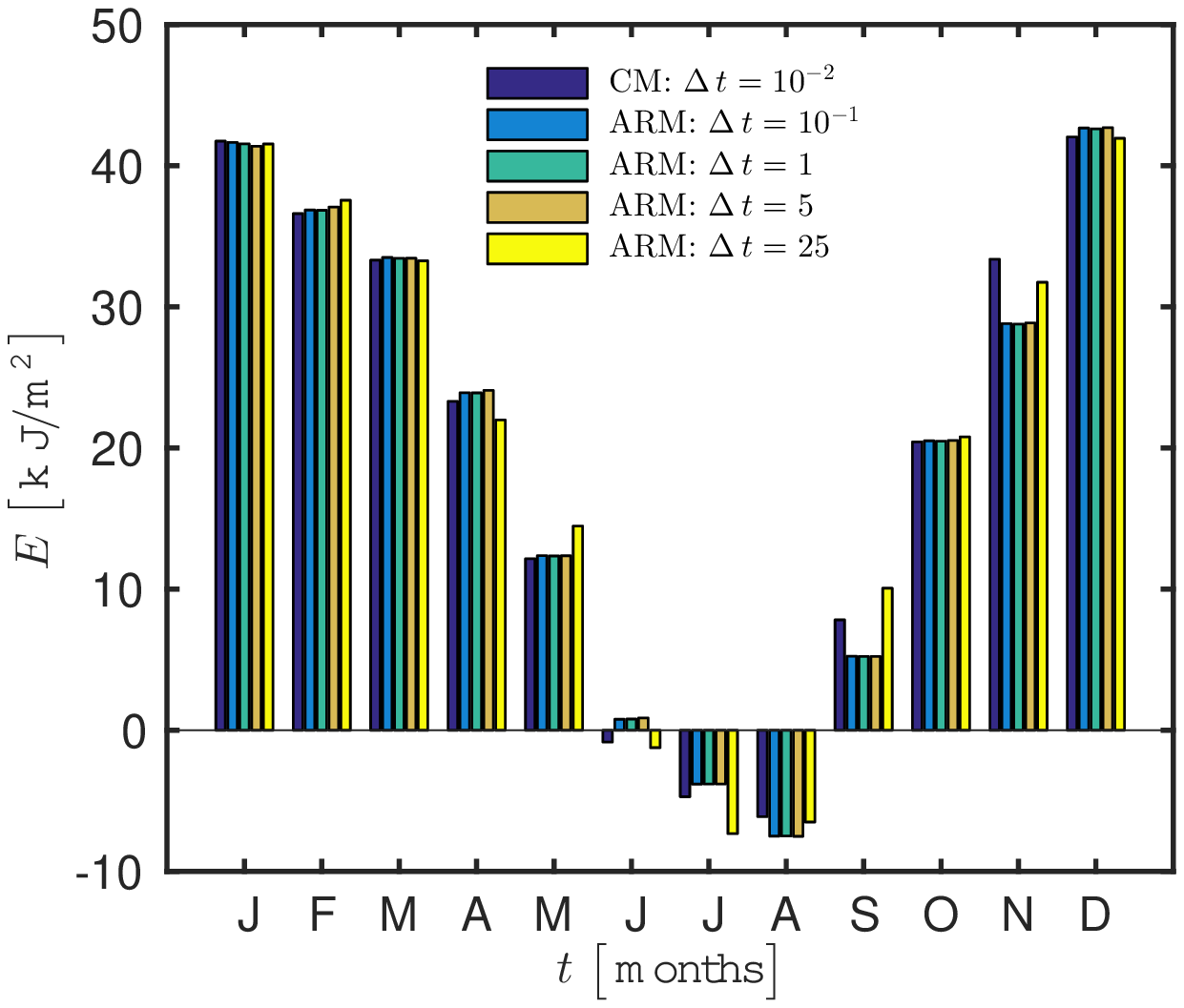}} 
	\caption{\small Conduction loads $E \ \bigl[\,\mathsf{J/m^2}\,\bigr]$ calculated for the whole year with the time-step $\Delta \,  t \egal 1 \ \mathsf{h}$ (\emph{a}) and with the time-averaging period  $\tau \egal 12 \ \mathsf{h}$ (\emph{b}).}
		\label{fig:Thermal_loads}
\end{figure}

However good results might look at a first glance, one should keep in mind the purpose of study. 
The error for the temperature is out of tolerance level, while for the heat flux it is still tolerable. 
This type of technique might be useful to make fast and efficient computations to evaluate the energy consumption problems. 
One of the main values, conduction loads $E \ \EUnit$, can be calculated for the whole year, \emph{i.e.} the total simulation time is $\mathds{T} \egal 365 \ \mathsf{days}$, to see the impact of time-step and time-averaging sizes. 
Figure~\ref{fig:Thermal_loads} shows the example of such calculations. 
As can be seen, the size of the time-step $\Delta \, t$ is more important for more accurate results than the time-averaging size $\tau$. 
Even with $\tau \egal 48 \ \mathsf{h} $ and $\Delta \, t \egal 1 \ \mathsf{h} $ one can get very close results to the reference. 
While, with $\tau \egal 12 \ \mathsf{h} $ and $\Delta \, t \egal 25 \ \mathsf{h} $ the discrepancy is obviously large. 
This can also be explained with higher errors of heat flux for bigger time-steps. 
Therefore, as a result of this study, it can be concluded that for fast and accurate simulations the most tolerable time-step size is $\Delta \, t \egal 1 \ \mathsf{h}$ and time-averaging period $\tau \egal 12 \ \mathsf{h}$. 

The comparison of complete and average reduced models results are displayed in Table~\ref{tab:HEAT_CMandRM_results}. 
\begin{table}[!ht]                                                              
	\centering                                                               
	\begin{tabular}{|l|c|c|}  
		\cline{2-3}                                                                                                          
		\multicolumn{1}{c|}{} & CM: $\Delta \, t \egal 1 \ \mathsf{h}$  & ARM: $\Delta \, t \egal 1 \ \mathsf{h}$ and $\tau \egal 12 \ \mathsf{h}$  \\   		                                                 
		\hline\hline                                                                                                                                   
		$\eta_{\, \infty}^{\, T} \ \bigl[\,\%\,\bigr] $  & $ 2.1 \times 10^{\, -2}$ & $ 1.8 \times 10^{\, -2}$\\
		\hline  
		$\eta_{\, \infty}^{\, J} \ \bigl[\,\%\,\bigr] $  & $1.85$ &   $4.25$\\
		\hline  
		$\varrho_{\, \text{CPU}}\ \bigl[\,\%\,\bigr]$ &\multicolumn{2}{c|}{$51.7$}\\		
		\hline\hline 
	\end{tabular}   
	\bigskip  \caption{\small The comparison of complete and average reduced model results. } 					
	\label{tab:HEAT_CMandRM_results}                                                       
\end{table} 
Following the above conclusion, average reduced model results are shown only for $\Delta \, t \egal 1 \ \mathsf{h}$ and $\tau \egal 12 \ \mathsf{h}$. 
It can be seen that the reduction of the computational time is almost $50 \, \%$ for the same level of error. 
The next step is to study the application of the average reduced model to the problem of the heat and mass transfer. 
\section{Heat and mass transfer}\label{sec:HM_transfer}
This section of the article is dedicated to study the reliability and efficiency of the ARM approach to model the coupled heat and mass transfer. 
The section follows the same structure as the previous one. 
\subsection{Complete Model}
The mathematical model can be expressed by the system of two coupled partial differential equations with respect to two unknowns $T \ \bigl[\,\mathsf{K}\,\bigr]$ and $\theta \ \bigl[\,\varnothing\,\bigr]$ \citep{mendes2002}. 
The equations are defined by the spacial $\Omega_{\, x} \egal [ \, 0, \, \ell \, ]$ and time $\Omega_{\, t} \egal [ \, 0, \, \mathds{T}\, ]$ domains, where $\ell\ \bigl[\,\mathsf{m}\,\bigr]$ is the thickness of a material and $\mathds{T} \ \bigl[\,\mathsf{h}\,\bigr]$ is the final time:
\begin{subequations}
	\label{eq:HM_physical_model}
	\begin{align}
	\rho_{\,2} \cdot \pd{\theta}{t} & \, = \, \pd{}{x} \, \biggl(\, D_{\,\theta} \cdot \pd{\theta}{x} \, + \, D_{\,T} \cdot \pd{T}{x} \,\biggr) \,, \\[4pt]
	\cTs \cdot  \pd{T}{t} &\, = \, \pd{}{x} \, \biggl(\, k_{\,T} \cdot  \pd{T}{x} \, \biggr)  
	\, + \, L_{\,12}^{\,\circ} \cdot \pd{}{x} \, \biggl(\,  k_{\,TM} \cdot  \pd{\theta}{x} \, \biggr)  \,,
	\end{align}
\end{subequations}
where $\rho_{\,2} \ \rhoUnit$ is the specific mass of liquid water,
$D_{\,\theta} \, \left(\,\theta\,\right) \ \DthetaUnit$ is the diffusion coefficient under the moisture gradient, 
$D_{\,T} \ \DTUnit$ is the diffusion coefficient under the temperature gradient, 
$\cTs \ \cQUnit$ is the heat storage coefficient, 
$k_{\,T} \ \kQUnit$ is the thermal conductivity of the material, 
$L_{\,12}^{\,\circ} \ \hUnit$ is the latent heat of vaporization,  
and $k_{\,TM} \ \kTMUnit$ is the vapor transfer coefficient under the moisture gradient. 
The temperature of a material at surfaces is measured during the experiment. 
Therefore, a \textsc{Dirichlet}--type of boundary conditions are imposed for temperature and moisture content:
\begin{equation*}
T \, \left(\,x \egal \left\{\,0, \,\ell\,\right\} , \, t \,\right) \egal  T_{\, \infty}^{\, L,\, R} \, \left(\,t \,\right) \,, \quad
\theta \, \left(\,x \egal \left\{\,0, \,\ell\,\right\} , \, t \,\right) \egal  \theta_{\, \infty}^{\, L,\, R} \, \left(\,t \,\right) \,,
\end{equation*}
where $\ell \, [\sf m]$ is the thickness of the wall and superscripts L and R correspond to the external ($x \egal 0$) and internal ($x \egal \ell$) boundary conditions, respectively.  
For the initial conditions, the moisture level and temperature within the material are considered to be uniform. 
As it was done for the heat transfer, it is interesting to observe the total output flux, which can be defined as follows:
\begin{equation}
J_{\, qm} \ \eqdef \ J_{\, q} \plus J_{\, m}\,.
\end{equation}
where $J_{\, q}$ is given in Eq.~\eqref{eq:heat_flux} and 
\begin{equation*}
J_{\, m} \ \eqdef  \moins L_{\,1 \,2} \cdot k_{\, TM} \cdot \pd{\theta}{x} \, \Bigr|_{\, x \egal \ell} \,.
\end{equation*}
Another important result to observe is the conduction loads $E \ \EUnit$: 
\begin{equation}\label{eq:HM_E}
E \egal \int_{t_{\,1}}^{t_{\,2}} \, J_{\, qm} \, \mathrm{d} \, t\, ,
\end{equation}
where $(\, t_{\,2} \moins t_{\,1}\, )$ is equal to one month. 
In addition to those values, also the standard thermal resistance of the material, denoted as $R_{\, o}$, is computed: 
\begin{equation*} 
R_{\, o} \, \eqdef \, \dfrac{\ell}{\kTs \, \left(\, \theta \, =\, 0\,\right)}  \, .
\end{equation*}
It is compared to the effective averaged thermal resistance with the total output flux :
\begin{equation}\label{eq:Thermal_resistances}
R \, \eqdef \, \int_{t_{\,1}}^{t_{\,2}} \, \Biggl|\, \dfrac{T \,\left(\, x \, =\,\ell,\, t \,\right) - T \,\left(\, x \, =\,0,\, t \,\right)}{J_{\, qm}} \,\Biggr| \, \mathrm{d} \, t\, ,
\end{equation}
where $(\, t_{\,2} \moins t_{\,1}\, )$ is equal to one day. \\
The model~\eqref{eq:HM_physical_model} is solved in its dimensionless form and is written for moisture content -- $v$ and temperature -- $u$ as:
\begin{subequations}
\label{eq:HM_dimless_model}
\begin{align}
\pd{v}{t^{\, \star} }  &\egal \FoM \cdot \pd{}{x^{\, \star} } \, \left(\, D_{\,\theta}^{\, \star}\, \left(\,v\,\right) \, \cdot \pd{v}{x^{\, \star}} \plus \gamma \cdot  
D_{\,T}^{\, \star} \cdot \pd{u}{x^{\, \star}} \,\right)\,, \\[4pt]
\cTs^{\, \star} \, \left(\,v\,\right) \, \cdot \pd{u}{t^{\, \star}}  &\egal \FoT \cdot \Biggl( \pd{}{x^{\, \star} }  \biggl( \kTs^{\, \star}\, \left(\,v\,\right) \, \cdot \pd{u}{x^{\, \star}}\biggr) 	
\,+ \,\delta \cdot \pd{}{x^{\, \star} }  \biggl( \kTMs^{\, \star} \cdot \pd{v}{x^{\, \star}} \biggr) \Biggr) \,,
\end{align}
\end{subequations}
where the superscript $^{\,\star}$ represents a dimensionless value of a variable. 

The dimensionless material properties $D_{\,\theta}^{\, \star} \, \left(\,v\,\right) $, $\cTs^{\, \star} \, \left(\,v\,\right) $ and $\kTs^{\, \star} \, \left(\,v\,\right)$ are taken as the moisture content dependent first order polynomials:
\begin{equation*}
D_{\,\theta}^{\, \star} \ : \ v \ \mapsto d_{\,1} \plus d_{\,0} \cdot \,\left( \,v \moins v_{\,0} \, \right) \, , \qquad
\cTs^{\, \star} \ : \ v \ \mapsto c_{\,1} \plus c_{\,0} \cdot v  \, , \qquad
\kTs^{\, \star} \ : \ v \ \mapsto k_{\,1} \plus k_{\,0} \cdot v \, ,
\end{equation*} 
where $d_{\,0}, \, d_{\,1}, \,v_{\,0}, \,  c_{\,0}, \, c_{\,1}, k_{\,0} \ \text{and} \ k_{\,1} \, \in \mathds{R}$. 
Dimensionless $D^{\, \star}_{\, T}$ and $\kTMs^{\, \star}$ are taken as constant values, \emph{i.e.} independent from temperature or moisture content. 	
The initial conditions at $t^{\, \star} \egal 0 $ are $u_{\, 0} \egal v_{\, 0} \egal 1$ for $\forall \, x^{\, \star} \, \in \, \bigl[\,0 \,,\, 1\,\bigr]$.
The boundary conditions at the surfaces $x^{\, \star} \egal \left\{\,0, \,1\,\right\}$ are defined as: 
\begin{equation*}
u \, \left(\,x \egal \left\{\,0, \,1\,\right\} , \, t^{\, \star} \,\right) \egal  u_{\, \infty}^{\, L,\, R} \, \left(\,t^{\, \star} \,\right) \,, \quad
v \, \left(\,x \egal \left\{\,0, \,1\,\right\} , \, t^{\, \star} \,\right) \egal  v_{\, \infty}^{\, L,\, R} \, \left(\,t^{\, \star} \,\right) \,.
\end{equation*}
More information about formulation of the dimensionless model can be found in the previous work \citep{abdykarim2019}. 

\subsection{Average Reduced Model}\label{sec:reduced_model_HM}
Following the same methodology, presented in the Section~\ref{sec:reduced_model_Heat}, this section demonstrates the derivation of the ARM for heat and mass transfer. 
The only added difference is the pair of empirical models for $u' \, \left(\,\mathbf{P_{\, u}}, \, x, \, t\,\right)$ and $v' \, \left(\,\mathbf{P_{\, v}}, \, x, \, t\,\right)$. 
In the following, the methodology of the Section~\ref{sec:obtaining_RM} is extended and applied to the model~\eqref{eq:HM_dimless_model}. 
\subsubsection{Obtaining an ARM}	
Solutions for the temperature and moisture content equations are taken as a sum of the time-averaged and fluctuating values: 
\begin{subequations}
	\label{eq:HM_RM}
	\begin{align}
	\overline{\pd{\left(\,\overline{v} \plus v' \,\right)}{t^{\, \star} } } &\egal \overline{\FoM \cdot \pd{}{x^{\, \star} } \, \left(\, D_{\,\theta}^{\, \star}\, \left(\,\overline{v} \plus v'\,\right) \, \cdot \pd{\left(\,\overline{v} \plus v' \,\right)}{x^{\, \star}} \plus \gamma \cdot  
	D_{\,T}^{\, \star} \cdot \pd{\left(\,\overline{u} \plus u' \,\right)}{x^{\, \star}} \,\right)}\,,  \label{eq:HM_RM_mass} \\[4pt]
\overline{	\cTs^{\, \star} \, \left(\,\overline{v} \plus v'\,\right) \, \cdot \pd{\left(\,\overline{u} \plus u' \,\right)}{t^{\, \star}} } 
&\egal 
	\overline{\FoT \cdot \Biggl( \pd{}{x^{\, \star} }  \biggl( \kTs^{\, \star}\, \left(\,\overline{v} \plus v'\,\right) \, \cdot \pd{\left(\,\overline{u} \plus u' \,\right)}{x^{\, \star}}\biggr) 	
	\,+ \,\delta \cdot \pd{}{x^{\, \star} }  \biggl( \kTMs^{\, \star} \cdot \pd{\left(\,\overline{v} \plus v' \,\right)}{x^{\, \star}} \biggr) \Biggr)} \,.	\label{eq:HM_RM_heat}			
	\end{align}
\end{subequations}
One can consider each equation separately. 
The Eq.~\eqref{eq:HM_RM_mass} is averaged as a reduced model in the following way:
\allowdisplaybreaks
\begin{align*}
	\overline{\pd{\,\overline{v}}{t^{\, \star} } } \plus	\overline{\pd{\,v'}{t^{\, \star} } }  &\egal \overline{\FoM \cdot \pd{}{x^{\, \star} } \, 
		\left(\, \bigl(\, d_{\,1} \plus d_{\,0} \cdot \,\left(\,\overline{v} \plus v' \moins v_{\,0}\,\right) \,\bigr) \, 
		\cdot 
	\biggl(\, \pd{\,\overline{v}}{x^{\, \star}} \plus \pd{\,v'}{x^{\, \star}}\,\biggr) 
	\plus \gamma \cdot  
		D_{\,T}^{\, \star} \cdot 
		\biggl(\, \pd{\,\overline{u}}{x^{\, \star}} \plus \pd{\,u'}{x^{\, \star}}\,\biggr) 
		 \,\right)}\,,   \\[4pt]
		&\egal 
			\overline{\FoM \cdot \pd{}{x^{\, \star} } \left(\,
				d_{\,1}	\cdot \pd{\,\overline{v}}{x^{\, \star}}\,\right)} 
			\plus 
			\overline{\FoM \cdot \pd{}{x^{\, \star} } \left(\,
				d_{\,0} \cdot \overline{v} \cdot \pd{\,\overline{v}}{x^{\, \star}}\,\right)}			
			\plus 
			\overline{\FoM \cdot \pd{}{x^{\, \star} } \left(\,  
				d_{\,0} \cdot v' \cdot \pd{\,\overline{v}}{x^{\, \star}}\,\right)} \\[4pt]
			& \moins	
			\overline{\FoM \cdot \pd{}{x^{\, \star} } \left(\,
				d_{\,0}	\cdot v_{\,0} \cdot \pd{\,\overline{v}}{x^{\, \star}}\,\right)}
			\plus  
			\overline{\FoM \cdot \pd{}{x^{\, \star} } \left(\,
			 	d_{\,1}	\cdot \pd{\,v'}{x^{\, \star}}\,\right)} \plus 
			\plus  
			\overline{\FoM \cdot \pd{}{x^{\, \star} } \left(\,
			 	d_{\,0} \cdot \overline{v} \cdot \pd{\,v'}{x^{\, \star}}\,\right)}	\\[4pt]
			 &  \plus 
			 \overline{\FoM \cdot \pd{}{x^{\, \star} } \left(\,  
			 	d_{\,0} \cdot v' \cdot \pd{\,v'}{x^{\, \star}}\,\right)} 
			 \moins	
			 \overline{\FoM \cdot \pd{}{x^{\, \star} } \left(\,
			 	d_{\,0}	\cdot v_{\,0} \cdot \pd{\,v'}{x^{\, \star}}\,\right)} \\[4pt]
			&\plus \overline{\FoM \cdot \pd{}{x^{\, \star} } \, 
				\left(\, \gamma \cdot  
			D_{\,T}^{\, \star} \cdot 
			\biggl(\, \pd{\,\overline{u}}{x^{\, \star}} \plus \pd{\,u'}{x^{\, \star}}\,\biggr) 
			\,\right)}\,,   
\end{align*}
after applying the averaging rules one obtains next results:		
		\begin{equation*}
		\pd{\,\overline{v}}{t^{\, \star} }  
		\egal \FoM \cdot \pd{}{x^{\, \star} } \, 
		\left(\,D_{\,\theta}^{\, \star}\, \left(\,\overline{v} \,\right)\, 
		\cdot 
		\pd{\,\overline{v}}{x^{\, \star}} 
		\plus  \gamma \cdot  
		D_{\,T}^{\, \star} \cdot 
		\pd{\,\overline{u}}{x^{\, \star}} \,\right)
		\plus 
		\mathds{S}_{\, v} \,\left(\,x, \, t\,\right) \,,
	\end{equation*} 
	where 
	\begin{equation}\label{eq:S_v}
	\mathds{S}_{\, v} \,\left(\,x, \, t\,\right) 
	\ \eqdef \  \FoM \cdot \pd{}{x^{\, \star} } \left(\,  
	d_{\,0} \cdot \overline{ v' \cdot \pd{\,v'}{x^{\, \star}}}\,\right) \,.
	\end{equation}
After almost same manipulations, the Eq.~\eqref{eq:HM_RM_heat} is written as:
\allowdisplaybreaks
\begin{equation*}
	\cTs^{\, \star} \, \left(\,\overline{v} \,\right) \, \cdot \pd{\left(\,\overline{u} \,\right)}{t^{\, \star}} 
 \egal 
 \FoT \cdot 
 \Biggl(\, \pd{}{x^{\, \star} } \biggl( \kTs^{\, \star}\, \left(\,\overline{v}\,\right) \, \cdot \pd{\,\overline{u}}{x^{\, \star}} \biggr) 	
 \plus 
 \delta \cdot \pd{}{x^{\, \star}}  \biggl( \kTMs^{\, \star} \cdot \pd{\,\overline{v}}{x^{\, \star}} \biggr) 
 \Biggr) 
		\plus 
		\mathds{S}_{\, u} \,\left(\,x, \, t\,\right) \,,
	\end{equation*} 
	where 
	\begin{equation}\label{eq:S_u}
	\mathds{S}_{\, u} \,\left(\,x, \, t\,\right) 
	\ \eqdef \  \FoT \cdot \pd{}{x^{\, \star} } \biggl( k_{\,0} \cdot \overline{ v'\cdot \pd{\,u'}{x^{\, \star}}}  \biggr)  
	\moins 
	c_{\,0} \cdot \overline{ v'\cdot \pd{\,u'}{t^{\, \star}}}  \,.
	\end{equation}
The boundary conditions are averaged as follows: 
\begin{equation*}
\overline{u} \, \left(\,x \egal \left\{\,0, \,1\,\right\} , \, t^{\, \star} \,\right) \egal  \overline{u_{\, \infty}}^{\, L,\, R} \, \left(\,t^{\, \star} \,\right) \,, \quad
\overline{v} \, \left(\,x \egal \left\{\,0, \,1\,\right\} , \, t^{\, \star} \,\right) \egal  \overline{v_{\, \infty}}^{\, L,\, R} \, \left(\,t^{\, \star} \,\right) \,.
\end{equation*}
As can be seen, the system to be solved has the same shape with added source terms, given in Eqs.~\eqref{eq:S_v} and~\eqref{eq:S_u}, which are dependent only on space $x$ and time $t$ variables. 
\subsubsection{Empirical Model of high frequency values}\label{sec:empir_model_HM}
Empirical models for the heat and mass transfer are chosen following the same methodology, described in the Section~\ref{sec:Methodology_empir_model}. 
One presents fluctuating values for temperature and moisture content in general mathematical models in forms as:
\begin{equation}\label{eq:gen_empirical_model_HM}
u^{\, \prime}  \ :\ \left(\,\mathbf{P_{\, u}}, \, x, \, t\,\right) \ \mapsto \ f \, \left(\,\mathbf{P_{\, u}}, \, x, \, t\,\right) \, , \quad
v^{\, \prime}  \ :\ \left(\,\mathbf{P_{\, v}}, \, x, \, t\,\right) \ \mapsto \ g \, \left(\,\mathbf{P_{\, v}}, \, x, \, t\,\right) \, ,
\end{equation}
where $\mathbf{P_{\, u}}$ and $\mathbf{P_{\, v}}$ are sets of parameters. 
The functions $f \, \left(\, \cdot \, \right)$ and $g \, \left(\, \cdot \, \right)$ should be designed in such a way, that the averaging of them should be in accordance with Eq.~\eqref{eq:mean_u'}:
\begin{equation*}
\overline{f \, \left(\, \cdot \, \right)} \egal 0 \, , \quad \overline{g \, \left(\, \cdot \, \right)} \egal 0 \, .
\end{equation*}
It should be noted that the offline procedure is performed to the both values simultaneously. 
The calculation of an error $\varepsilon_{\, 2}^{\, uv}$, which needs to be optimized, is adapted as:
\begin{equation}\label{eq:err_empirical_model_HM}
\varepsilon_{\, 2}^{\, uv} \, (\,\xs\,)\  \eqdef  \ \omega_{\, u} \cdot \varepsilon_{\,2}^{\, u} \plus \omega_{\, v} \cdot \varepsilon_{\,2}^{\, v} \, ,
\end{equation}  
where
\begin{equation*}
	\varepsilon_{\,2}^{\, u} \, (\,\xs\,)\ \eqdef  \ \bigl|\bigl|\, \,u_{\, \text{ref}} \, (\,\xs,\,\ts\,) \moins \widetilde{u}\, (\,\xs,\,\ts\,) \, \bigr|\bigr|_{\, 2}\, ,  \qquad 
	\varepsilon_{\,2}^{\, v} \, (\,\xs\,)\ \eqdef \ \bigl|\bigl|\, \,v_{\, \text{ref}} \, (\,\xs,\,\ts\,) \moins \widetilde{v}\, (\,\xs,\,\ts\,) \, \bigr|\bigr|_{\, 2}\, ,
\end{equation*}
are errors of each transfer and
\begin{equation*}
\omega_{\, u} \  \eqdef  \ \dfrac{1}{\max \, \left\{ \, u_{\, \text{ref}}\,(\,\xs \,,\ts\,) \, \right\} 
	\moins \min\, \left\{ \, u_{\, \text{ref}}\,(\,\xs \,,\ts\,)\, \right\} \, }\, ,  \qquad 
\omega_{\, v} \  \eqdef  \ \dfrac{1}{\max \, \left\{ \, v_{\, \text{ref}}\,(\,\xs \,,\ts\,) \, \right\} 
	\moins \min\, \left\{ \, v_{\, \text{ref}}\,(\,\xs \,,\ts\,)\, \right\} \, } \, ,
\end{equation*}  
are weights for each error in order to eliminate the scaling differences. 
\subsection{Extension of the STS method to a coupled system of nonlinear equations}\label{sec:num_methods_HM}
In order to discretize the system of equations~\eqref{eq:HM_dimless_model} the STS RKL method is written in the same way as Eq.~\eqref{eq:STS_general_heat_equation_discretisation} but with vector notation:
\begin{equation}\label{eq:STS_general_coupled_case_discretisation}
{\bf u^{\, k+1}} \egal {\bf \mathsf{P}}_{\, N_{\, \mathrm{S}}} \bigl(\, \Delta \, t_{\, \mathrm{S}} ,\, \bf A \, \bigr) \cdot {\bf u^{\, k}} \, ,
\end{equation}
where ${\bf u} \eqdef \bigl(\, u_{\, 1}, \, u_{\, 2}, \, \ldots, \, u_{\, n}\, \bigr)$ is the vector, which includes all $n$ variables, $\bf A$ is the discretization matrix and $\Delta \, t_{\, \mathrm{S}}$ is the super-time-step.
The value for the $\Delta \, t_{\, \mathrm{S}}$ should be found with corresponding $\Delta \, \ts^{\, \star}_{\, \text{exp}}$ the time-step for the \textsc{Euler} explicit scheme, which must satisfy the CFL stability condition of the whole system of equations. 
More details on how to obtain the super-time-step can be found in previous works \citep{abdykarim2019, abdykarim2019b}. 

\subsection{Case study}\label{sec:case_study_HM}
This section is aimed to analyze the reliability and fidelity \citep{clark2010ancient, kavetski2010ancient} of the ARM for heat and moisture transfer. 
For this purpose, one needs to follow the methodology starting with the offline procedure. 
The latter provides parameters for empirical model candidates.
Based on observations of an error and a computational time, the most suitable empirical model shall be chosen. 
After that, the online procedure is carried out for a longer simulation time.
A parametric study, in comparison with the complete model, helps to identify the optimal pair of time-averaging period and time-step sizes.
Later, those parameters are implemented into the ARM to perform comparison with the experimental observations. 
The data is obtained in investigations for drying of a rammed earth wall \citep{chabriac2014mesure,soudani2016,soudani2017} and the period of two years is observed. 
When fidelity of the model is proved, the ARM is used to predict the physical performance of the studied wall.  
\subsubsection{Presentation of the case study}
The drying of a rammed earth wall during its first year after installation is simulated. 
The experimental data is obtained from the observations of a house located in Saint-Antoine-l$'$Abbaye, in Is$\grave{\text{e}}$re, South-Eastern France \citep{soudani2017}. 
In this house, several walls are built with the rammed earth (RE) material, and for the sake of clarity only the South wall data is taken in this study. 
	\begin{figure}[!ht]
		\centering
	\subfigure[]{\includegraphics[width=.45\textwidth]{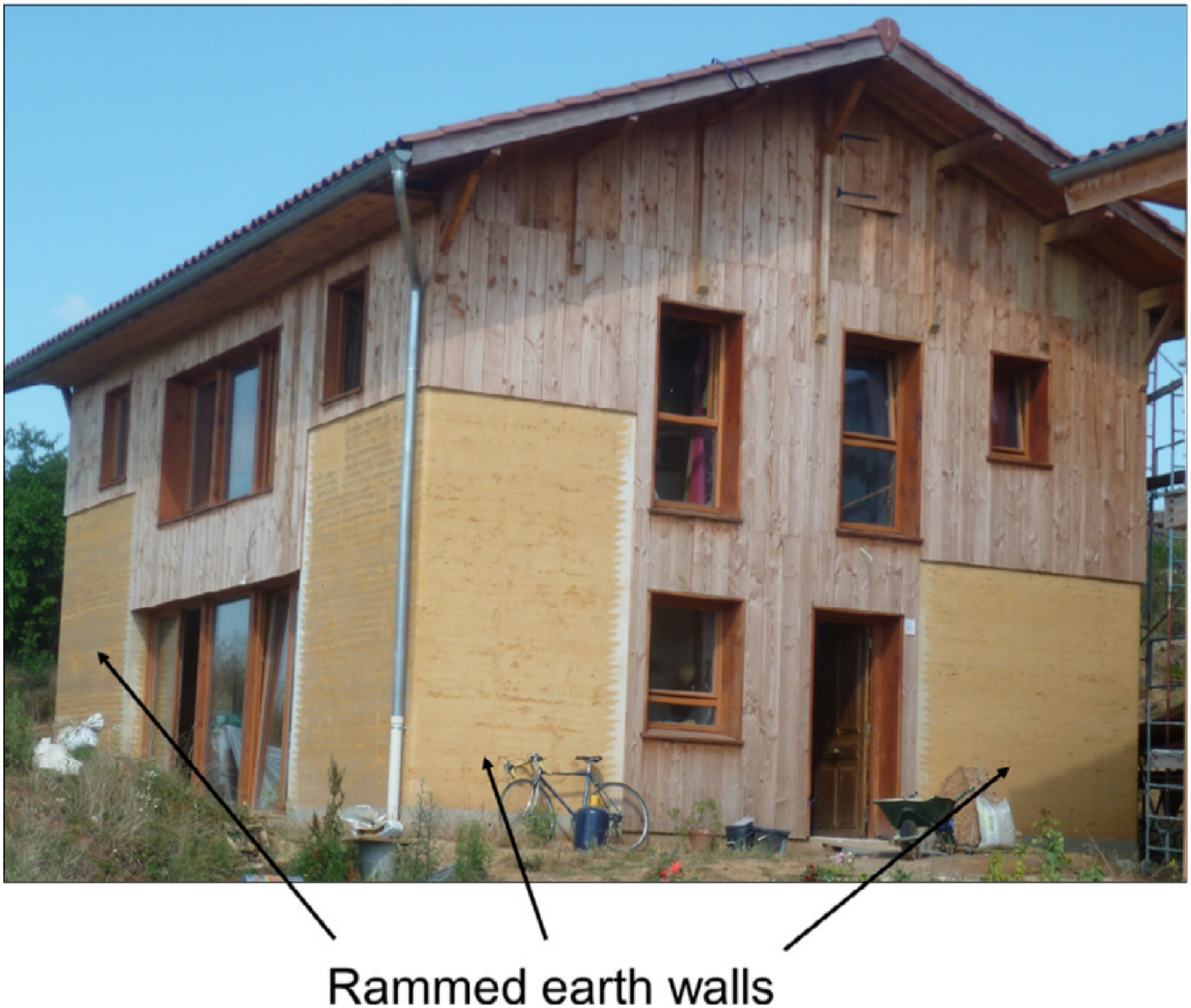}} \hspace{0.2cm}
	\subfigure[]{\includegraphics[width=.45\textwidth]{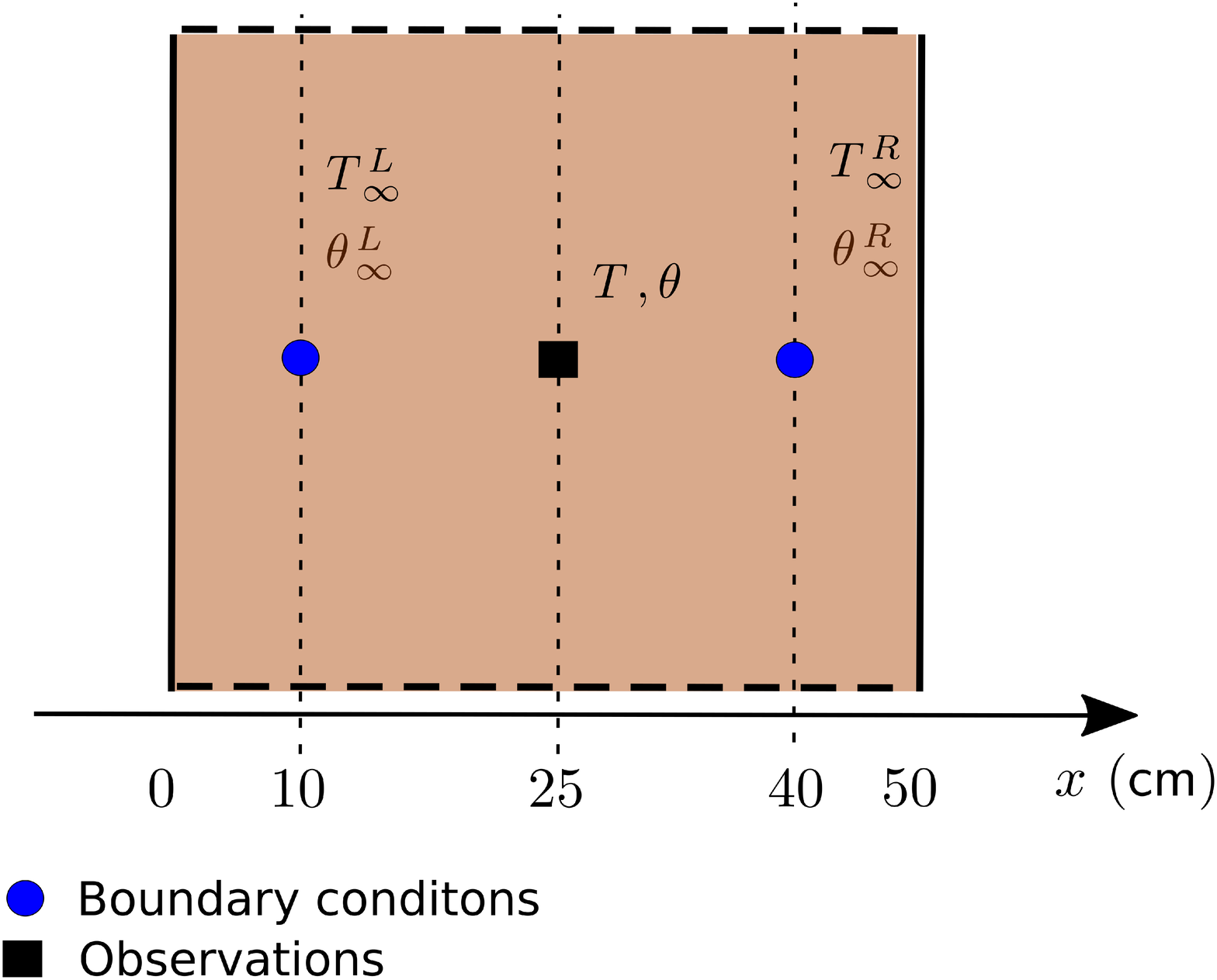}} 
		\caption{House in Saint-Antoine-l'Abbaye (\emph{a}) and sensors' locations in the wall (\emph{b}) \citep{soudani2017}.}
		\label{fig:pics}	
	\end{figure}
	One-dimensional simulations are executed for the RE wall $\ell_{\, \text{RE}} \, = \, 0.3 \,{\sf m} $ in width. 
	The study is focused on the phenomena inside the material. 
	It should be noted that the variations of the temperature and moisture content are measured at $10 \, \sf{cm}$ from the inside and outside surfaces of the wall (see Figure~\ref{fig:pics} (\emph{b})). 
	Therefore, the boundary conditions are taken as \textsc{Dirichlet}--type for the shorter width of a wall ( $0.3 \,{\sf m}$ instead of $0.5 \,{\sf m}$ ). 
	This also permits to avoid additional uncertainties in the physical model on the surface transfer coefficients appearing in \textsc{Robin}--type boundary conditions. 
	\begin{figure}[!ht]
		\centering
	\subfigure[]{\includegraphics[width=.45\textwidth]{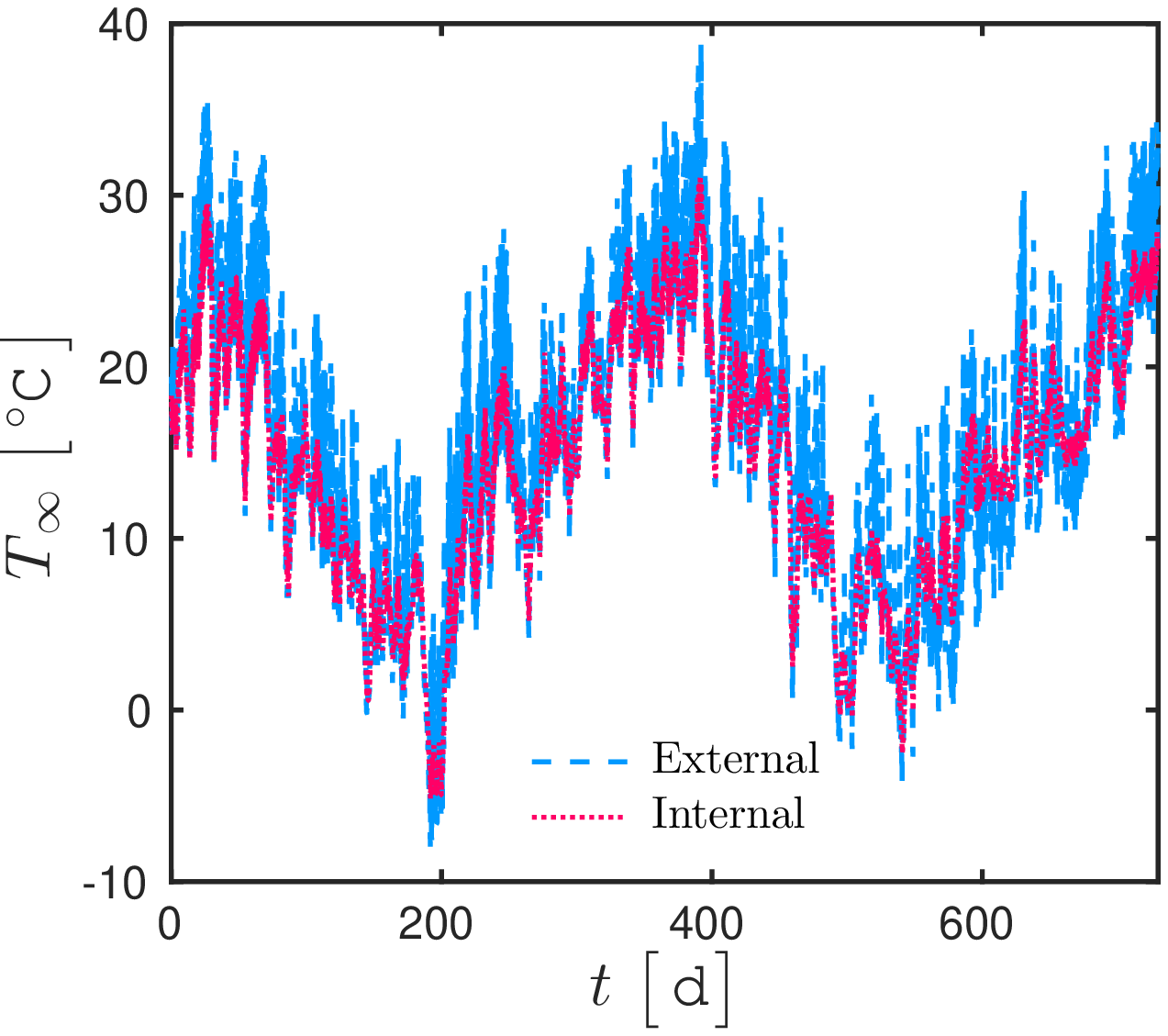}} \hspace{0.2cm}
	\subfigure[]{\includegraphics[width=.45\textwidth]{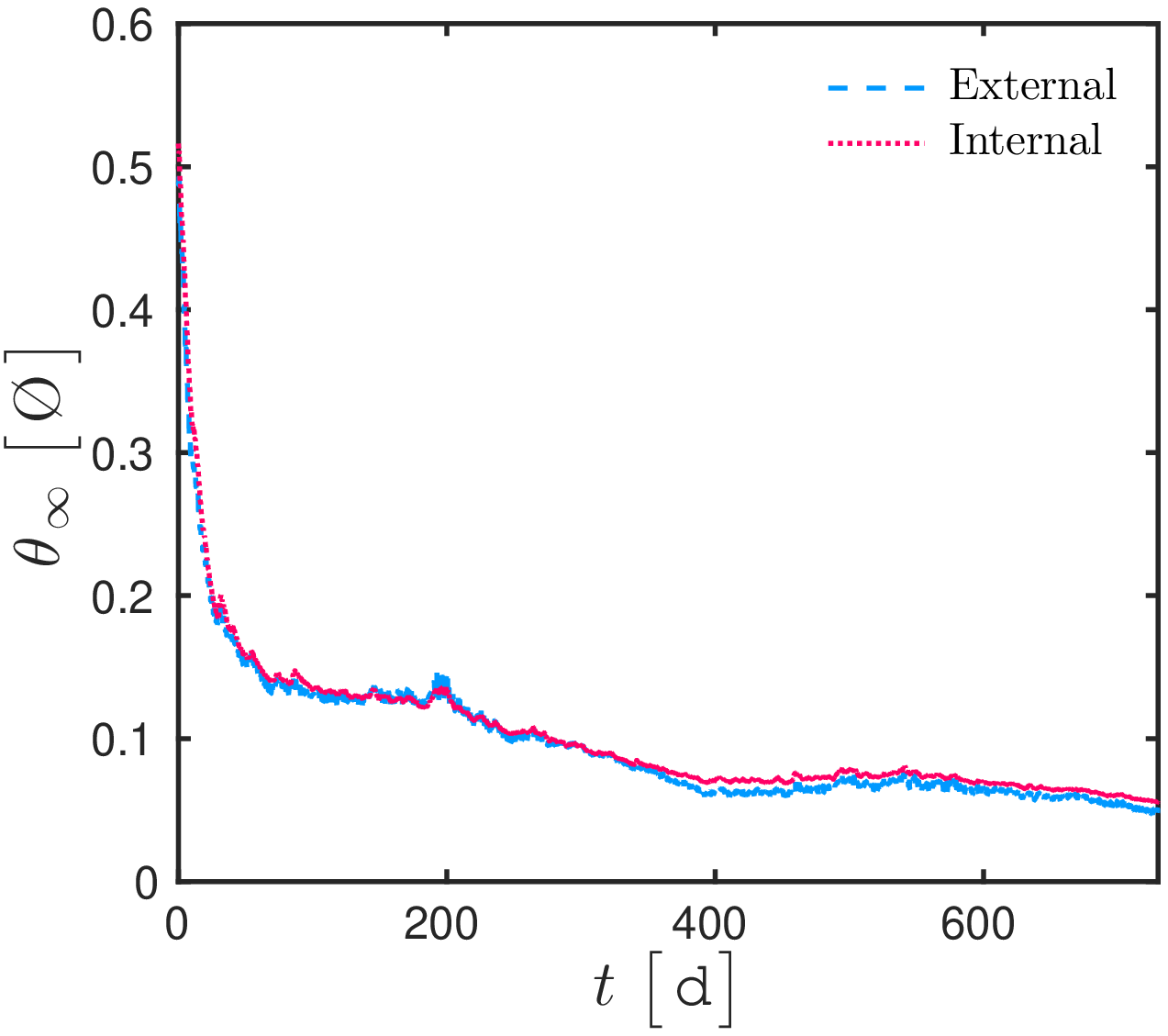}} 
		\caption{External and internal boundary conditions together with the experimental data for the middle of the material during two years after installation.}
		\label{fig:BC_RE_2years}	
	\end{figure}
Total simulation time is $\tau \, = \, 730 \,{\sf d} \,$, \emph{i.e.} the two years after installation of the wall (starts from July). 	
The material properties for the RE are obtained in the previous works \citep{soudani2014etude, soudani2016, soudani2017}. 
The specific heat capacity and dry mass are equal to
$c_{\,0} \egal 648  \ \cOUnit$ and 
$ \rho_{\,0} \egal 1730 \ \rhoUnit$ respectively. 
The diffusion coefficients under temperature and moisture gradient are equal to
$D_{\,T} \egal 10^{\,-10} \ \DTUnit$ and 
$D_{\,\theta} \, \left(\,\theta\,\right) \egal 10^{\,-7} \plus 2.4 \cdot 10^{\,-9} \cdot \,(\,\theta \moins 0.1\,) \ \DthetaUnit$. 
The function of the thermal conductivity $k_{\,T}$ as a function of $\theta$ is expressed as 
$k_{\,T} \, \left(\,\theta\,\right) \egal 0.6 \plus  5 \cdot \theta \ \kQUnit$. 
The value of the vapor transfer coefficient under a moisture gradient is equal to 
$k_{\,TM} \egal 4 \cdot 10^{\,-18} \ \kTMUnit$. 
The values of the thermo-physical constants are as follows:  
$c_{\,2} \egal 4185.5 \ \cQUnit$, 
$L_{\,12}^{\,\circ} \egal 2.5\cdot 10^{\,6} \ \hUnit$, 
$\rho_{\,2} \egal 10^{\,3} \ \rhoUnit$, 
$R_{\,1} \egal 2 \cdot 10^{\,-3} \ \RUnit$. 
The initial conditions are $\theta_{\,i} \, = \, 0.53 \ \bigl[\,\varnothing\,\bigr]$ and $T_{\,i} \, = \, 300.15 \,\bigl[\,{\mathsf K }\,\bigr]\,$.
\subsubsection{Offline procedure: candidates for empirical models}\label{sec:HM_empir_model}
The empirical models (EM) to describe the fluctuating values for heat and mass transfer are inspired from the study in the Section~\ref{sec:heat_eq} and an analytical solution of diffusion equation \citep{ozisik2002boundary}. 
The proposal is made for three EMs. 
As boundary conditions (see Figure~\ref{fig:BC_RE_2years}) for the moisture content are much less fluctuated, the EM of the mass transfer is based on analytical solution and is the same for all three models. 
Let us define the EM of the mass transfer as: 
\begin{equation}\label{eq:HM_empirical_model_v}
v^{\, \prime} \ :\ \left(\,x, \, t\,\right) \ \mapsto \Biggl(\, v_{\, 0} \plus v_{\, 1}\cdot \cos\, \left(\,\dfrac{x}{x_{\, v}}\,\right) \plus v_{\, 2}\cdot \sin\, \left(\,\dfrac{x}{x_{\, v}}\,\right) \,\Biggr) \, \cdot \,  \sin \,\Biggl(\,\frac{2 \pi}{\tau}  \cdot t\,\Biggr) \, ,
\end{equation}
where $\left(\, v_{\, 0}, \, v_{\, 1}, \, v_{\, 2}, \, x_{\, v} \,\right) \, \in \,\mathds{R}^{\, 4}$ are parameters to be estimated and optimized for each time-averaging period $\tau$ in offline procedure. 
In order to model the fluctuated values of the heat transfer three different EMs is proposed. 
The first candidate to model the temperature fluctuating values is based on analytical solution too:
\begin{equation}\label{eq:HM_empirical_model_I}
u^{\, \prime} \ :\ \left(\,x, \, t\,\right) \ \mapsto \Biggl(\, u_{\, 0} \plus u_{\, 1}\cdot \cos\, \left(\,\dfrac{x}{x_{\, u}}\,\right) \plus u_{\, 2}\cdot \sin\, \left(\,\dfrac{x}{x_{\, u}}\,\right) \,\Biggr) \, \cdot \,  \sin \,\Biggl(\,\frac{2 \pi}{\tau}  \cdot t\,\Biggr) \, ,
\end{equation}
where $\left(\,u_{\, 0}, \, u_{\, 1}, \, u_{\, 2}, \, x_{\, u}\,\right) \, \in \,\mathds{R}^{\, 4}$  are parameters to be optimized. 
So, the Model I is a set of Eq.~\eqref{eq:HM_empirical_model_v} and Eq.~\eqref{eq:HM_empirical_model_I}.  
The second model for the temperature is based on the study in Section~\ref{sec:Heat_empir_model}: 
\begin{equation}\label{eq:HM_empirical_model_II}
u'  \ :\ \left(\,x, \, t\,\right) \ \mapsto \ e^{\moins \dfrac{\bigl( \, x \moins x_{\, o} \, \bigr)}{\ell_{\, o}} } \cdot \sin \,\Biggl(\,\frac{2 \pi}{\tau}  \cdot t\,\Biggr) \, ,
\end{equation}
where the parameters $\left(\,x_{\, o}, \, \ell_{\, o} \,\right)\, \in \,\mathds{R}^{\, 2}$ are also optimized according to the period $\tau$. 
The Model II is a pair of Eq.~\eqref{eq:HM_empirical_model_v} and Eq.~\eqref{eq:HM_empirical_model_II}. 
The third candidate for $u'$ is taken as the sum of two exponentials:
\begin{equation}\label{eq:HM_empirical_model_III}
u'  \ :\ \left(\,x, \, t\,\right) \ \mapsto \ \Biggl(\, u_{\, 1} \, \cdot \, e^{\moins \dfrac{\bigl( \, x \moins x_{\, a} \, \bigr)}{\ell_{\, a}} } \plus  u_{\, 2} \, \cdot \, e^{\moins \dfrac{\bigl( \, x \moins x_{\, b} \, \bigr)}{\ell_{\, b}} }  \,\Biggr)\cdot \sin \,\Biggl(\,\frac{2 \pi}{\tau}  \cdot t\,\Biggr) \, ,
\end{equation}
where the parameters $\left(\,u_{\, 1}, \, x_{\, a}, \, \ell_{\, a}, \,u_{\, 2}, \, x_{\, b}, \, \ell_{\, b} \,\right) \, \in \,\mathds{R}^{\, 6}$ are optimized and thus, the Model III is a combination of Eq.~\eqref{eq:HM_empirical_model_v} end Eq.~\eqref{eq:HM_empirical_model_III}.
\begin{table}[!ht]  
	\caption{\small Parameters of empirical model pairs in relation to time-averaging periods $\tau \egal [\,6, \, 12, \, 24, \, 48\,] \ \mathsf{h}$.}                                            
	\bigskip 
	\label{tab:empir_model_HM}                                                                  
	\small
	\begin{tabular}{|c||c|c|c|c||c|c|c|c|}  
		\cline{2-9}                                                                                
		 \multicolumn{1}{ c|| }{\it{\textbf{M I}}} & \multicolumn{4}{ c|| }{$v^{\, \prime}$}  &  \multicolumn{4}{ c| }{$u^{\, \prime}$} \\                                              
	\hline \hline 
		 $\tau$ & $v_{\, 0}$& $v_{\, 1}$& $v_{\, 2}$ &$x_{\, v}$& $u_{\, 0}$ &$u_{\, 1}$  &$u_{\, 2}$  & $x_{\, u}$\\                                           
		\hline \hline                                                                         
		 $6$ &$8.2 \times 10^{\, -3}$& $-8.3 \times 10^{\, -3}$& $2.6 \times 10^{\, -4}$ & $2.9 \times 10^{\, -2}$& $6.4 \times 10^{\, -3}$ &$-4.7 \times 10^{\, -3}$ & $-5.8 \times 10^{\, -3}$& $1.6$  \\
	\hline 
		$12$ &$1.1 \times 10^{\, -2}$ & $-1.1 \times 10^{\, -2}$& $-5.7 \times 10^{\, -2}$& $-1.3 \times 10^{\, -2}$& $6.6 \times 10^{\, -4}$ &$5.6 \times 10^{\, -3}$ & $-3.1 \times 10^{\, -3}$& $10$  \\
		\hline 
		$24$ &$-1.9 \times 10^{\, -2}$ & $1.9 \times 10^{\, -2}$& $4.9 \times 10^{\, -3}$& $5.8 \times 10^{\, -1}$& $-4.3 \times 10^{\, -3}$ &$8.1 \times 10^{\, -3}$ & $-2.7 \times 10^{\, -3}$& $6.2 \times 10^{\, -2}$  \\
		\hline 
		$48$ &$2.5 \times 10^{\, -2}$ & $-2.3 \times 10^{\, -2}$& $-1.5 \times 10^{\, -1}$& $3 \times 10^{\, -2}$& $-2.1 \times 10^{\, -2}$ &$1.6 \times 10^{\, -2}$ & $1.9 \times 10^{\, -2}$& $2.1$  \\
		\hline \hline 
			\end{tabular} 
	\begin{tabular}{|c||c|c|c|c||c|c|}  
		\cline{2-7}                                                                                
		\multicolumn{1}{ c|| }{\it{\textbf{M II}}} & \multicolumn{4}{ c|| }{$v^{\, \prime}$}  &  \multicolumn{2}{ c| }{$u^{\, \prime}$} \\                                              
		\hline \hline 
		$\tau$  & $a_{\, 0}$& $a_{\, 1}$& $b_{\, 1}$ &$w$& \multicolumn{1}{ c| }{$x_{\, o}$} & \multicolumn{1}{ c| }{$\ell_{\, o}$}\\                                           
		\hline \hline                                                                         
		$6$& $4.2 \times 10^{\, -3}$& $ -4.6 \times 10^{\, -3}$& $ 2.4 \times 10^{\, -2}$ &$4.6 \times 10^{\, -2}$ & \multicolumn{1}{ c| }{$0.9$} & \multicolumn{1}{ c| }{$-11.2$}  \\
		\hline 
		$12$ & $2 \times 10^{\, -4}$& $7.6 \times 10^{\, -5}$& $-3.7 \times 10^{\, -1}$ &$2.6 \times 10^{\, -3}$ & \multicolumn{1}{ c| }{$1.16$} & \multicolumn{1}{ c| }{$-12.3$}  \\
		\hline 
		$24$ & $ 5.1\times 10^{\, -4}$& $ 4.9\times 10^{\, -4}$& $-4.1 \times 10^{\, -1}$ &$4.9 \times 10^{\, -3}$ & $0.45$ & $-12.1$  \\
		\hline 
		$48$ & $ 1.1\times 10^{\, -3}$& $1.1 \times 10^{\, -3}$& $-3.5 \times 10^{\, -1}$ &$1.3 \times 10^{\, -2}$ & $1.48$ & $-12.8$  \\
		\hline \hline 
	\end{tabular} \\
		\begin{tabular}{|c||c|c|c|c||c|c|c|c|}  
			\cline{2-9}                                                                                
			\multicolumn{1}{ c|| }{\it{\textbf{M III}}} & \multicolumn{4}{ c|| }{$v^{\, \prime}$}  &  \multicolumn{4}{ c| }{$u^{\, \prime}$} \\                                              
			\hline \hline 
			$\tau$ & $a_{\, 0}$& $a_{\, 1}$& $b_{\, 1}$ &$x_{\, v}$ &$u_{\, 1}$  &$ \ell_{\, a}$ &$u_{\, 2}$  &$\ell_{\, b}$ \\                    
			\hline \hline                                                                         
			$6$ & $-7.1 \times 10^{\, -4}$& $-4.3 \times 10^{\, -5}$& $7.1 \times 10^{\, -4}$ &$3.1 $ &$-3.3 \times 10^{\, -3}$ &$-9.5 \times 10^{\, -1}$ &$2.5 \times 10^{\, -3}$ &$1.1 \times 10^{\, -3}$ \\		
			\hline 
			$12$ 	& $1.6 \times 10^{\, -2}$& $-1.4 \times 10^{\, -2}$& $-8.3 \times 10^{\, -3}$ &$-6.7 \times 10^{\, -3}$ &$-9.3 \times 10^{\, -1}$ &$-10 $ &$9.2 \times 10^{\, -1}$ &$-8.4 $ \\
			\hline 
			$24$ & $1.2 \times 10^{\, -2}$& $-1.1 \times 10^{\, -2}$& $-1.6 \times 10^{\, -3}$ &$6.8 \times 10^{\, -1}$ &$-8.1 \times 10^{\, -3}$ &$-2.7 \times 10^{\, -1} $ &$9.2 \times 10^{\, -3}$ &$5.6\times 10^{\, -1}$ \\	
			\hline 
			$48$ & $3.2 \times 10^{\, -3}$& $-9.5 \times 10^{\, -4}$& $-4.5 \times 10^{\, -1}$ &$1 \times 10^{\, -2}$ &$4.2 \times 10^{\, -3}$ &$-2.6 \times 10^{\, -1} $ &$-2.4 \times 10^{\, -3}$ &$1.2$ \\
			\hline \hline 
		\end{tabular} 			
		\end{table}	 
By following the same methodology as in the Sections~\ref{sec:Heat_empir_model} and \ref{sec:empir_model_HM}, the optimization process is carried out for all three EM candidates. 
The simulation time is chosen as $\mathds{T} \egal 240 \, \sf{h}$. 
It should be noted that the offline procedures are run for each pair of models separately, thereby the coefficients for $v^{\, \prime} \,\left(\,x, \, t\,\right)\,$ in model~\eqref{eq:HM_empirical_model_v} are different for each case. 
The parameters $x_{\, a}$ and $x_{\, b}$ of the Model III are taken to be equal to zero. 
All optimized parameters of empirical models are collected in the Table~\ref{tab:empir_model_HM}. 	                                                             
\begin{table}[!ht]  
	\centering    
	\caption{\small Computational time and the errors $\varepsilon_{\, 2}$ between simulation results of the dimensionless complete model and average reduced models.}                                            
	\bigskip 
	\label{tab:empir_model_HM_err}                                                                  
	\small
	\begin{tabular}{|c||c|c||c||c|}  
		\cline{2-5}                                                                                
		\multicolumn{1}{c||}{Model}  & $\varepsilon_{\, 2}^{\, v} \ [\,\sf{\varnothing}\,] $ &$\varepsilon_{\, 2}^{\, u} \ [\,\sf{\varnothing}\,] $ &$\varepsilon_{\, 2}^{\, uv} \ [\,\sf{\varnothing}\,]$ & $\varrho_{\, \text{\tiny CPU}}^{\, \text{\tiny day}} $ \\                                      
		\hline \hline 
		\multicolumn{5}{c}{$\tau \egal 6$}  \\                                              
		\hline \hline                                                                   
		\it{M I}  &  $2.9 \times 10^{\, -2}$ & $2 \times 10^{\, -1}$ &  $29.78$& 3.48  \\
		\hline  \hline                                                                   
		\it{M II}  & $8.2 \times 10^{\, -2}$& $2 \times 10^{\, -1}$ &  $30.55$ & 3.04 \\
		\hline \hline                                                                        
		\it{M III} & $7.8 \times 10^{\, -1}$& $2 \times 10^{\, -1}$ &  $30.51$ & 3.37\\	
		\hline \hline 
		\multicolumn{5}{c}{$\tau \egal 12$}  \\                                              
		\hline \hline                                                                   
		\it{M I}  &  $9.6 \times 10^{\, -2}$ & $1.01$ &  $148.06$& 3.17  \\
		\hline  \hline                                                                   
		\it{M II}  & $9 \times 10^{\, -2}$& $1.01$ &  $147.36$ & 2.33 \\
		\hline \hline                                                                        
		\it{M III} & $5.4 \times 10^{\, -1}$& $1.43$ &  $215.91$ & 2.59\\								
		\hline \hline                                                                                                                                     
	\end{tabular}  
	\qquad 
	\begin{tabular}{|c||c|c||c||c|}  
		\cline{2-5}                                                                                
		\multicolumn{1}{c||}{Model}  & $\varepsilon_{\, 2}^{\, v} \ [\,\sf{\varnothing}\,] $ &$\varepsilon_{\, 2}^{\, u} \ [\,\sf{\varnothing}\,] $ &$\varepsilon_{\, 2}^{\, uv} \ [\,\sf{\varnothing}\,]$ & $\varrho_{\, \text{\tiny CPU}}^{\, \text{\tiny day}} $ \\                                              
		\hline \hline  
		\multicolumn{5}{c}{$\tau \egal 24$}  \\                                              
		\hline \hline                                                                   
		\it{M I}  &  $1.4 \times 10^{\, -1}$ & $7.9 \times 10^{\, -1}$ &  $117.47$& 2.26  \\
		\hline  \hline                                                                   
		\it{M II}  & $1.5 \times 10^{\, -1}$& $7.9 \times 10^{\, -1}$ &  $117.95$ & 2.18 \\
		\hline \hline                                                                        
		\it{M III} & $1.5 \times 10^{\, -1}$& $8 \times 10^{\, -1}$ &  $118.37$ & 2.37\\								
		\hline \hline  
		\multicolumn{5}{c}{$\tau \egal 48$}  \\                                              
		\hline \hline                                                                   
		\it{M I}  &  $3.1 \times 10^{\, -1}$ & $1.18$ &  $176.01$& 2.01  \\
		\hline  \hline                                                                   
		\it{M II}  & $2.7 \times 10^{\, -1}$& $1.18$ &  $175.61$ & 1.95 \\
		\hline \hline                                                                        
		\it{M III} & $2.6 \times 10^{\, -1}$& $1.19$ &  $176.47$ & 2.27\\								
		\hline \hline  		                                                                                                                                    
	\end{tabular}                                                                                                                                          
\end{table}

The Table~\ref{tab:empir_model_HM_err} displays the errors $\varepsilon_{\, 2}$ computed for the heat and moisture transfers separately as well as the total error as in Eq.~\eqref{eq:err_empirical_model_HM}.
In addition to that, $\varrho_{\, \text{\tiny CPU}}^{\, \text{\tiny day}}$ the computational time ratio per day required by each of the simulations is calculated as in Eq.~\eqref{eq:varrho_CPUday}.  
It can be clearly seen that the Model II is the fastest performing candidate, while the error varies according to the time-averaging size $\tau$.  
By comparing the errors and computational time it can be concluded that the Model II presents the best compromise than any other candidate.	
\subsubsection{Results and discussion for online procedure}\label{sec:HM_results}
\begin{figure}[!ht]
	\centering
	\subfigure[]{\includegraphics[width=.45\textwidth]{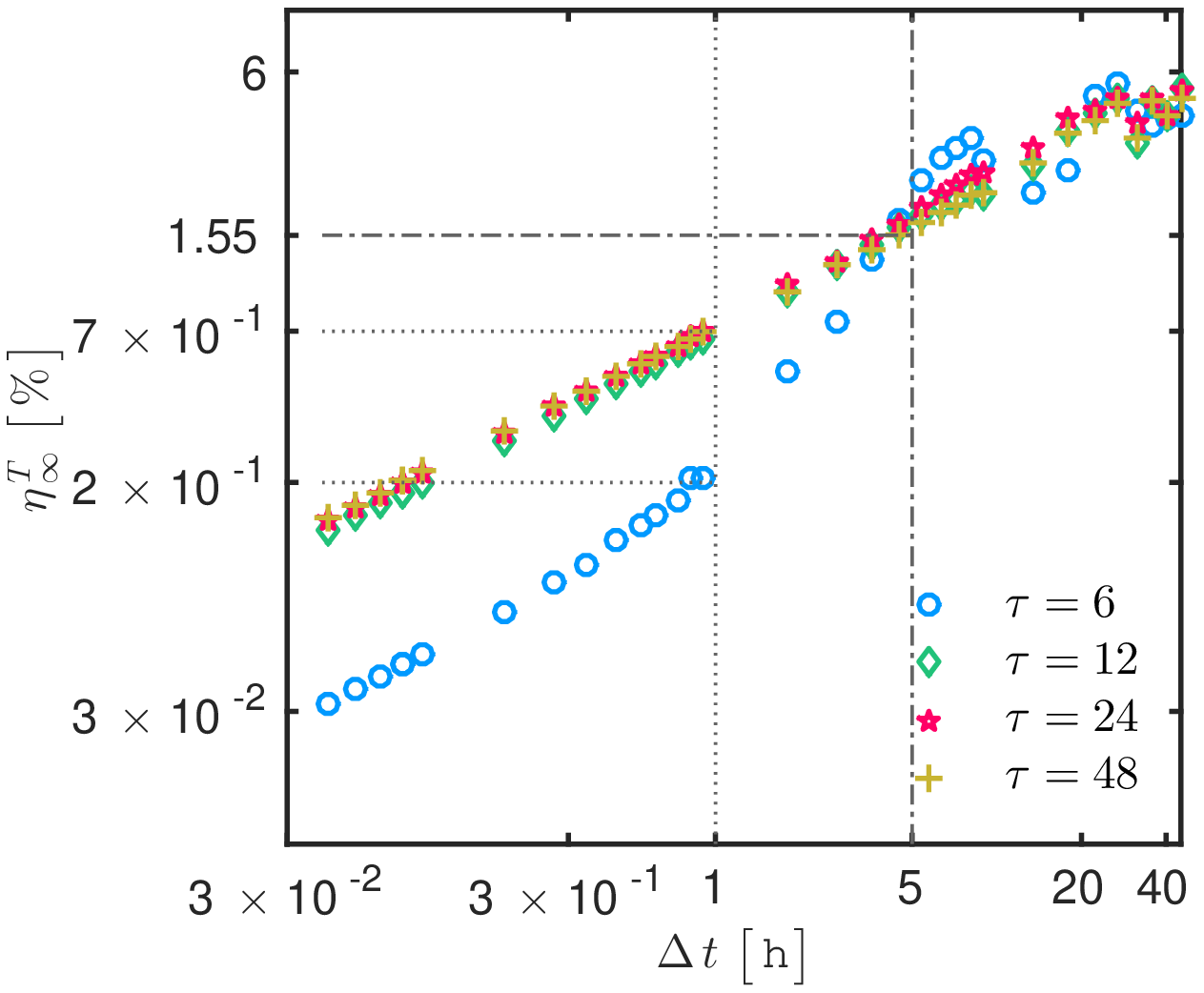}} \hspace{0.2cm}
	\subfigure[]{\includegraphics[width=.45\textwidth]{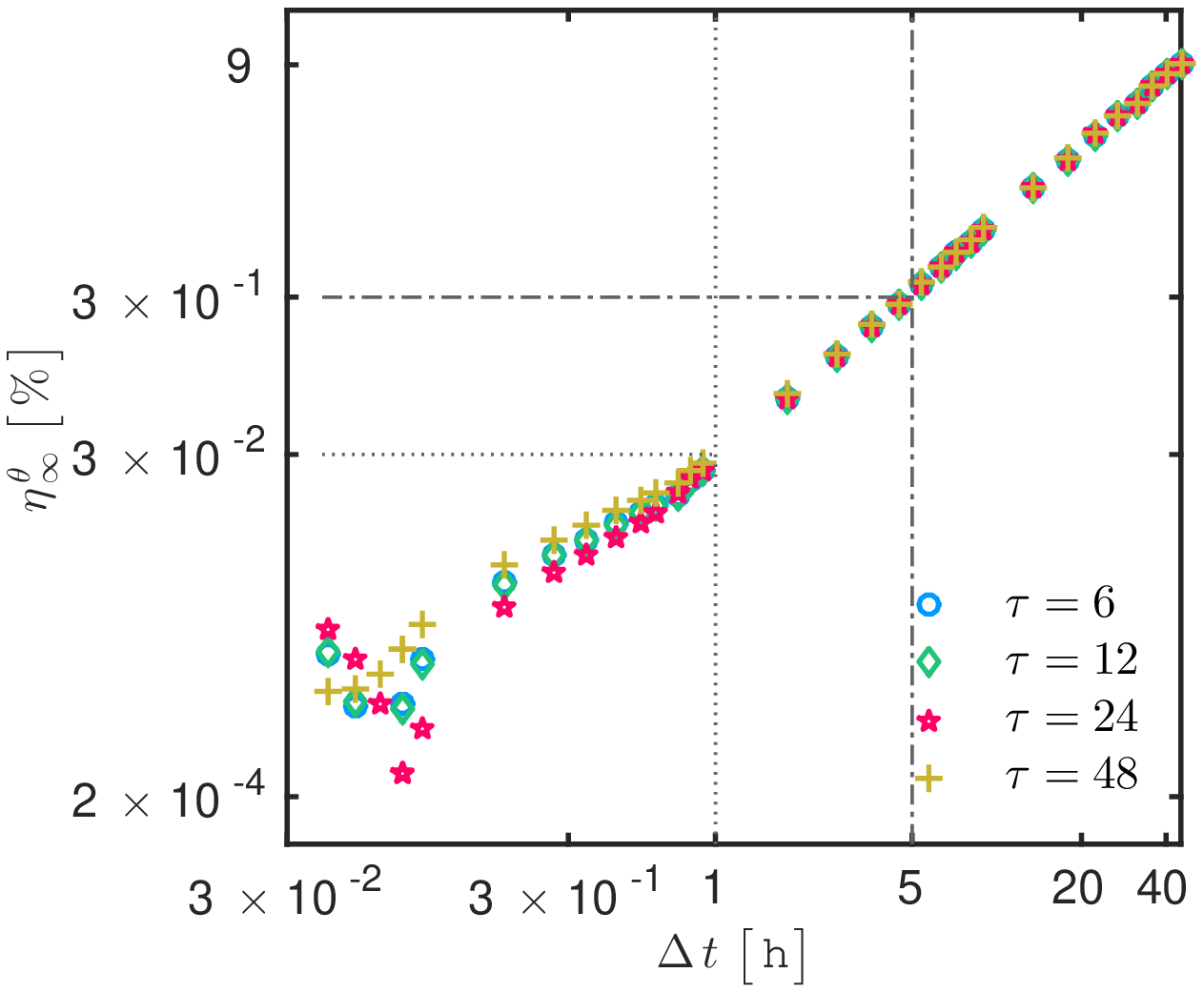}} \\
	\subfigure[]{\includegraphics[width=.45\textwidth]{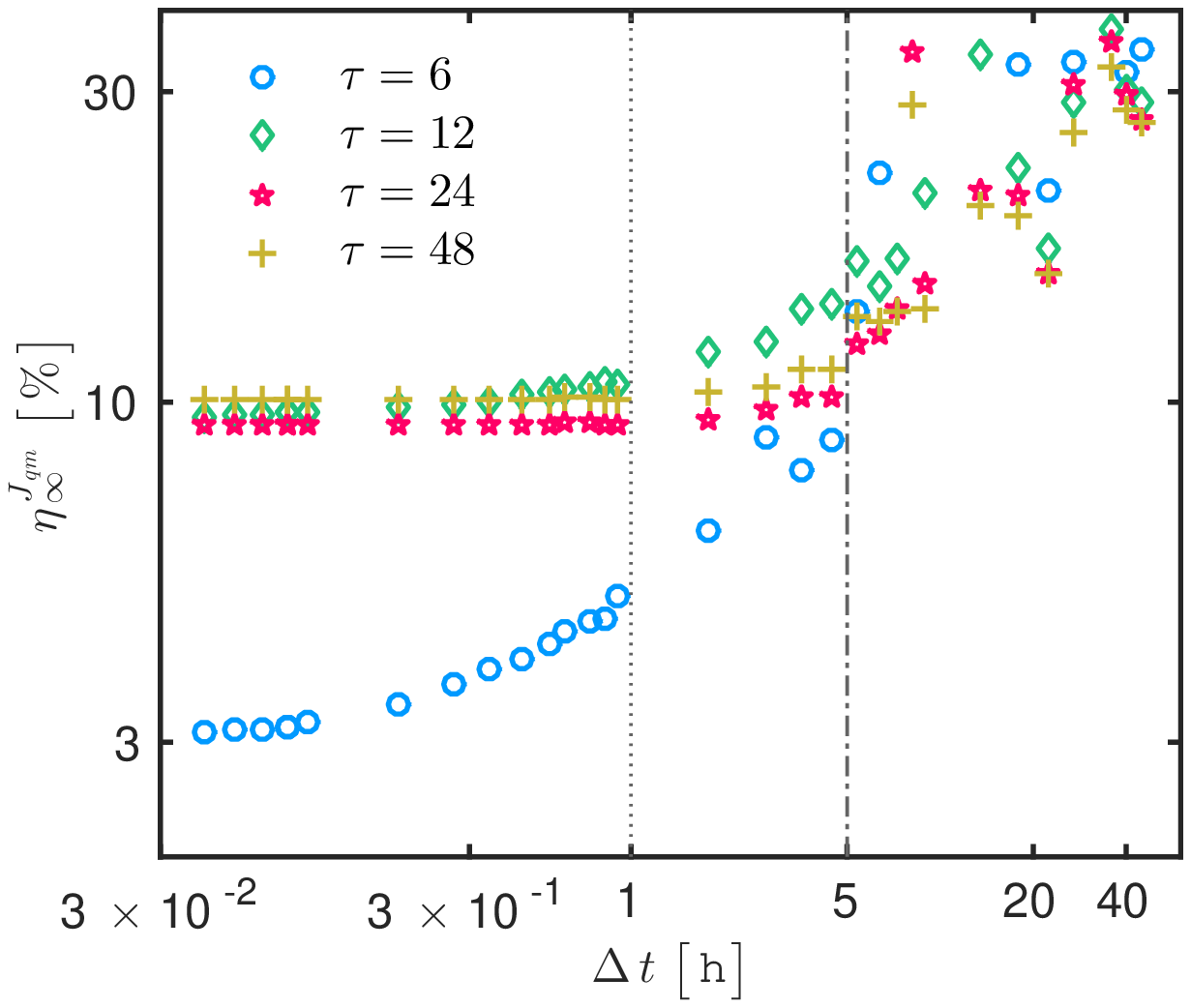}} \hspace{0.2cm}
	\subfigure[]{\includegraphics[width=.45\textwidth]{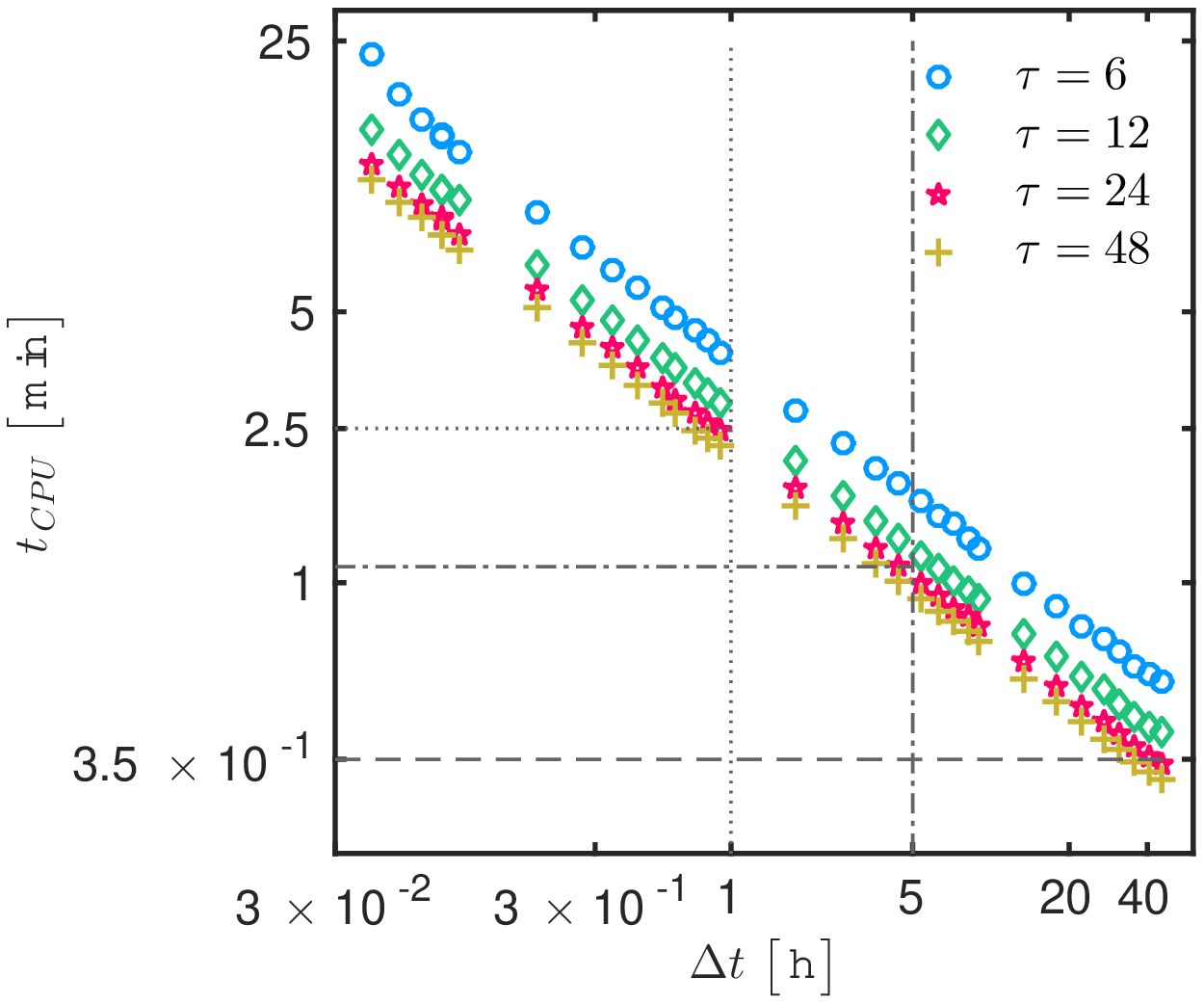}} 
	\caption{\small Average reduced model in comparison with the complete model:  $\eta_{\, \infty}$ errors of temperature (\emph{a}), moisture content (\emph{b}) and total output flux (\emph{c}). CPU time to solve average reduced heat equation (\emph{d}). }
	\label{fig:HM_CMvsARM}
\end{figure}

\paragraph{\small On the importance of period $\tau$ and time-step size $\Delta \, t$. }
The chosen empirical Model II is now implemented into online procedure. 
The next validation step is to discover the influence of important parameters.   
For this purpose both complete and average reduced models are run for the simulation time $\mathds{T} \egal 150 \,\mathsf{d}$. 
The Figure~\ref{fig:HM_CMvsARM} presents the $\eta_{\, \infty}$ errors and the computational time of ARM with different time-averaging periods.  
The errors are computed not only for various $\tau$ but also for the time-step $\Delta \, t$. 
It can be seen that even with the time-step $\Delta \, t \egal 5 \, \sf{h}$ the global relative errors for temperature and the moisture content remain in the tolerable range below $2 \, \%$.
The simulations themselves take only few minutes and the CPU time can be reduced for almost $100$ times as the size of the time-step increases. 
One can further observe the $\eta_{\, \infty}$ error of the output flux in Figure~\ref{fig:HM_CMvsARM} (\emph{c}). 
It can be seen that for the time-step sizes $\Delta \, t \, \in \, [\,1 ; \, 5\,]\, \sf{h}$ the error is tolerable. 
Moreover, it can be observed that the most trustworthy time-averaging period is still $\tau \egal 6 \, \sf{h}$ as for the heat transfer case in the Section~\ref{sec:Results_heat}. 
\paragraph{\small Comparison with experimental data.}
Based on the above study, as the next step, the performance of the ARM is evaluated in comparison with the experimental data of two years \citep{soudani2017}. 
It should be noted that the experimental data, used in the study, has not been compared to the numerical results in its original publications. 
Therefore, it is important to determine the uncertainties of experimental measurements before starting any comparison with numerical simulations. 
The formula to compute the uncertainty of a measurement is given below as \citep{taylor1997introduction}:  
\begin{equation}\label{eq:uncertainty_of_measument}
\left(\, \sigma_{\, m}^{\, T, \, \theta}\, \right)^{\, 2} \egal \left(\, \sigma_{\, S}^{\, T, \, \theta}\, \right)^{\, 2} \plus \left(\, \sigma_{\, P}^{\, T, \, \theta}\, \right)^{\, 2} \, ,
\end{equation}
where $\sigma_{\, S}^{\, T, \, \theta}$ are the accuracy of sensors used for the experiments and $\sigma_{\, P}^{\, T, \, \theta}$ are the positioning sensitivities, evaluated numerically. 
The former is relative to an absolute standard, so the values are calculated with the characteristics of the sensor to measure water content \texttt{Campbell CS616}, with the accuracy of $\sigma_{\, S}^{\, \star \, \theta} \egal \pm \, 2.5 \, \%$ of volumetric water content, and temperature sensors \texttt{Campbell CS215}, the accuracy of which is $\sigma_{\, S}^{\, \star \, T} \egal\pm \, 1.5 \, \%$ of temperature in $^\circ \sf{C}$: 
\begin{equation}\label{eq:uncertainty_measument_sensors}
\sigma_{\, S}^{\, \theta} \egal \sigma_{\, S}^{\, \star \, \theta} \, \cdot \, \theta_{\, m}  \, , \qquad
\sigma_{\, S}^{\, T} \egal  \sigma_{\, S}^{\, \star \, T} \, \cdot \, T_{\, m}  \, ,
\end{equation}
where $\theta_{\, m}$ and $T_{\, m}$ are the data obtained from monitoring changes in the middle of a material. 
The latter depends on the accuracy of a placement which can be taken as $\Delta \, x_{\, P} \egal 0.5 \, {\sf cm}$ and consequently the uncertainty can be evaluated in the following way: \begin{equation}\label{eq:uncertainty_measument_position}
\sigma_{\, P}^{\, \theta} \egal \pd{\theta}{x} \, \cdot \, \Delta \, x_{\, P}  \, , \qquad
\sigma_{\, S}^{\, T} \egal  \pd{T}{x} \, \cdot \, \Delta \, x_{\, P}  \, .
\end{equation}
Thereafter, corresponding uncertainty of measurement values for the water content and temperature are determined with the following equalities:
\begin{equation}\label{eq:uncertainty_measument_values}
\theta_{\, m}^{\, \pm} \egal \theta_{\, m} \, \pm \, \sigma_{\, m}^{\, \theta}  \, , \qquad
T_{\, m}^{\, \pm} \egal T_{\, m} \, \pm \, \sigma_{\, m}^{\, T} \, .
\end{equation}

The final simulation time is set as $\mathds{T} \egal 730 \ \sf{days}$, \emph{i.e.} two years. 
The fidelity of the ARM is now evaluated by setting up the time-averaging period as $\tau \egal 6 \, \sf{h}$. 
The time-step size is $\Delta \, t \egal 1 \, \sf{h}$ for both complete and average reduced models. 

\begin{figure}[!ht]
	\centering
	\subfigure[]{\includegraphics[width=.45\textwidth]{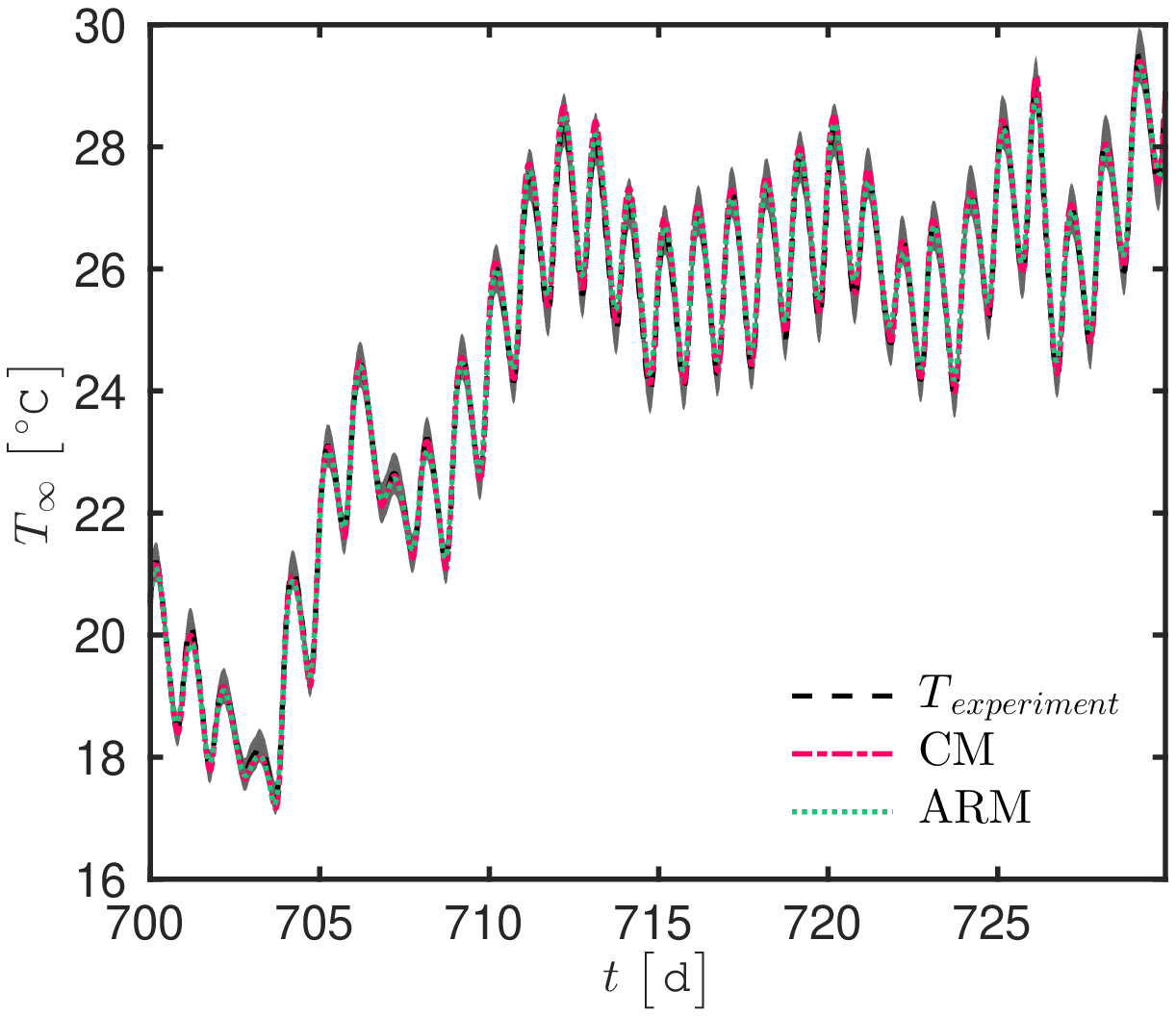}} \hspace{0.2cm}
	\subfigure[]{\includegraphics[width=.45\textwidth]{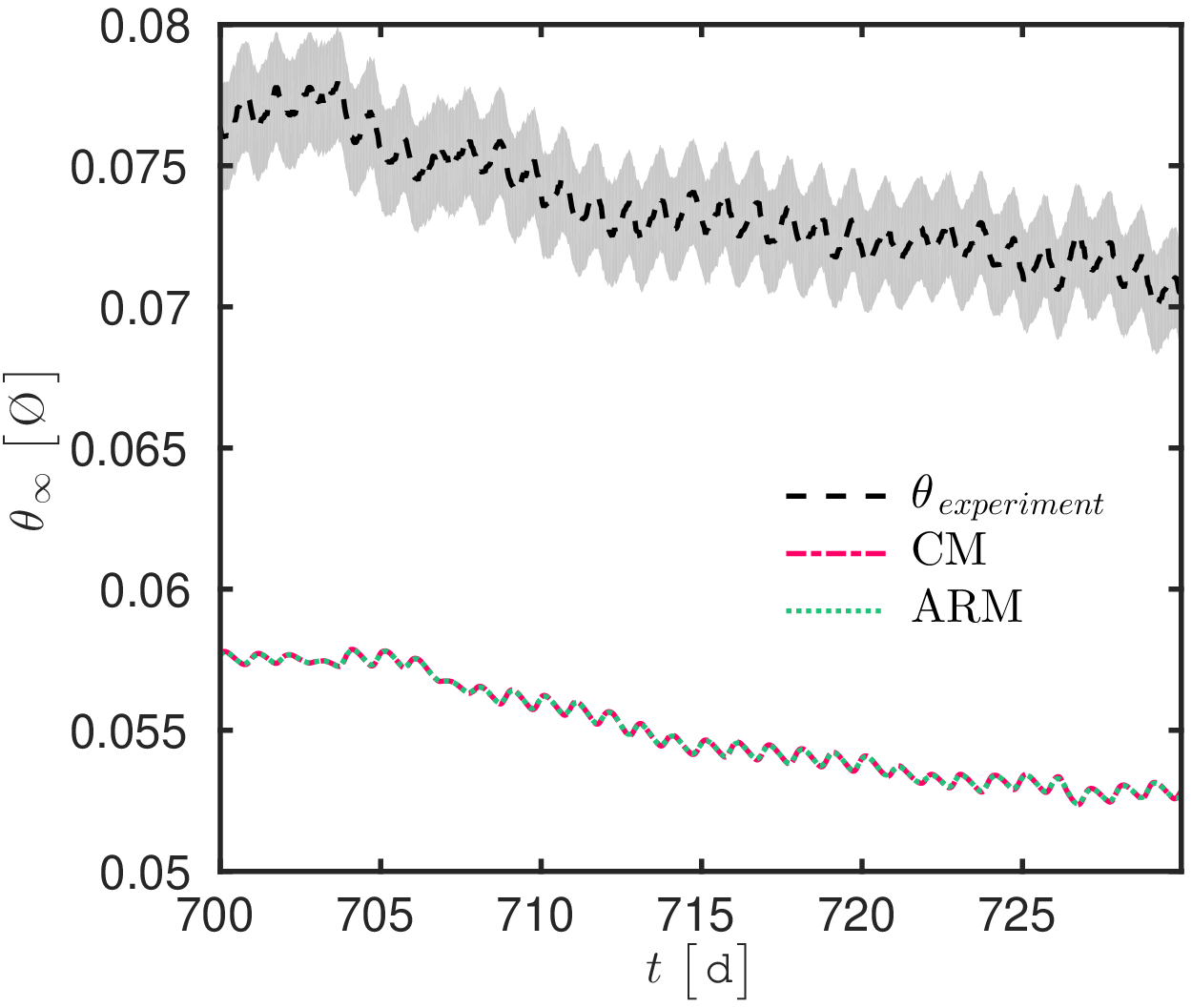}} \\
	\subfigure[]{\includegraphics[width=.45\textwidth]{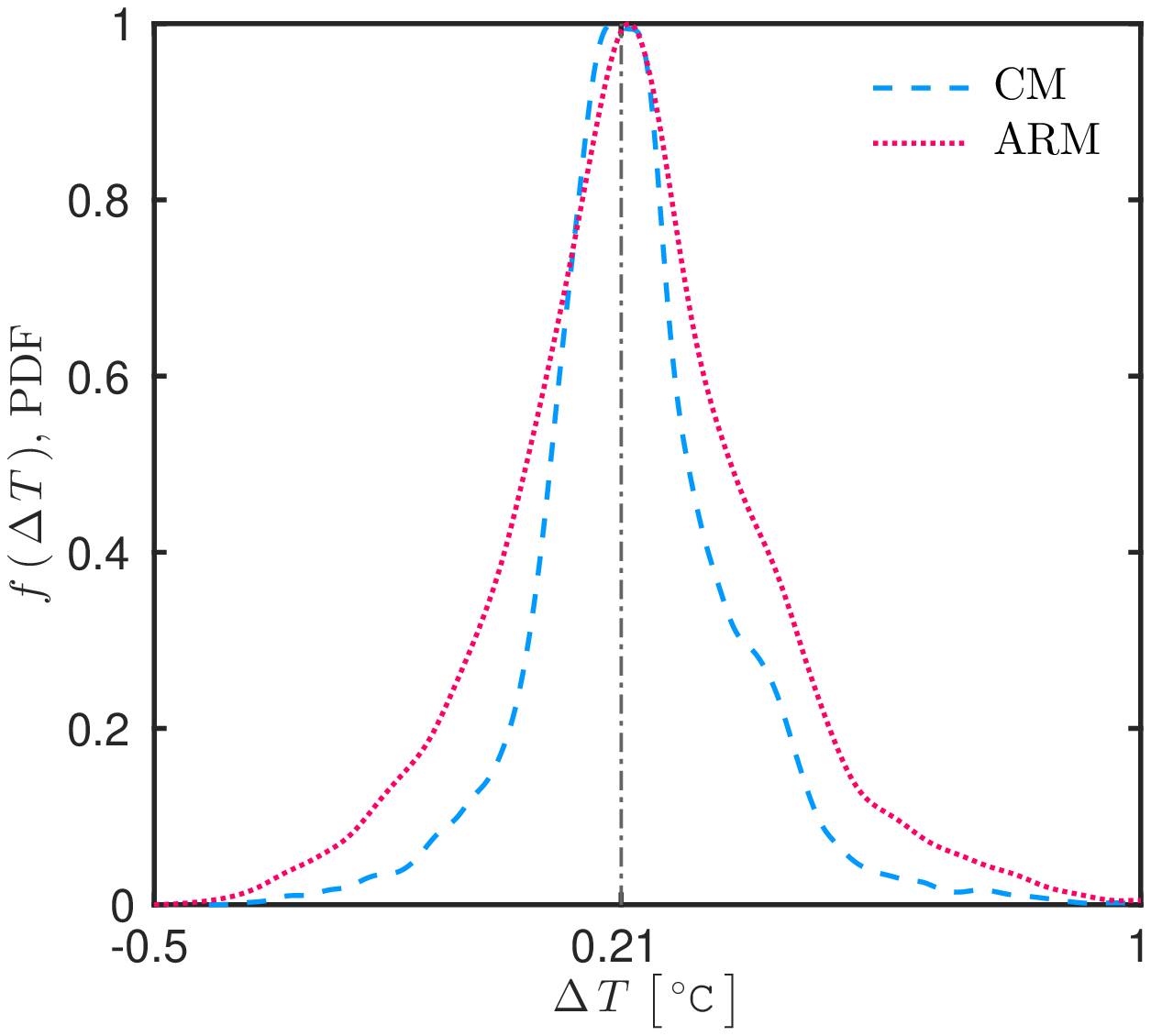}} \hspace{0.2cm}
	\subfigure[]{\includegraphics[width=.45\textwidth]{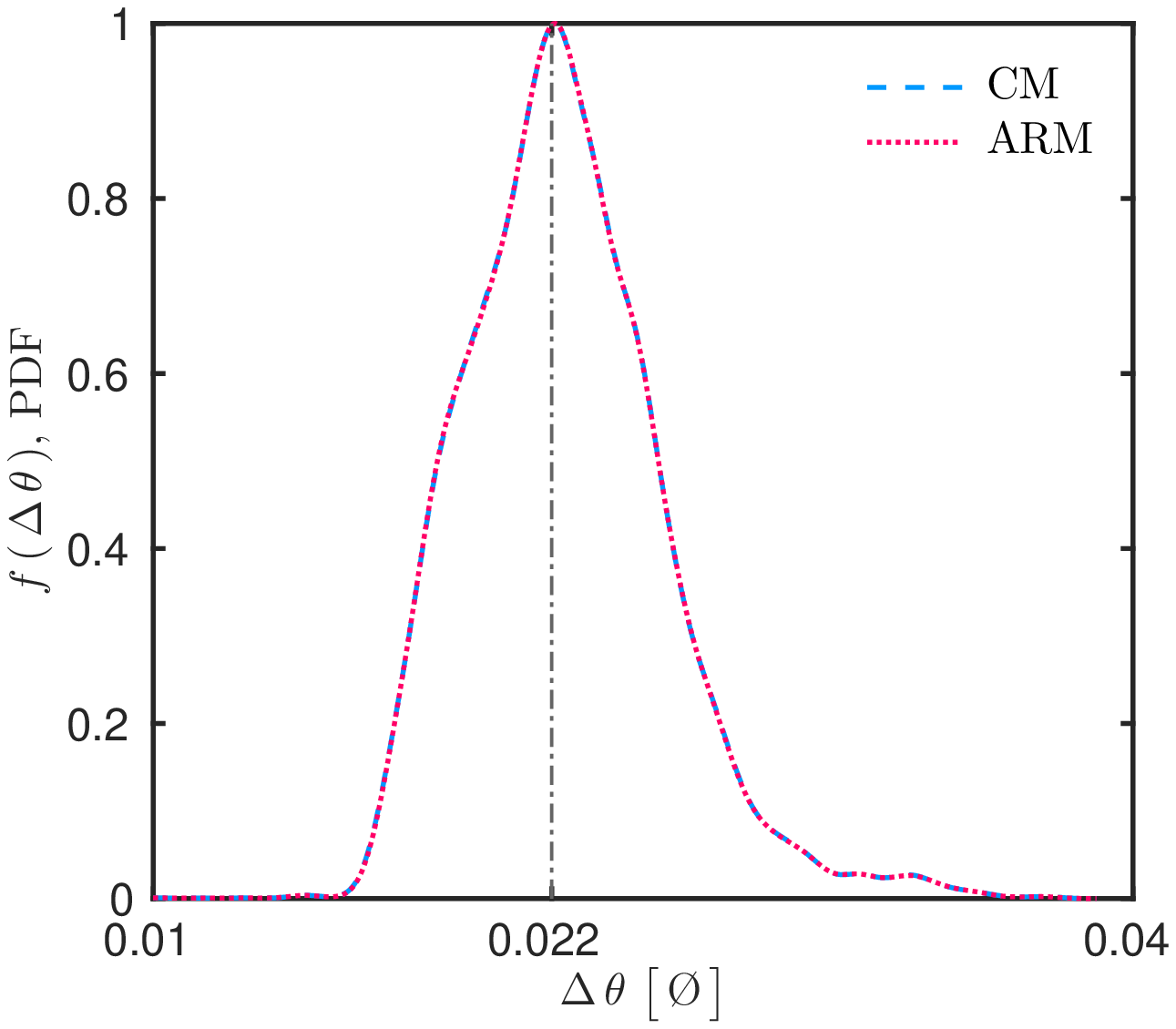}}
	\caption{\small Uncertainty of measurements and time evolutions for the last month of the two years' period of the temperature (\emph{a}) and mass content (\emph{b}) in the middle of the wall in comparison with the experimental data with the complete and average reduced models results with $\Delta \, t \egal 1 \, \sf{h}$ and $\tau \egal 6 \, \sf{h}$. 
	The distribution functions $f$ are computed for the temperature (\emph{c}) and mass content (\emph{d}). }
	\label{fig:HM_Lt_dt1}
\end{figure}

In Figures~\ref{fig:HM_Lt_dt1} (\emph{a}) and (\emph{b}) one can observe time evolutions of the temperature and moisture content together with the uncertainty of measurements. 
The plots are given only for the last month of the period $\mathds{T}$ in order to show the graph in more details. 
It can be seen that the phenomena are modeled in accordance with the general behavior. 
Moderate errors can be due to the lack of information about material properties or other unnoticed phenomena, which are not included into the physical model. 
Nonetheless, the difference in prediction of the moisture content is tolerable in a given configuration and scale of study. 
Further, one can observe the Figures~\ref{fig:HM_Lt_dt1} (\emph{c}) and (\emph{d}). 
The distribution function $f$ permits to determine whether numerical results over or under predict the experimental data. 
It can be seen that the numerical models slightly overestimate the experimental results for both phenomena. 
The discrepancy of function $f$ for temperature evolution computed with CM and ARM is due to averaging. 

At this point, it is worth recalling the purpose of the given study. 
The above accuracy is obtained, first of all, thanks to the Super--Time--Stepping (STS) numerical method, whose reliability is proved in the previous works \citep{meyer2014, abdykarim2019} and can be seen in Figures~\ref{fig:HM_CMvsARM} (\emph{a} - \emph{b}). 
It can be observed, that the enlarged time-step sizes do not disturb the stability of results. 
So, the STS method permits to overcome numerical restriction of the stability condition, while the physical restriction to the time grid is relaxed with the ARM. 
The combination of those two methods allows to justify the implementation of big time-steps as $\Delta \, t \egal 1 \, \sf{h}$. 

In order to further evaluate the efficiency of the ARM in relation to the complete model, the Table~\ref{tab:HM_CMandRM_results} is presented. 
The global relative error is in the same range for both CM and ARM with the time-step size $\Delta \, t \egal 1 \ \mathsf{h}$. 
It can be seen that the predictions are more accurate for the heat transfer, the error being smaller for CM.  
In terms of the CPU time, it is shown, that by maintaining the same order of the error, it is possible to cut the computational effort by almost $74 \, \%$. 
Here it can be mentioned that initially the STS numerical method itself allows to cut the computational effort in comparison to standard \textsc{Euler} explicit scheme. 
The ratio $\varrho_{\, \text{CPU}}\ \egal \dfrac{t_{\, \text{CPU}}^{\, \text{RKL1}}}{t_{\, \text{CPU}}^{\, \textsc{Euler}}} \cdot 100 \, \% \egal 34.8 \, \%$, \emph{i.e.} saving of $65 \, \%$, according to previous work dedicated to a study of the fidelity of the STS methods in application to heat and mass transfer phenomena \citep{abdykarim2019}. 
So, in addition to the advantages of the numerical method in comparison to standard approaches, with the ARM it is possible to reach even faster computational efficiency. 
This conclusion is supported with the ratio $\varrho_{\, \text{\tiny CPU}}^{\, \text{\tiny day}}$, which shows that simulation for one astronomical day can be performed almost four times faster with the ARM being $1.4 \,\left[\,{\sf s/d}\,\right]$ in contrast to $5.5 \,\left[\,{\sf s/d}\,\right]$ of the complete model. 
\begin{table}[!ht]                                                              
	\centering                                                               
	\begin{tabular}{|l|c|c|}  
		\cline{2-3}                                                                                                          
		\multicolumn{1}{c|}{} & CM: $\Delta \, t \egal 1 \ \mathsf{h}$  & ARM: $\Delta \, t \egal 1 \ \mathsf{h}$ and $\tau \egal 6 \ \mathsf{h}$  \\   		                                                 
		\hline\hline                                                                                                                                   
		$\eta_{\, \infty}^{\, T} \ \bigl[\,\%\,\bigr] $  & $ 7.2 \times 10^{\, -3} $ & $ 7.9 \times 10^{\, -3} $\\
		\hline  
		$\eta_{\, \infty}^{\, \theta} \ \bigl[\,\%\,\bigr] $  & $4.8 \times 10^{\, -2}$ &   $4.8\times 10^{\, -2}$\\
		\hline  
		$\varrho_{\, \text{\tiny CPU}}^{\, \text{\tiny day}} \,\left[\,{\sf s/d}\,\right]$ &$5.5$ &$1.4$\\			
		\hline  
		$\varrho_{\, \text{CPU}}\ \bigl[\,\%\,\bigr]$ &\multicolumn{2}{c|}{$26.2$}\\	
		\hline\hline                                            
	\end{tabular}  
	\bigskip  \caption{\small The comparison of complete and average reduced model results.} 					
	\label{tab:HM_CMandRM_results}                                                       
\end{table} 

Since fidelity of the numerical model is proved, it can be used in order to compute the physical values such as conductions loads, given in Eq.~\eqref{eq:HM_E} and thermal resistance, given in Eq.~\eqref{eq:Thermal_resistances}.  
These computations are necessary for several reasons, including observations about the influence of the mass transfer on the energy efficiency. 
Moreover, simulations can be performed efficiently and considerably fast as was shown above.  
			\begin{figure}[!ht]
				\centering
	\subfigure[]{\includegraphics[width=.9\textwidth]{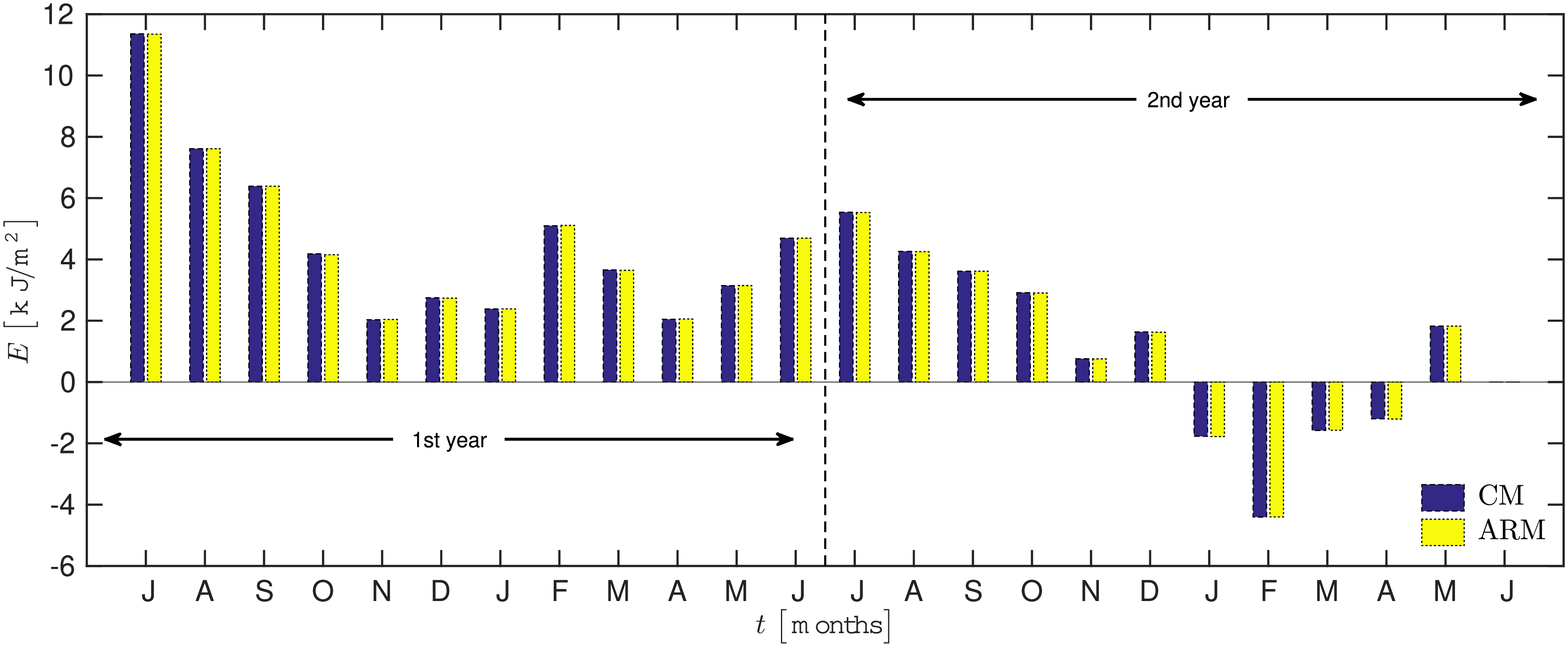}} \\
	\subfigure[]{\includegraphics[width=.45\textwidth]{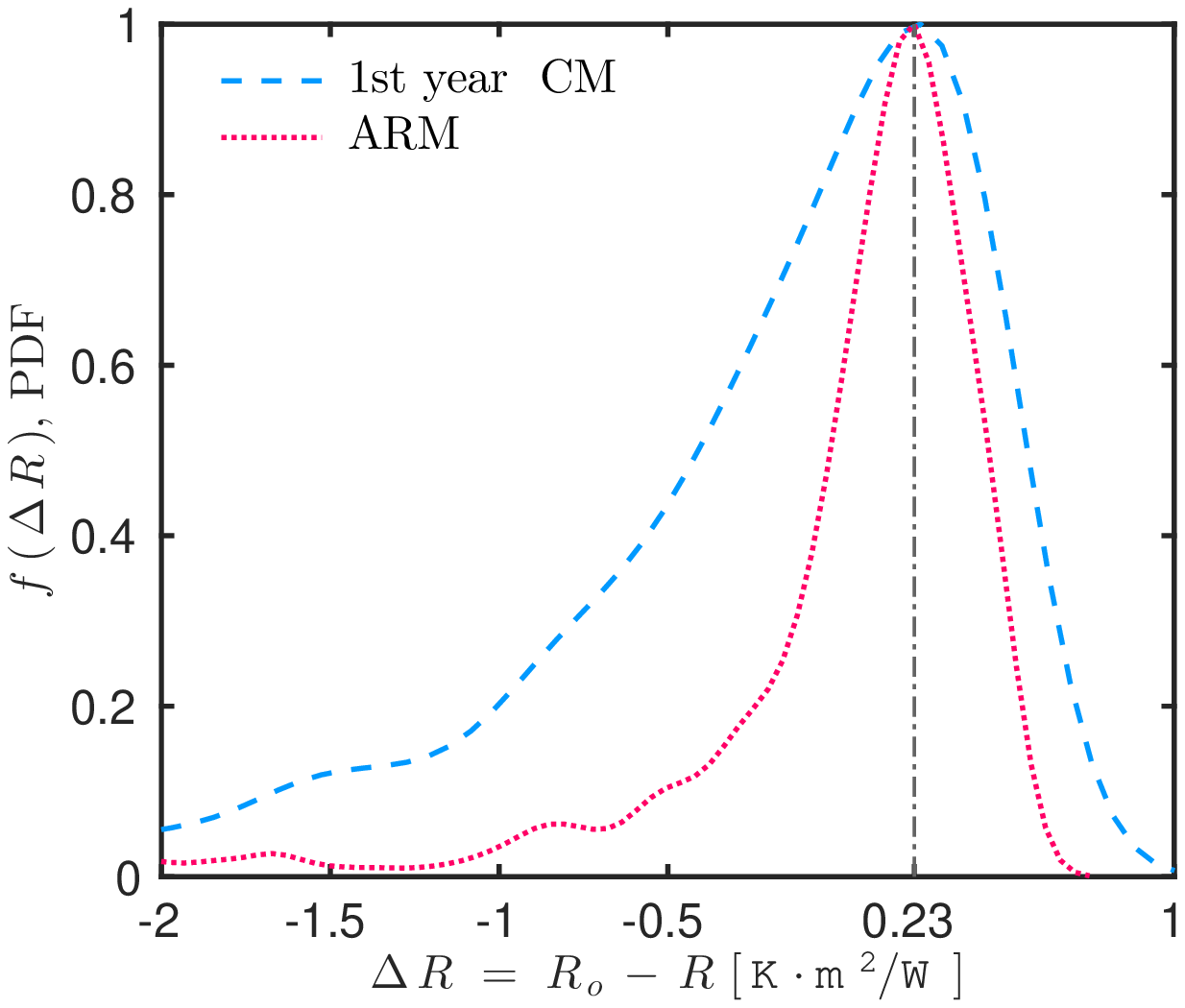}} \hspace{0.2cm}
	\subfigure[]{\includegraphics[width=.45\textwidth]{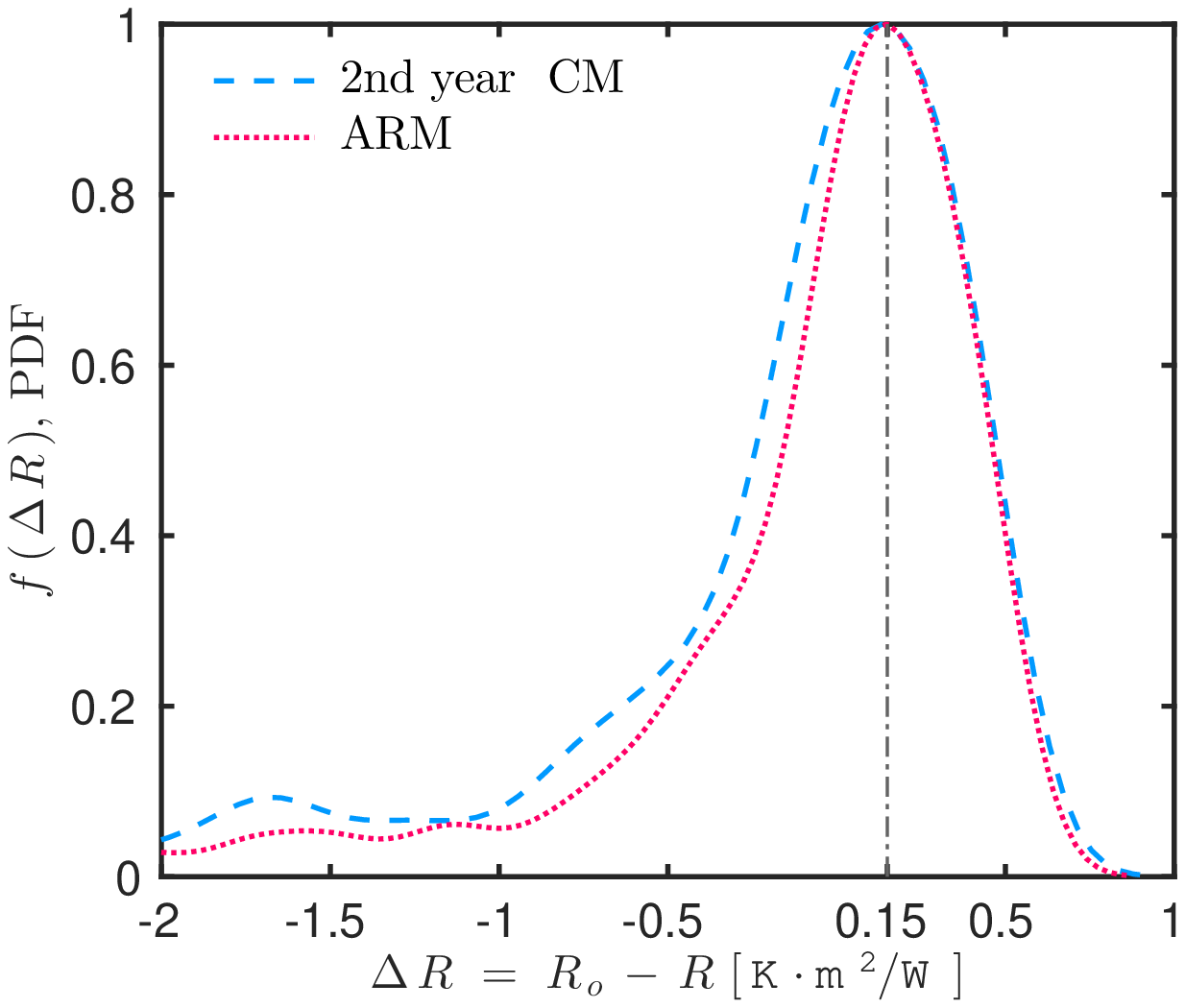}}		
				\caption{\small Conduction loads $E \ \bigl[\,\mathsf{J/m^2}\,\bigr]$ (\emph{a}) and the distribution functions $f$ of the effective averaged thermal resistance $R$ (\emph{b}) calculated for two years with the complete and average reduced models with $\Delta \, t \egal 1 \, \sf{h}$ and $\tau \egal 6 \, \sf{h}$.}
				\label{fig:HM_ER_MI_2years_dt}
			\end{figure}
From the Figure~\ref{fig:HM_ER_MI_2years_dt} (\emph{a}) one can observe that both complete and average reduced models can predict the conductions loads $E$ of the wall throughout two years. 
The predictions are accurate even with the time step size $\Delta \, t \egal 1 \, \sf{h}$ for both models. 
The ARM is time-averaged with $\tau \egal 6 \, \sf{h}$ as was discussed earlier. 
Instead of plotting the time evolution of the thermal resistance, the distribution functions $f$ of the effective averaged thermal resistance $R$ in comparison with the standard thermal resistance of the material $R_{\, o}$ is presented. 
One of the purposes of such comparison lies in the fact, that the value of $R_{\, o}$ is used in regulatory standards. 
As the $R_{\, o}$ is usually evaluated based on constant material properties, there might be some overestimations. 
The latter can lead to misinterpretations of the general behavior of the wall and cause any further related difficulties. 
Moreover, the given case study concerns the rammed earth wall which passes through the drying process at the beginning of installation. 
Therefore, the thermal resistance needs to be compared for each year separately.   
From the Figures~\ref{fig:HM_ER_MI_2years_dt} (\emph{b} - \emph{c}) it can be seen that, in general, the overestimation of the phenomenon is observed. 
These results coincide with the previous observations for heat and mass content.  
Additionally, the discrepancy with the standard value is less for the second year. 
This can be explained with the more stable conditions of the wall itself. 
To sum up, this study has shown that the time-averaging methodology is relevant and applicable to perform efficient studies of complex physical phenomena.   
\section{Conclusion}\label{sec:conclusion}
Various simulation tools are widely used to predict the behavior of building components. 
However, the classical numerical approaches have certain limitations mainly based on the choice of a computational time grid. 
One can distinguish two types of restrictions as a numerical and physical. 
The former one can be relaxed with innovative numerical methods, one of them being the Super--Time--Stepping (STS) \citep{alexiades1996, meyer2014, abdykarim2019}. 
The design of the method permits to attain explicit formulation while overcoming stability restriction of standard ones. 
In this way, a tolerable accuracy can be reached with a relatively scarce time grid. 
Nonetheless, the physical restriction limits the choice of a big time-step size.
This article proposes the methodology to overcome it. 

The general idea is to relax the physical restriction on a time scale imposed by a characteristic time of boundary conditions. 
The signal is time-averaged with certain periods and further implemented into simulations. 
Application of an averaged signal allows to overcome the requirement of a fine time grid to model the highly fluctuating boundary conditions.
The solution itself is decomposed as a sum of averaged values and high frequency fluctuating values. 
Such methodology is called as Average Reduced Model (ARM). 

ARM is tested and studied for a heat diffusion equation and for a coupled heat and mass transfer. 
Each of the phenomena is discussed separately in the Sections~\ref{sec:heat_eq} and \ref{sec:HM_transfer}, which follow the similar structure. 
After describing the complete mathematical model in both physical and dimensionless formulations, the methodology of the ARM is presented. 
The latter is given in more details for the case of the heat diffusion equation. 
The numerical method, implemented in this article, is the STS \textsc{Runge--Kutta--Legendre} method of the first order \citep{meyer2012, abdykarim2019}. 
Case studies are divided into offline and online procedure parts, that enables to understand the advantages of the ARM. 
 
The results of the case studies, presented in the Sections~\ref{sec:case_study_heat} and \ref{sec:case_study_HM}, show benefits of the ARM. 
First of all, it is proved that the methodology is accurate and reliable to model both simple and coupled phenomena. 
In addition to that, the physical phenomena such as heat flux, conduction loads and thermal resistance can be predicted accurately for long-term studies. 
Secondly, the computational effort can be cut considerably. 
It is worth mentioning that the numerical scheme itself permits to save almost $65 \, \%$ of the CPU time in comparison to standard explicit scheme \citep{abdykarim2019}. 
In this article, it is shown that the ARM permits to obtain additional CPU time savings, summing to $50 \, \%$ for heat transfer and almost $74 \, \%$ for heat and mass transfer.  
The efficiency of the ARM to model complex phenomena is proved by its ability to preserve the tolerable rate of error in a much shorter computational time. 
Lastly, the time-averaging of boundary conditions and their consequent implementation into the numerical model with the time-step size $\Delta \, t \egal 1 \, \sf{h}$ justifies the initial purpose of the methodology. 
It can be clearly seen, that it is possible to overcome the restriction on the time-step size not only numerically and also physically without losing important information. 

To sum up, the article proposes substantial study of an innovative methodology which confirms the possibility to implement scarce time grids into numerical models of heat and mass transfer in building porous material without breaking the physical reality of the phenomena. 
Further studies should be carried out in higher dimensions, for different materials and boundary conditions.  
\section*{Acknowledgements}
This work was partly funded by the French Environment and Energy Management Agency (ADEME), Technical Center for Buildings (CSTB) and Saint Gobain Isover. 
The authors also acknowledge the Junior Chair Research program {\it "Building performance assessment, evaluation and enhancement "} from the University of Savoie Mont Blanc in collaboration with the French Atomic and Alternative Energy Center (INES/CEA) and Scientific and Technical Center for Building (CSTB). 
The authors also would like to thank Dr. \textsc{A. Fabbri} and Dr. \textsc{L. Soudani} for valued discussions on the experimental data. 

\section*{Data availability}
The data that support the findings of this study are available on request from the corresponding author. The data are not publicly available due to privacy restrictions.
\bibliographystyle{plainnat} 
\bibliography{biblio}
\end{document}